\documentclass[11pt,leqno,fleqn]{article}

\usepackage{tikz}
   
\usetikzlibrary{calc,automata,patterns,decorations,%
    decorations.pathmorphing,decorations.markings}
\usetikzlibrary{fadings}
\tikzset{>=latex}
\tikzset{->-/.style={decoration={
  markings,
  mark=at position #1 with {\arrow{>}}},postaction={decorate}}}

\usepackage{latexsym}
\usepackage{amsmath}
\usepackage{amsthm}
\usepackage{amssymb}
\usepackage{amsfonts}
\usepackage{alltt}

\usepackage[linesnumbered,ruled,vlined]{algorithm2e}
\let\oldnl\nl 
\newcommand{\nonl}{\renewcommand{\nl}{\let\nl\oldnl}}

\usepackage{amsbsy}
\usepackage[makeroom]{cancel}  

\usepackage{enumitem} 

\usepackage[colorlinks=true,
            citecolor=blue,filecolor=blue,linkcolor=blue,urlcolor=blue,  
            linktoc=page,
            pagebackref=true]{hyperref} 
\usepackage{url}      
\usepackage{afterpage}  
\usepackage{float} 
\floatstyle{boxed} 

\usepackage{times}
\usepackage{bm}   
\usepackage{mathrsfs}  
\usepackage{graphicx}

\usepackage{fancybox}

\newcommand{\Hide}[1]{}

\newif
\ifnote
\notetrue

\newif
\ifTR
\TRtrue

\newtheorem{theorem}{Theorem}
\newtheorem{lemma}[theorem]{Lemma}
\newtheorem{corollary}[theorem]{Corollary}
\newtheorem{proposition}[theorem]{Proposition}

\newtheorem{restxxx}[theorem]{Restriction}

\newtheorem{agreexxx}[theorem]{Agreement}

\newtheorem{termxxx}[theorem]{Terminology}

\newtheorem{notxxx}[theorem]{Notation}

\newtheorem{assumxxx}[theorem]{Assumption}

\newtheorem{convenxxx}[theorem]{Convention}

\newtheorem{exaxxx}[theorem]{Example}
\newenvironment{example}{\begin{exaxxx}\rm}{\hfill\QED\end{exaxxx}}
\newtheorem{exexxx}[theorem]{Exercise}

\newtheorem{remxxx}[theorem]{Remark}
\newenvironment{remark}{\begin{remxxx}\rm}{\hfill\QED\end{remxxx}}
\newtheorem{openxxx}[theorem]{Open Problem}
\newtheorem{conjxxx}[theorem]{Conjecture}
\newenvironment{conjecture}{\begin{conjxxx}\rm}{\hfill\QED\end{conjxxx}}
\newtheorem{defxxx}[theorem]{Definition}
\newenvironment{definition}[1]{\begin{defxxx}[\emph{#1}]\rm}%
   {\hfill\QED\end{defxxx}}
\newtheorem{defxxxsansQED}[theorem]{Definition}
   {\end{defxxxsansQED}}

\newenvironment{sketch}
{\smallskip\noindent\ignorespaces\textit{Proof Sketch.}}
{\hfill\QED\medskip}
\newtheorem{Prxxx}[theorem]{Proof}
{\end{Prxxx}} 

\newenvironment{custommargins}[2]%
  {\addtolength{\leftskip}{#1}\addtolength{\rightskip}{#2}}{\par}

\usepackage[margin=0pt,justification=centerlast,font=small,format=hang,%
labelfont=bf,up,textfont=rm,up]{caption}

\newcommand{\Set}[1]{\{ #1 \}}
\newcommand{\SET}[1]{\bigl\{ #1 \bigr\}}
\newcommand{\cycles}[1]{\mathsf{cycles}(#1)}
\newcommand{\inmost}[1]{\mathsf{innermost}\text{-}\mathsf{cycle}(#1)}

\newcommand{\vv}{\mathbf{V}}
\newcommand{\ee}{\mathbf{E}}
\newcommand{\G}{{\cal G}}
\newcommand{\A}{{\cal A}}
\newcommand{\B}{{\cal B}}

\newcommand{\HH}{{\cal H}}

\newcommand{\bigO}[1]{{\cal O}\bigl(#1\bigr)} 
\newcommand{\bigOO}[1]{{\cal O}(#1)} 

\newcommand{\Let}[3]%
    {\textbf{\textsf{let}}\ {#1}\,{#2}\ \textbf{\textsf{in}}\;{#3}\,}
\newcommand{\Try}[3]%
    {\textbf{\textsf{try}}\ {#1} {#2}\ \textbf{\textsf{in}}\;{#3}\;}
\newcommand{\Mix}[3]%
    {\textbf{\textsf{mix}}\ {#1} {#2}\ \textbf{\textsf{in}}\;{#3}\;}
\newcommand{\LET}[3]%
    {\textbf{\textsf{let}}^{\bm{*}}\ {#1} {#2}\ \textbf{\textsf{in}}\;{#3}\;}
\newcommand{\Letrec}[3]%
    {\textbf{\textsf{letrec}}\ {#1} {#2}\ \textbf{\textsf{in}}\;{#3}\;}

\newcommand{\bridges}[2]{{\partial}_{\,#1}(#2)}  
\newcommand{\bbridges}[2]{{\partial}_{\,#1}\bigl(#2\bigr)}

\newcommand{\Bridges}[2]{\overline{\partial}_{\,#1}\bigl(#2\bigr)}

\newcommand{\degreeSym}{\mathit{deg}} 

\newcommand{\degr}[2]{{\degreeSym}_{#1}(#2)}

\newcommand{\ie}{\textit{i.e.}}
\newcommand{\eg}{\textit{e.g.}}
\newcommand{\QED}{{\Large $\square$}} 

\newcommand{\nats}{\mathbb{N}}

\newcommand{\nreals}{\mathbb{R}_{+}}

\newcommand{\size}[1]{|\,#1\,|}  
\newcommand{\ssize}[1]{\bigl|\,#1\,\bigr|}

\newcommand{\set}[1]{\overline{#1}}

\newcommand{\OutF}[1]{\mathsf{OuterFace}(#1)}
\newcommand{\OutPlan}[2]{{{#1}\text{-}\mathsf{outerplanarity}}(#2)}

\newcommand{\merge}[2]{ \mathsf{merge}(#1,#2) }

\newcommand{\CC}{\mathscr{C}}
\newcommand{\DD}{\mathscr{D}}

\newcommand{\levelSym}{\mathsf{level}}
\newcommand{\level}[1]{\levelSym(#1)}

\newcommand{\KS}{\mathsf{KS}}

\begin{document}


\setcounter{page}{1}     
\setcounter{tocdepth}{1} 
\ifTR
  \pagenumbering{roman} 
\else
\fi

\title{Efficient Reassembling of Three-Regular Planar Graphs} 
\author{Assaf Kfoury%
           \thanks{Partially supported by NSF awards CCF-0820138
           and CNS-1135722.} \\
        Boston University \\
        \ifTR Boston, Massachusetts \\ 
        \href{mailto:kfoury@bu.edu}{kfoury{@}bu.edu}
        \else \fi
\and   Benjamin Sisson%
           \footnotemark[1] \\
        Boston University   \\
        \ifTR Boston, Massachusetts \\ 
        \href{mailto:bmsisson@bu.edu}{bmsisson{@}bu.edu}
        \else \fi
}

\ifTR
   \date{\today}
\else
   \date{} %
\fi
\maketitle
  \ifTR
     \thispagestyle{empty} 
  \else
  \fi

\ifTR
    \tableofcontents
    \newpage
\else
    \vspace{-.2in}
\fi

  \begin{abstract}
  \addcontentsline{toc}{section}{Abstract}

\noindent
A \emph{reassembling} of a simple graph $G = (V,E)$ is an abstraction
of a problem arising in earlier studies of network
analysis~\cite{BestKfoury:dsl11,Kfoury:sblp11,Kfoury:SCP2014,%
SouleBestKfouryLapets:eoolt11}. There are
several equivalent definitions of graph reassembling; in this report we
use a definition which makes it closest to the notion of \emph{graph
  carving}. A reassembling is a rooted binary tree whose nodes are subsets of
$V$ and whose leaf nodes are singleton sets, with each of the latter
containing a distinct vertex of $G$.  The parent of two nodes in the
reassembling is the union of the two children's vertex sets. The root
node of the reassembling is the full set $V$. The \emph{edge-boundary
  degree} of a node in the reassembling is the number of edges in $G$
that connect vertices in the node's set to vertices not in the
node's set. A reassembling's \emph{$\alpha$-measure} is the largest edge-boundary
degree of any node in the reassembling. A reassembling of $G$ is
\emph{$\alpha$-optimal} if its $\alpha$-measure is the minimum among all
$\alpha$-measures of $G$'s reassemblings.

\medskip
\noindent
The problem of finding an $\alpha$-optimal reassembling of a
simple graph in general was already shown to be
NP-hard~\cite[among others]{kfoury+mirzaei:2017,kfoury+mirzaei:2017B}.
\Hide{
  In previous papers, graph reassembling and minimum-cutwidth linear
arrangement (minimum cutwidth graph carving) were shown to be
polynomial reducible to each other. In the general case, determining
whether a particular graph has a carving width of at most k is an
NP-complete problem. Carving widths of size $k$ and reassemblies with
alpha measures of $k$ are also polynomial reducible to each other.}

\medskip
\noindent
In this report we present an algorithm which, given a $3$-regular
plane graph $G = (V,E)$ as input, returns a reassembling of $G$ with
an $\alpha$-measure independent of $n = \size{V}$ and upper-bounded by
$2k$, where $k$ is the \emph{edge-outerplanarity} of $G$.
(Edge-outerplanarity is distinct but closely related to the usual
notion of outerplanarity; as with outerplanarity, for a fixed
edge-outerplanarity $k$, the number $n$ of vertices can be arbitrarily
large.) Our algorithm runs in linear time $\bigOO{n}$.  Moreover, we
construct a class of $3$-regular plane graphs for which this
$\alpha$-measure is optimal, by proving that $2k$ is the lower bound on
the $\alpha$-measure of any reassembling of a graph in that class.


  \end{abstract}

  \newpage
  \pagenumbering{arabic}  

\section{Introduction}
\label{sect:intro}
We skip repeating the informal definition of \emph{graph reassembling}
given in the abstract above; it is further elaborated and made
formally precise in Section~\ref{sect:reassembling}, where we also
spell out the connection with \emph{graph carving}.

\textbf{Background and Motivation.}
Besides questions of optimization and the variations which it
naturally suggests, \emph{graph reassembling} is an abstraction 
of an operation carried out by programs in a domain-specific language
for the design of flow networks
\cite{BestKfoury:dsl11,Kfoury:sblp11,Kfoury:SCP2014,%
SouleBestKfouryLapets:eoolt11}. 
In network \emph{reassembling}, the network is taken apart 
and reassembled in an order determined by the designer.

Underlying a flow network is a directed graph, where vertices and edges are
assigned various attributes that regulate flow through the network.%
   \footnote{Such networks are typically more complex 
             than the capacited directed graphs that algorithms
             for max-flow (and other related quantities) and its
             generalizations (\eg, multicommodity max-flow) operate on.}
Programs for flow-network design are meant to connect network
components in such a way that \emph{typings at their interfaces},
\ie, formally specified properties at their common boundaries, 
are satisfied. Network typings guarantee there are no conflicting
data types when different components are connected, and insure that
desirable properties of safe and secure operation are not violated by
these connections, \ie, they are \emph{invariant properties} of the
whole network construction.

A typing $\tau$ for a \emph{network component} $X$ (or \emph{vertex cluster}
$X$ in this report's terminology) formally expresses a constraining
relationship between the variables denoting the outer ports of $X$ (or
the edge-boundary $\bridges{}{X}$ in this report). The smaller the set
of outer ports of $X$ is, the easier it is to formulate the typing
$\tau$ and to test whether it is compatible with the typing $\tau'$ of
another network component $Y$. Although every outer port of $X$ is
directed, as input port or output port, the complexity of the
formulation of $\tau$ depends only on the number of outer ports (or
$\size{\bridges{}{X}}$ in this report), not on their directions. 

If $\delta$ is a uniform upper bound on the number of outer ports of all
network components, the time complexity of reassembling the network
without violating any component typing $\tau$ can be made linear in
the size $n$ of the completed network and exponential in the bound $\delta$
-- not counting the pre-processing time $f(n)$ to determine an
appropriate reassembling order. Hence, the smaller are $\delta$ and $f(n)$,
the more efficient is the construction of the entire network. From
this follows the importance of minimizing the pre-processing time
$f(n)$ for finding a reassembling strategy that also minimizes the
bound $\delta$ (or \emph{$\alpha$-measure} in this report). If the reassembling
order minimizes the quantity $\delta$ among all possible reassemblings,
we say that the reassembling is \emph{$\alpha$-optimal}.

\textbf{Main Results.}
Let $G$ be the underlying graph of a flow network as described above,
where we ignore direction on edges.  While the problem of finding an
$\alpha$-optimal reassembling of $G$ in general is NP-hard~\cite[among
  others]{kfoury+mirzaei:2017,kfoury+mirzaei:2017B}, we show in this
report that the problem is solvable in linear time for $3$-regular
planar graphs within an upper bound on the $\alpha$-measure which is
independent of graph size. Specifically, we prove the existence of a linear-time
algorithm which, given an arbitrary $3$-regular plane graph
$G = (V,E)$, returns a reassembling of $G$ with an $\alpha$-measure
$\leqslant 2k$ and independent of $n = \size{V}$, where $k$ is
the \emph{edge-outerplanarity} of $G$. The significance of the parameter
$k$ comes from the fact that $n$ does not depend on it, and indeed,
for a fixed value of \emph{edge-outerplanarity}, the size $n$ of $G$
can be arbitrarily large.

Although our algorithm does not return $\alpha$-optimal reassemblings
for all $3$-regular plane graphs in general, we prove that it does return
$\alpha$-optimal reassemblings for a significant class of $3$-regular
plane graphs. This class of graphs satisfies a certain ``high density
condition'' (spelled out clearly in Section~\ref{sect:optimality-of-KS});
given any $3$-regular plane graph $G = (V,E)$ not satisfying this ``density
condition'', our algorithm may return a sub-optimal reassembling, though
its $\alpha$-measure is guaranteed not to exceed twice the graph's
\emph{edge-outerplanarity} $k$.

\textbf{Organization of the Report.}
Sections~\ref{sect:reassembling}
and~\ref{sect:plane-and-planar} are background material. They set the stage
for the rest of the report, making precise many of the terms
we use throughtout. Several of the concepts are closely related to familiar
ones, \eg, \emph{edge outerplanarity} in connection with standard
\emph{outerplanarity} (herein called \emph{vertex outerplanarity}),
and their differences are clearly explained as they make a difference
in our analysis.

Our algorithm, which we call $\KS$ for lack of a better
name, is presented in Section~\ref{sect:algorithm-KS}. It proceeds
by a long exhaustive (and exhausting!) case analysis of possible
configurations of subgraphs in $3$-regular plane graphs.
Section~\ref{sect:algorithm-KS} also
includes a proof of correctness and a complexity analysis of $\KS$;
the proof is elementary in that it does not invoke any deep theorem
from elsewhere in graph theory. The pseudocode of $\KS$ is included
in Appendix~\ref{appendix:pseudocode}, and a full Python implementation
is available for download from the website
\href{http://cs-people.bu.edu/bmsisson/}{Graph Reassembling}.%
   \footnote{\texttt{http://cs-people.bu.edu/bmsisson/}}

Section~\ref{sect:optimality-of-KS} defines a class of $3$-regular plane graphs for
which our algorithm $\KS$ returns $\alpha$-optimal reassemblings.
Section~\ref{sect:optimality-of-KS} also defines a ``density condition''
on the topology of $3$-regular plane graphs which, if satisfied, guarantees
that the returned reassemblings are $\alpha$-optimal.

The concluding Section~\ref{sect:conclusion} explains how results on
\emph{graph carving} can be transferred to results on \emph{graph
  reassembling}, and vice-versa. Outside the forementioned
  problems of network design and analysis, Section~\ref{sect:conclusion}
  also includes an application of our results to a flow problem,
  namely, the existence of a fixed-parameter linear-time algorithm
  for maximum flow in planar graphs in general (not restricted to $3$-regular).


\section{Graph Reassembling}
\label{sect:reassembling}

We refer to the vertices and edges of a graph $G$ by writing $\vv(G)$
and $\ee(G)$. If $G$ is simple, an edge is uniquely identified by the
two-element set of its endpoints $\Set{v,w}$, which we also write as
$\set{v\,w}$.

There are several equivalent definitions of graph
reassembling~\cite{kfoury+mirzaei:2017}. We here
use a definition which makes it closest to the notion of graph 
\emph{carving}~\cite{seymour1994call} and requires the preliminary notion
of a binary tree, also defined in a way that makes the connection with carving
easier.%
  \footnote{Our definition of \emph{binary tree} is unusual but more
     convenient for our analysis. It is the same as 
     \emph{full binary merging} in~\cite{qian-ping-gu2008}.}

We take a (rooted, unordered) \emph{binary tree} $\B$ over a set
$V=\Set{v_1,\ldots,v_n}$ where $n\geqslant 1$ to be a collection of
$(2n-1)$ non-empty subsets of $V$ -- these are the nodes of $\B$ --
satisfying three conditions:
\begin{enumerate}[itemsep=0pt,parsep=2pt,topsep=5pt,partopsep=0pt] 
\item For every $v\in V$, the singleton set $\Set{v}$ is in $\B$. 
      These are the $n$ \emph{leaf} nodes of $\B$.
\item The full set $V$ is in $\B$. This is the \emph{root} node of $\B$.
\item Every node $X\in\B - \Set{V}$ other than the root has a unique 
      sibling $Y\in\B-\Set{V}$ 
      such that: $X\cap Y=\varnothing$ and $(X\cup Y)\in\B$. 
      This is a property of every of the $(2n - 2)$ nodes that are not the root.
\end{enumerate}
We also call $\B$ a \emph{binary reassembling} of $V$, or also 
a \emph{binary reassembling} of $G$ when $V = \vv(G)$.%
   \footnote{A binary reassembling in our sense mimicks what is called 
   ``agglomerative, or bottom-up, hierarchical clustering'' in data mining.}
To denote the \emph{reassembling of graph $G$} according to
$\B$, we write $(G,\B)$. Depending on the context, we may
refer to the nodes of $\B$ as \emph{vertex clusters} of $G$.%
    \footnote{To keep them apart, we reserve the words ``node'' and ``branch''
              for the tree $\B$, and the words ``vertex''
              and ``edge'' for the graph $G$.}

If $X$ and $Y$ are disjoint subsets of $\vv(G)$,
we write ${\bridges{G}{X,Y}}$ to denote the set of all edges connecting
$X$ and $Y$, each such edge with one endpoint in $X$ and one endpoint in $Y$:
\[
   {\bridges{G}{X,Y}}\ \triangleq
   \ \Set{\,\set{v\,w}\in\ee(G)\,|\,v\in X\text{ and } w\in Y\,}.
\]
If $G$ is clear from the context, we just write ${\bridges{}{X,Y}}$.
We also write ${\bridges{}{X}}$ to denote ${\bridges{}{X,\ee(G)-X}}$.

There are different ways of optimizing the reassembling of $G$,
depending on the measure we choose on it. For $X\subseteq\vv(G)$, the
\emph{edge-boundary size} of $X$ in $G$ is 
$\Bridges{G}{X} \triangleq \size{\bridges{G}{X}}$. If $G$ is clear from the
context, we write $\Bridges{}{X}$ instead of $\Bridges{G}{X}$.
If $X$ is a singleton set $\Set{v}$, then
$\Bridges{}{\Set{v}}$ is simply $\degr{}{v}$, the degree of $v$.  
What we call the \emph{$\alpha$-measure} of the reassembling $(G,\B)$
is defined by:
\[
    \alpha(G,\B) \triangleq 
     \ \max\, \SET{\,\Bridges{}{X}\; \bigl| \; X\in\B \,}.
\]
We say the reassembling $(G,\B)$ is \emph{$\alpha$-optimal} iff: 
\[
  \alpha (G,\B)\ =\ \min\,
  \SET{\,\alpha (G,{\B}')\;\bigl|
  \;{\B}'\text{ is a binary reassembling of $G$}\,} .
\]
For other ways of optimizing graph reassembling relative
to other measures, consult the earlier~\cite{kfoury+mirzaei:2017},
none used in this report. 

\subsection{Connections with Graph Carving}
\label{sect:graph-carving}

Graph reassembling is essentially a different name for
\emph{graph carving} \cite{seymour1994call}, although the 
former is perhaps better understood as a bottom-up process (of rebuilding a
graph back to its original form) whereas the latter is a top-bottom
process (of repeatedly bipartitioning a graph's vertex set). To be
more precise, a carving is defined relative to a
\emph{routing tree} (sometimes called a \emph{call-routing tree})
$T$ for $G$, which is an \emph{unrooted} tree where 
every internal node has degree $= 3$.  The number of leaf nodes in $T$
is $n = \size{\vv(G)}$, the number of internal nodes in $T$ is
$(n-2)$, and the number of branches in $T$ is $(2n-3)$.  The leaf
nodes of $T$ correspond to the vertices of $G$ and, for every branch
$e$ of $T$, deleting $e$ yields two trees whose leaf nodes define a
bipartition of the vertices of $G$; we say that the edge cut in $G$
corresponding to this bipartition is \emph{induced} by $e$.

\newcommand{\width}[1]{\mathit{width}(#1)}
\newcommand{\cw}[1]{\mathit{cw}(#1)}

To refer to the carving of $G$ relative to the routing tree
$T$, we here write $(G,T)$. The measure $\width{G,T}$ 
is the maximum size of an edge cut in $G$ that
is induced by a branch $e$ of $T$. The \emph{carving-width} of $G$,
denoted $\cw{G}$, is the minimum width of all possible carvings of $G$:
\[
   \cw{G}\ \triangleq\ \min\,\SET{\,\width{G,T}\;
        \bigl|\;\text{$T$ is a routing tree for $G$}\,} .
\]
We say the carving $(G,T)$ is \emph{optimal} iff  $\width{G,T} = \cw{G}$.

Given a carving $(G,T)$, we obtain a reassembling $(G,\B)$ by turning
$T$ into a rooted binary tree; namely, by introducing a fresh node to
be the root labelled with the entire set $V = \vv(G)$, deleting one of
the branches $\set{X\,Y}$ of $T$, and introducing two new branches:
from the root (now labelled with $V$) to each of $X$ and $Y$.  We
associate with every internal node $X$ the union of the two vertex
sets associated with $X$'s two children.  There are $(2n-3)$ different
ways of obtaining such $(G,\B)$ from $(G,T)$, one for each branch in
$T$. We say $(G,\B)$ is one of the binary reassemblings \emph{induced}
by the carving $(G,T)$. Note that $(G,\B)$ may or may not be
$\alpha$-optimal, even if the carving $(G,T)$ from which it is induced
is optimal; however, at least one of the $(2n-3)$ reassemblings
$(G,\B)$ induced by an optimal carving $(G,T)$ is $\alpha$-optimal.

Conversely, given a reassembling $(G,\B)$ of $G$, where $\B$ has $n$
leaf nodes, $(n-1)$ internal nodes, and $(2n-2)$ branches, we
(uniquely) obtain a carving $(G,T)$ by deleting the two branches from
the root node of $\B$ to its two children $X$ and $Y$, deleting the
root node of $\B$, and introducing a new branch $\set{X\,Y}$. We now
ignore the association of every internal node of $\B$ with a subset of
$\vv(G)$, but the one-one correspondence between $\B$'s leaf nodes and
$\vv(G)$ is preserved. We say the reassembling $(G,\B)$
(uniquely) \emph{induces} the carving $(G,T)$. It is easy to see that
if $(G,\B)$ is $\alpha$-optimal, then the induced carving $(G,T)$ is
optimal.

\medskip
The following is a consequence of the preceding discussion.

\begin{proposition} 
\label{fact:reassembling+carving}
\label{prop:reassembling+carving}
For an arbitrary graph $G$, the following hold:
\begin{enumerate}[itemsep=1pt,parsep=2pt,topsep=2pt,partopsep=0pt] 
\item If $(G,T)$ is an optimal carving, 
      then $(G,T)$ induces (not uniquely) an $\alpha$-optimal reassembling $(G,\B)$.
\item If $(G,\B)$ is an $\alpha$-optimal reassembling,
      then $(G,\B)$ induces (uniquely) an optimal carving $(G,T)$. 
\end{enumerate}
The construction of $(G,T)$ from $(G,\B)$ is carried out in constant time, 
the contruction of $(G,\B)$ from $(G,T)$ is carried out in linear
time.
\end{proposition} 

\subsection{Further Preliminary Notions}
\label{sect:further-preliminary}

In the opening paragraphs of Section~\ref{sect:reassembling} we
restricted reassemblings to simple graphs (no multi-edges, no
self-loops).  The definitions can be generalized in the obvious way
to multigraphs, where multi-edges and self-loops are allowed, which
will be encountered in Section~\ref{sect:algorithm-KS} and later. However,
we agree that the degree of a vertex $v$ in a multigraph $G$
omits self-loops, \ie, 
\[
   \degr{}{v} \triangleq
   \ssize{\Set{\,w\,|\,\text{there is $e\in\ee(G)$ whose endpoints
          are $\Set{v,w}$ with $v\neq w$}}} .
\]
Let ${\B}_1$ and ${\B}_2$ be binary reassemblings of the sets $V_1$ and
$V_2$, respectively. By our definition, $V_1$ and $V_2$ are
the \emph{roots} of ${\B}_1$ and ${\B}_2$.  
If $V_1\cap V_2 =\varnothing$, we can construct a new binary
reassembling ${\B}$ whose root is $V_1\cup V_2$, whose two children
are ${\B}_1$ and ${\B}_2$. 

The earlier definition of graph reassembling $(G,\B)$ is in fact
a \emph{total reassembling} because $\B$ is defined over the full
set of vertices $\vv(G)$. If $V'$ is a subset of $\vv(G)$ and
$\B'$ is a binary tree over $V'$, then $(G,\B')$ is a
\emph{partial reassembling} of $G$. A total reassembling is
a special case of a partial reassembling. The notions of `total
reassembling' and `partial reassembling' apply equally well to
multigraphs. Unless stated otherwise, `reassembling' means `total
reassembling', and `graph' means `simple graph'.

The next proposition is used in several places later in this report.
Let $\B$ be a binary reassembling of a simple graph $G$, with $V = \vv(G)$.
Consider a node $X\in \B-\Set{V}$ other than the root $V$, and its
sibling $Y\in \B-\Set{V}$, so that also $X\cap Y =\varnothing$. We write
$\merge{X}{Y}$ to denote the node $(X\cup Y)\in\B$ which is the common
parent of $X$ and $Y$. We say $\merge{X}{Y}$ has \emph{degree $m$} iff
$\Bridges{}{X,Y} = m \geqslant 0$.

\begin{proposition}
\label{lem:no-degree-0-merges}
\label{prop:no-degree-0-merges}
If $\B$ is a binary reassembling of a simple connected graph $G$, then there is
a binary reassembling $\B'$ of $G$ such that:
\begin{itemize}[itemsep=0pt,parsep=2pt,topsep=5pt,partopsep=0pt]
\item every $\merge{X}{Y} = (X\cup Y)$ of two sibling nodes in $X, Y\in {\B}'$ has
      degree $\geqslant 1$, and
\item $\alpha (G,{\B}') \leqslant \alpha (G,{\B})$.
\end{itemize}
In words, we can assume that, whenever two vertex clusters $X$ and $Y$
are merged in a reassembling, there is at least one edge connecting
$X$ and $Y$; put differently, merging two vertex clusters $X$ and $Y$ that
do not have at least one edge connecting them is a wasted step in the
reassembling.
\end{proposition}

\begin{proof}
Suppose the given binary reassembling ${\B}$ contains $p\geqslant 1$
degree-$0$ merges of two siblings.
It suffices to show how we can construct a new binary reassembling ${\B}'$
from ${\B}$ such that ${\B}'$ contains $(p-1)$ degree-$0$ merges of two siblings.

Consider a degree-$0$ merge of two siblings $X_1,X_2\in\B$ in the binary tree $\B$ 
such that every node/merge containing $\merge{X_1}{X_2} = (X_1\cup X_2)$ has
degree $\geqslant 1$, \ie, $\merge{X_1}{X_2}$ is the closest to the root
$\vv(G)$ among all degree-$0$ merges in $\B$. Let $X_3, X_4, \ldots, X_{\ell}$ be
the nodes in $\B$ such that:
\begin{itemize}[itemsep=0pt,parsep=5pt,topsep=5pt,partopsep=0pt]
\item[] $X_3$ is the sibling of $(X_1\cup X_2)$,
\item[] $X_4$ is the sibling of $(X_1\cup X_2\cup X_3)$,
\item[] $\ldots$
\item[] $X_{\ell}$ is the sibling of $(X_1\cup X_2\cup X_3\cup\cdots\cup X_{{\ell}-1})$,
\end{itemize}
and $X_1\cup X_2\cup \cdots \cup X_{\ell} = V$. By assumption, the merge
$(X_1\cup X_2\cup\cdots\cup X_{k})$ has degree $\geqslant 1$ for every
$k \geqslant 3$. Define the quantity $m_{i,j}$ as follows:
\[
    m_{i,j} \ \triangleq\ \Bridges{}{X_i,X_j}
\]
for all $1\leqslant i < j \leqslant \ell$. By assumption $m_{1,2} = 0$, and
$m_{1,3} \geqslant 1$ or $m_{2,3} \geqslant 1$ (or both). With no loss of generality,
let $m_{1,3} \geqslant 1$ and let $k\geqslant 3$ be the smallest index such
that $m_{2,k} \geqslant 1$. Such an index $k$ must exist since $G$ is connected.
The desired $\B'$ is obtained by re-arranging the nodes/merges above $X_1$
as follows:
\begin{itemize}[itemsep=0pt,parsep=5pt,topsep=5pt,partopsep=0pt]
\item[] $X_2$ is merged with $X_k$ (bypassing $X_3,\ldots,X_{k-1}$),
\item[] $X_1$ is merged with $X_3$ (bypassing $X_2$),
\item[] $(X_1\cup X_3)$ is merged with $X_4$,
\item[] $\ldots$
\item[] $(X_1\cup X_3\cup\cdots\cup X_{k-2})$ is merged with $X_{k-1}$,
\item[] $(X_1\cup X_3\cup\cdots\cup X_{k-1})$ is merged with $(X_2\cup X_{k})$,
\item[] $(X_1\cup X_2 \cup X_3\cup\cdots\cup X_{k})$ is merged with $X_{k+1}$,
\item[] $\ldots$ .
\end{itemize}
It is now easy to check that the number of $0$-degree merges in $\B'$
is one less than the number of $0$-degree merges in $\B$ and that 
$\alpha (G,{\B}') \leqslant \alpha (G,{\B})$.
\end{proof}

\section{Plane and Planar Graphs}
\label{sect:plane-and-planar}

We recall standard notions and properties of planar graphs, some adapted to
our own needs. These are used explicitly in later sections or else explain
the background in which to place our analysis.

A graph $G$ is a \emph{plane graph} if it is drawn on the plane
without any edge crossings. A graph $G$ is a \emph{planar graph} if
it is isomorphic to a plane graph; \ie, it is embeddable in the plane
in such a way that its edges intersect only at their endpoints.

Given a plane graph $G$, a \emph{face} is a maximal region $F$ of the 
plane such that $x, y \in F$ implies that $x$ and $y$ can be joined by
a curve which does not meet any edge of $G$. The unique
unbounded face of $G$ is the \emph{exterior}, or \emph{outer face},
of $G$. The edges of $G$ that are incident with a face $F$ form
the \emph{boundary} of $F$.

We distinguish between two kinds of edges,
\emph{bounding} and \emph{non-bounding}. An edge $e$ of a plane graph $G$
is \emph{bounding} if $e$ is on the boundary of two adjacent faces of $G$,
otherwise $e$ is \emph{non-bounding}. Bounding and non-bounding edges
are illustrated in Figure~\ref{fig:non-regular-plane-graph} and again
in Figure~\ref{fig:three-regular-plane-graph}.

\begin{proposition}
\label{prop:bounding-vs-bounding}
If $G$ is a biconnected plane graph, then every edge in $\ee(G)$ is a
bounding edge and there are no non-bounding edges in $G$.
\end{proposition}

\begin{sketch} 
If $G$ is biconnected, then every vertex is on a cycle.
\Hide{
More
specifically, from an arbitrary vertex $v_0$, start a walk along any
edge $\set{v_0\,v_1}$. From $v_1$, continue the walk to a vertex $v_2$
such that $\set{v_1\,v_2}$ is the clockwise neighbor of
$\set{v_0\,v_1}$;
\ie, by rotating $\set{v_0\,v_1}$ clockwise around $v_1$, we can align
the two edges $\set{v_0\,v_1}$ and $\set{v_1\,v_2}$ without
encountering any edge with $v_1$ as one of its endpoints. From $v_2$,
continue to a vertex $v_3$ such that $\set{v_2\,v_3}$ is the clockwise
neighbor of $\set{v_1\,v_2}$, and again from $v_3$ to $v_4$ in a
similar way, etc. Because every vertex has degree $\geqslant 2$, the
walk is bound to reach the initial vertex $v_0$ after traversing
finitely many edges.
}
If every vertex is on a cycle, it is easy to see
that every edge is a bounding edge. Obvious details omitted.
\end{sketch} 

A natural measure on a planar graph is its outerplanarity
index. Informally, if a planar graph $G$ is given with one of its
plane embeddings $G'$, then the \emph{outerplanarity} of $G'$ (not that of
$G$) is the number of times that all the vertices on the outer face
(together with all their incident edges) have to be removed in order
to obtain the empty graph. The \emph{outerplanarity index} of $G$ is the
minimum of the outerplanarities of all the plane embeddings $G'$ of
$G$. Deciding whether an arbitrary graph is planar can be carried out in
linear time $\bigOO{n}$ and, if it is planar, a plane embedding of it 
can also be carried out in linear time~\cite{patrignani2013}. 
Given a planar graph $G$, the outerplanarity index $k$ of $G$ and a 
$k$-outerplanar embedding of $G$ in the plane can be computed in
time $\bigOO{n^2}$, and a $4$-approximation of
its outerplanarity index can be computed in linear time~\cite{Kammer2007}.

For easier bookkeeping, we use a modified definition
of outerplanarity, called \emph{edge-outerplanarity} and already
used by others~\cite{bentz2009}. There is a close relationship between this
modified notion and the standard notion (Theorem 4 in
Section 5.1 in~\cite{bentz2009}). In the case of three-regular plane graphs,
the relationship is much easier to state:
\emph{vertex-outerplanarity} (the standard notion of \emph{outerplanarity})
and \emph{edge-outerplanarity} are ``almost the same''
(Proposition~\ref{prop:modified} below).

\begin{definition}{Edge Outerplanarity}
\label{def:modified}
Let $G$ be a plane graph. If $\ee(G) =\varnothing$ and
$G$ is a graph of isolated vertices, the \emph{edge outerplanarity} of $G$
$= 0$. If $\ee(G) \neq\varnothing$, we pose $G_0 = G$ and 
define $K_0 = L_0\uplus M_0$ as the set of edges lying on $\OutF{G_0}$,
where the edges in $L_0$ are bounding and the edges in $M_0$ are non-bounding.

For every $i>0$, we define $G_i$ as the plane graph obtained
after deleting all the edges in $K_0 \cup \cdots \cup K_{i-1}$ from the
initial $G$ and $K_i = L_i\uplus M_i$ the set of edges lying on
$\OutF{G_i}$, where the edges in $L_i$ are bounding and the edges in $M_i$
are non-bounding.

The \emph{edge outerplanarity} of $G$, denoted $\OutPlan{E}{G}$, 
is the least integer $k$ such that $G_{k}$ is a graph without edges,
\ie, the edge outerplanarity of $G_{k}$ is $0$. This process of peeling
off the edges lying on the outer face $k$ times produces a $k$-block
partition of $\ee(G)$, namely, $\Set{K_0,\ldots,K_{k-1}}$.%
   \footnote{There is an unessential difference between our definition here and
     the definition in~\cite{bentz2009}. In Section 2.2 of that
     reference, ``a $k$-edge-outerplanar graph is a planar graph
     having an embedding with \emph{at most} $k$ layers of edges.'' In
     our presentation, we limit the definition to plane graphs and say ``a
     $k$-edge-outerplanar plane graph has \emph{exactly} $k$ layers of
     edges.'' Our version simplifies a few things later.}
\end{definition}

An example of the decomposition of a plane graph $G$
according to Definition~\ref{def:modified} is shown in
Figure~\ref{fig:three-regular-plane-graph}.

Let $G$ be a finite simple graph and $K\subseteq \ee(G)$.
Let $G'$ be the subgraph of $G$ defined by:
\begin{alignat*}{8}
   & \ee(G') \ &&=\ && K \quad\text{and}\quad
   && \vv(G')   &&=  && \Set{\,v\;|\; v\in\Set{v_1,v_2}\subseteq\vv(G)
                     \text{ \ and \ }\set{v_1\, v_2}\in K}.
\end{alignat*}
We say $G'$ is the \emph{subgraph of $G$ induced by $K$} and write $G[K]$ to
denote it.

As usual, a \emph{simple cycle} in $G$ is a closed walk with no repeated
vertices. A simple cycle $C$ in $G$ is a \emph{chordless cycle} if
no two distinct vertices on $C$ are connected by an 
edge that does not itself belong to $C$.

A \emph{cactus} (plural: \emph{cacti}) is a connected graph in which
any two simple cycles have at most one vertex in common. Hence, in a
cactus, every simple cycle is chordless, and every cactus is a planar
graph.  An (unrooted) tree, a connected acyclic graph, is a special
case of a cactus.

\begin{proposition} 
\label{prop:partition}
\label{lem:partition}
Let $G$ be a plane graph with $\OutPlan{E}{G} = k$. Let
$\Set{K_0, \ldots, K_{k-1}}$ be the $k$-block partition of $\ee(G)$,
with $K_i = L_i\uplus M_i$ for every $0\leqslant i\leqslant k-1$,
as specified in Definition~\ref{def:modified}. For every
$i\in\Set{0,\ldots, k-1}$:
\begin{enumerate}[itemsep=1pt,parsep=4pt,topsep=2pt,partopsep=0pt] 
\item 
  $G[K_i]$ is a finite collection of cacti.
\item 
  $G[L_i]$ is a finite collection of simple cycles.
\item
  $G[M_i]$ is a finite collection of trees.
\end{enumerate}
\end{proposition} 

\begin{proof}
  We provide some of the details for part 1, the proofs for parts 2 and 3 are
  just as straightforward.
  By the definitions, we have for every $i \in \Set{0,1, \ldots, k-1}$:
\[
   \OutPlan{E}{G_i} = k-i .
\]
A straightforward induction on $i = 0,\ldots,k-1$, shows that:
\[
  \OutPlan{E}{G_i} = \OutPlan{E}{G_{i+1}} + \OutPlan{E}{G[K_i]} ,
\]
using the fact that $\ee(G_{i+1})$ is the set of all edges \emph{not}
lying on the outer face of $G_i$, and $K_i$ is the set of all edges
lying on the outer face of $G_i$. Hence:
\[
   \OutPlan{E}{G[K_i]} = \OutPlan{E}{G_{i}} - \OutPlan{E}{G_{i+1}}.
\]
Working backwards, $\OutPlan{E}{G_{k}} = 0$ and:
\[
\OutPlan{E}{G[K_{k-1}]} = \OutPlan{E}{G_{k-1}} = 1 ,
\]
which implies $G[K_{k-1}]$ is a finite collection of cacti.
Similarly, for every $i = 0,\ldots,k-1$:
\[
  \OutPlan{E}{G[K_{k-i}]} = \OutPlan{E}{G_{k-i}} - \OutPlan{E}{G_{k-i+1}} = 1,
\]
which implies $G[K_{k-i}]$ is a finite collection of cacti. 
\end{proof}

\subsection{Three-Regular Plane Graphs}

We specialize notions introduced earlier in Section~\ref{sect:plane-and-planar}
to the case of $3$-regular graphs. 

\begin{figure}[H] 
\begin{custommargins}{0cm}{0cm} 
\begin{centering}
   \input{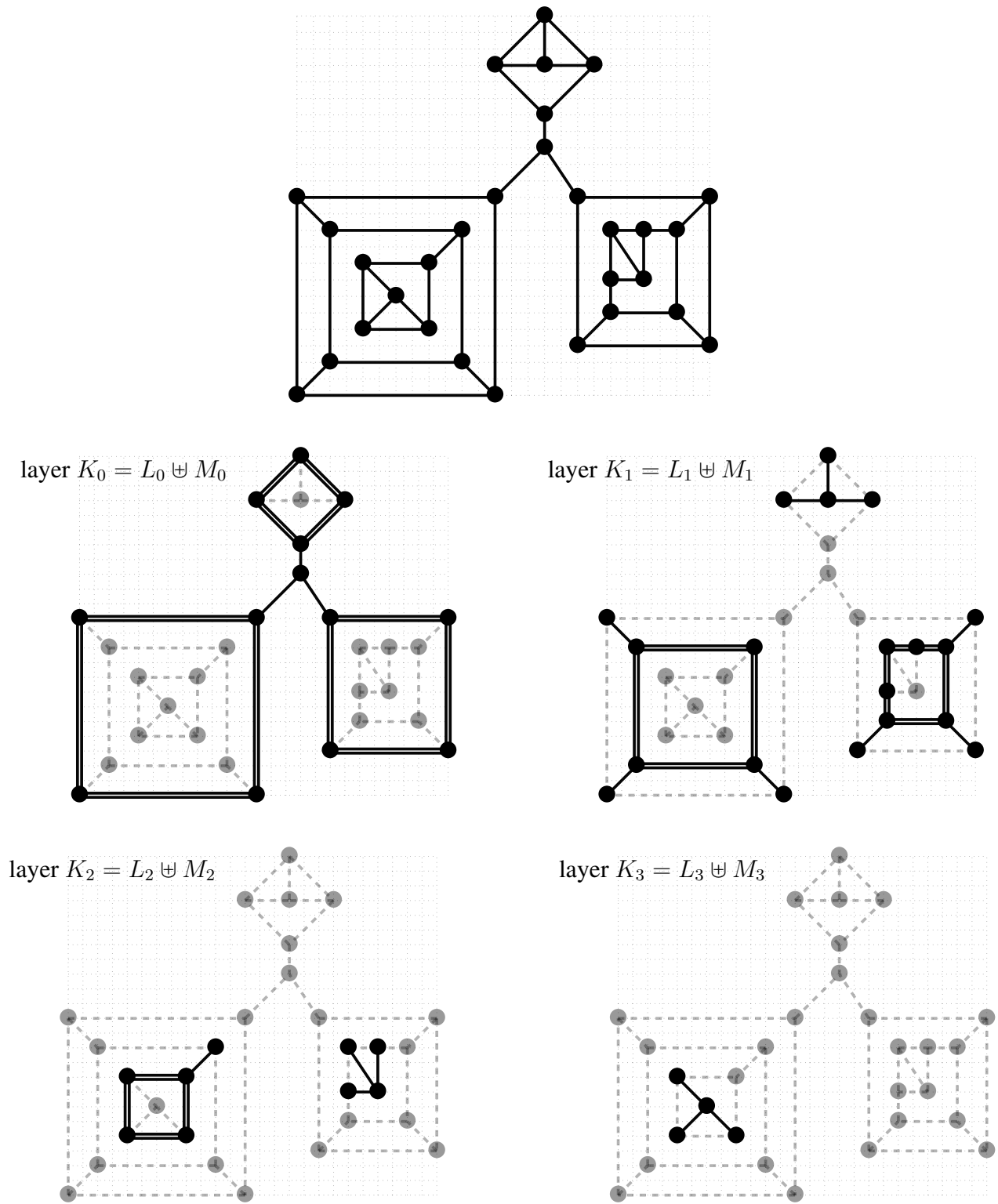}
\end{centering}
   \caption{A $3$-regular plane graph $G$ (at the top)
   with $\OutPlan{E}{G} = 4$
   and its $4$ layers of edges. Double-line edges in $K_i$
   belong to $L_i$, the \emph{level-$i$ cycles}, single-line edges in $K_i$
   belong to $M_i$, the \emph{level-$i$ inter-cycle trees} (ICT's),
   for $i\in\Set{0,1,2,3}$. This particular
   $G$ is decomposed into $6$ vertex-disjoint cycles and $11$ ICT's. 
   The three edges of the single level-$0$ ICT are \emph{non-bounding},
   all the other edges are \emph{bounding};
   cf. Proposition~\ref{prop:bounding-vs-bounding}.}
\label{fig:three-regular-plane-graph}
\end{custommargins}
\end{figure}

\begin{proposition} 
\label{prop:partition-for-3-regular}
Let $G$ be a $3$-regular plane graph with $\OutPlan{E}{G} = k\geqslant 1$.
Let $\Set{K_0, \ldots, K_{k-1}}$ be the $k$-block partition of $\ee(G)$,
with $K_i = L_i\uplus M_i$ for every $0\leqslant i\leqslant k-1$,
as specified in Definition~\ref{def:modified} and
Proposition~\ref{prop:partition}. We then have:
\begin{enumerate}[itemsep=1pt,parsep=2pt,topsep=2pt,partopsep=0pt]
\item
  The edges in $L_0\cup\cdots\cup L_{k-1}$ form vertex-disjoint cycles
  such that, for every such cycle $C$, the edges of $C$ are all in the
  same $L_i$ for some $0\leqslant i\leqslant k-1$.
\item
  The edges in $M_0\cup\cdots\cup M_{k-1}$ form trees, whose
  non-leaf vertices have degree $3$, such that for every such tree $T$,
  the edges of $T$ are all in the same $M_i$ for some $0\leqslant i\leqslant k-1$.
\item
  If in addition $G$ is biconnected, then $M_0 = \varnothing$.
\end{enumerate}
We can thus view $G$ as a finite collection of vertex-disjoint cycles
connected by trees. We thus call the latter \emph{inter-cycle trees}
(ICT's). For later reference, we call each $K_i$ a \emph{layer} of
$G$, which is partitioned into \emph{cycle edges} (those in $L_i$) and
\emph{ICT edges} (those in $M_i$). See Figure~\ref{fig:three-regular-plane-graph}
for an example (which is not biconnected).
\end{proposition}

\begin{proof}
Straightforward by inspection. All details omitted.
\end{proof}

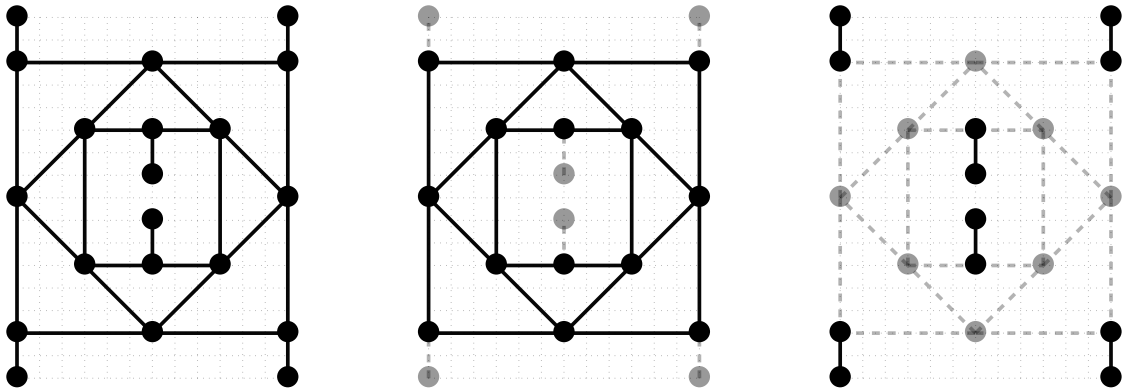
\begin{figure}[H] 
\begin{centering}

\noindent
%
\begin{tikzpicture}[scale=.3] 
       \newcommand\EdgeOpacity{[line width=1.4,black,opacity=.98]};
       \draw [help lines, dotted] (0, 0) grid (12,16);
       \coordinate (A) at (0,0);
       \coordinate (B) at (12,0);
       \coordinate (C) at (0,2);
       \coordinate (D) at (6,2);
       \coordinate (E) at (12,2);
       \coordinate (F) at (3,5);
       \coordinate (G) at (6,5);
       \coordinate (H) at (9,5);
       \coordinate (I) at (6,7);
       \coordinate (J) at (0,8);       
       \coordinate (K) at (12,8);
       \coordinate (L) at (6,9);       
       \coordinate (M) at (3,11);
       \coordinate (N) at (6,11);       
       \coordinate (O) at (9,11);
       \coordinate (P) at (0,14);       
       \coordinate (Q) at (6,14);
       \coordinate (R) at (12,14);
       \coordinate (S) at (0,16);
       \coordinate (T) at (12,16);            

       \draw \EdgeOpacity (A) -- (C) ;
       \draw \EdgeOpacity (B) -- (E) ;       
       \draw \EdgeOpacity (C) -- (D) ;
       \draw \EdgeOpacity (C) -- (J) ;
       \draw \EdgeOpacity (D) -- (E) ;
       \draw \EdgeOpacity (D) -- (F) ;
       \draw \EdgeOpacity (D) -- (H) ;
       \draw \EdgeOpacity (E) -- (K) ;
       \draw \EdgeOpacity (F) -- (G) ;
       \draw \EdgeOpacity (F) -- (J) ;
       \draw \EdgeOpacity (F) -- (M) ;
       \draw \EdgeOpacity (G) -- (H) ;
       \draw \EdgeOpacity (G) -- (I) ;
       \draw \EdgeOpacity (H) -- (K) ;
       \draw \EdgeOpacity (H) -- (O) ;       
       \draw \EdgeOpacity (J) -- (M) ;
       \draw \EdgeOpacity (J) -- (P) ;
       \draw \EdgeOpacity (K) -- (O) ;
       \draw \EdgeOpacity (K) -- (R) ;
       \draw \EdgeOpacity (L) -- (N) ;
       \draw \EdgeOpacity (M) -- (N) ;
       \draw \EdgeOpacity (M) -- (Q) ;
       \draw \EdgeOpacity (N) -- (O) ;
       \draw \EdgeOpacity (O) -- (Q) ;
       \draw \EdgeOpacity (P) -- (Q) ;
       \draw \EdgeOpacity (P) -- (S) ;
       \draw \EdgeOpacity (Q) -- (R) ;
       \draw \EdgeOpacity (R) -- (T) ;               

       \node at (A) {\huge $\bullet$};   
       \node at (B) {\huge $\bullet$}; 
       \node at (C) {\huge $\bullet$};
       \node at (D) {\huge $\bullet$};
       \node at (E) {\huge $\bullet$};
       \node at (F) {\huge $\bullet$};   
       \node at (G) {\huge $\bullet$}; 
       \node at (H) {\huge $\bullet$};
       \node at (I) {\huge $\bullet$};
       \node at (J) {\huge $\bullet$};
       \node at (K) {\huge $\bullet$};   
       \node at (L) {\huge $\bullet$}; 
       \node at (M) {\huge $\bullet$};
       \node at (N) {\huge $\bullet$};
       \node at (O) {\huge $\bullet$};
       \node at (P) {\huge $\bullet$};   
       \node at (Q) {\huge $\bullet$}; 
       \node at (R) {\huge $\bullet$};
       \node at (S) {\huge $\bullet$}; 
       \node at (T) {\huge $\bullet$};       
\end{tikzpicture}
\qquad\quad
\begin{tikzpicture}[scale=.3] 
       \newcommand\EdgeOpacity{[line width=1.4,black,opacity=.98]};
       \newcommand\EdgeOpacityA{[line width=1.4,dashed,black,opacity=.3]};
       \draw [help lines, dotted] (0, 0) grid (12,16);
       \coordinate (A) at (0,0);
       \coordinate (B) at (12,0);
       \coordinate (C) at (0,2);
       \coordinate (D) at (6,2);
       \coordinate (E) at (12,2);
       \coordinate (F) at (3,5);
       \coordinate (G) at (6,5);
       \coordinate (H) at (9,5);
       \coordinate (I) at (6,7);
       \coordinate (J) at (0,8);       
       \coordinate (K) at (12,8);
       \coordinate (L) at (6,9);       
       \coordinate (M) at (3,11);
       \coordinate (N) at (6,11);       
       \coordinate (O) at (9,11);
       \coordinate (P) at (0,14);       
       \coordinate (Q) at (6,14);
       \coordinate (R) at (12,14);
       \coordinate (S) at (0,16);
       \coordinate (T) at (12,16);            

       \draw \EdgeOpacityA (A) -- (C) ;
       \draw \EdgeOpacityA (B) -- (E) ;       
       \draw \EdgeOpacity (C) -- (D) ;
       \draw \EdgeOpacity (C) -- (J) ;
       \draw \EdgeOpacity (D) -- (E) ;
       \draw \EdgeOpacity (D) -- (F) ;
       \draw \EdgeOpacity (D) -- (H) ;
       \draw \EdgeOpacity (E) -- (K) ;
       \draw \EdgeOpacity (F) -- (G) ;
       \draw \EdgeOpacity (F) -- (J) ;
       \draw \EdgeOpacity (F) -- (M) ;
       \draw \EdgeOpacity (G) -- (H) ;
       \draw \EdgeOpacityA (G) -- (I) ;
       \draw \EdgeOpacity (H) -- (K) ;
       \draw \EdgeOpacity (H) -- (O) ;       
       \draw \EdgeOpacity (J) -- (M) ;
       \draw \EdgeOpacity (J) -- (P) ;
       \draw \EdgeOpacity (K) -- (O) ;
       \draw \EdgeOpacity (K) -- (R) ;
       \draw \EdgeOpacityA (L) -- (N) ;
       \draw \EdgeOpacity (M) -- (N) ;
       \draw \EdgeOpacity (M) -- (Q) ;
       \draw \EdgeOpacity (N) -- (O) ;
       \draw \EdgeOpacity (O) -- (Q) ;
       \draw \EdgeOpacity (P) -- (Q) ;
       \draw \EdgeOpacityA (P) -- (S) ;
       \draw \EdgeOpacity (Q) -- (R) ;
       \draw \EdgeOpacityA (R) -- (T) ;               

       \node at (A) [opacity = .4] {\huge $\bullet$};   
       \node at (B) [opacity = .4] {\huge $\bullet$}; 
       \node at (C) {\huge $\bullet$};
       \node at (D) {\huge $\bullet$};
       \node at (E) {\huge $\bullet$};
       \node at (F) {\huge $\bullet$};   
       \node at (G) {\huge $\bullet$}; 
       \node at (H) {\huge $\bullet$};
       \node at (I) [opacity = .4] {\huge $\bullet$};
       \node at (J) {\huge $\bullet$};
       \node at (K) {\huge $\bullet$};   
       \node at (L) [opacity = .4] {\huge $\bullet$}; 
       \node at (M) {\huge $\bullet$};
       \node at (N) {\huge $\bullet$};
       \node at (O) {\huge $\bullet$};
       \node at (P) {\huge $\bullet$};   
       \node at (Q) {\huge $\bullet$}; 
       \node at (R) {\huge $\bullet$};
       \node at (S) [opacity = .4] {\huge $\bullet$}; 
       \node at (T) [opacity = .4] {\huge $\bullet$};       
\end{tikzpicture}
\qquad\quad
\begin{tikzpicture}[scale=.3] 
       \newcommand\EdgeOpacity{[line width=1.4,black,opacity=.98]};
       \newcommand\EdgeOpacityA{[line width=1.4,dashed,black,opacity=.3]};
       \draw [help lines, dotted] (0, 0) grid (12,16);
       \coordinate (A) at (0,0);
       \coordinate (B) at (12,0);
       \coordinate (C) at (0,2);
       \coordinate (D) at (6,2);
       \coordinate (E) at (12,2);
       \coordinate (F) at (3,5);
       \coordinate (G) at (6,5);
       \coordinate (H) at (9,5);
       \coordinate (I) at (6,7);
       \coordinate (J) at (0,8);       
       \coordinate (K) at (12,8);
       \coordinate (L) at (6,9);       
       \coordinate (M) at (3,11);
       \coordinate (N) at (6,11);       
       \coordinate (O) at (9,11);
       \coordinate (P) at (0,14);       
       \coordinate (Q) at (6,14);
       \coordinate (R) at (12,14);
       \coordinate (S) at (0,16);
       \coordinate (T) at (12,16);            

       \draw \EdgeOpacity (A) -- (C) ;
       \draw \EdgeOpacity (B) -- (E) ;       
       \draw \EdgeOpacityA (C) -- (D) ;
       \draw \EdgeOpacityA (C) -- (J) ;
       \draw \EdgeOpacityA (D) -- (E) ;
       \draw \EdgeOpacityA (D) -- (F) ;
       \draw \EdgeOpacityA (D) -- (H) ;
       \draw \EdgeOpacityA (E) -- (K) ;
       \draw \EdgeOpacityA (F) -- (G) ;
       \draw \EdgeOpacityA (F) -- (J) ;
       \draw \EdgeOpacityA (F) -- (M) ;
       \draw \EdgeOpacityA (G) -- (H) ;
       \draw \EdgeOpacity (G) -- (I) ;
       \draw \EdgeOpacityA (H) -- (K) ;
       \draw \EdgeOpacityA (H) -- (O) ;       
       \draw \EdgeOpacityA (J) -- (M) ;
       \draw \EdgeOpacityA (J) -- (P) ;
       \draw \EdgeOpacityA (K) -- (O) ;
       \draw \EdgeOpacityA (K) -- (R) ;
       \draw \EdgeOpacity (L) -- (N) ;
       \draw \EdgeOpacityA (M) -- (N) ;
       \draw \EdgeOpacityA (M) -- (Q) ;
       \draw \EdgeOpacityA (N) -- (O) ;
       \draw \EdgeOpacityA (O) -- (Q) ;
       \draw \EdgeOpacityA (P) -- (Q) ;
       \draw \EdgeOpacity (P) -- (S) ;
       \draw \EdgeOpacityA (Q) -- (R) ;
       \draw \EdgeOpacity (R) -- (T) ;               

       \node at (A) {\huge $\bullet$};   
       \node at (B) {\huge $\bullet$}; 
       \node at (C) {\huge $\bullet$};
       \node at (D) [opacity = .4] {\huge $\bullet$};
       \node at (E) {\huge $\bullet$};
       \node at (F) [opacity = .4] {\huge $\bullet$};   
       \node at (G) {\huge $\bullet$}; 
       \node at (H) [opacity = .4] {\huge $\bullet$};
       \node at (I) {\huge $\bullet$};
       \node at (J) [opacity = .4] {\huge $\bullet$};
       \node at (K) [opacity = .4] {\huge $\bullet$};   
       \node at (L) {\huge $\bullet$}; 
       \node at (M) [opacity = .4] {\huge $\bullet$};
       \node at (N) {\huge $\bullet$};
       \node at (O) [opacity = .4] {\huge $\bullet$};
       \node at (P) {\huge $\bullet$};   
       \node at (Q) [opacity = .4] {\huge $\bullet$}; 
       \node at (R) {\huge $\bullet$};
       \node at (S) {\huge $\bullet$}; 
       \node at (T) {\huge $\bullet$};       
\end{tikzpicture} 

\end{centering}   
   \caption{A non-regular plane graph (on the left), 
   with its \emph{bounding edges} in
   boldface (in the middle), and its \emph{non-bounding edges}
   in boldface (on the right). Cf. Proposition~\ref{prop:bounding-vs-bounding}.}
\label{fig:non-regular-plane-graph}
\end{figure}

The preceding proposition is not true for
arbitrary plane graphs. Consider, for example, the non-regular
plane graph $G$ in Figure~\ref{fig:non-regular-plane-graph}:
The cycles formed by the edges in
$L_0\cup L_1\cup L_2$ are not vertex-disjoint.

In later sections we use the following definition. We identify
a simple cycle and an ICT by the edges it contains.

\begin{definition}{Levels in Three-Regular Plane Graphs}
  \label{def:levels}
  Let $G$ be a $3$-regular plane graph as in
  Definition~\ref{def:modified} and
  Propositions~\ref{prop:partition} and~\ref{prop:partition-for-3-regular}.
\begin{enumerate}[itemsep=1pt,parsep=2pt,topsep=2pt,partopsep=0pt]
\item Let $C$ be a cycle in the induced subgraph
  $G[L_0\cup\cdots\cup L_{k-1}]$ which therefore satisfies $C\subseteq L_i$
  for some $0\leqslant i\leqslant k-1$ by part 1 of
  Proposition~\ref{prop:partition-for-3-regular}. We define:
  \[
         \level{C} \triangleq i
  \]
\item Let $T$ be an ICT in the induced subgraph $G[M_0\cup\cdots\cup M_{k-1}]$
  which therefore satisfies $T\subseteq M_i$ for some
  $0\leqslant i\leqslant k-1$ by part 2 of
  Proposition~\ref{prop:partition-for-3-regular}. We define:
  \[
         \level{T} \triangleq i
  \]
\end{enumerate}
Assume $G$ is biconnected (part 3 in
Proposition~\ref{prop:partition-for-3-regular}).
It is easy to see that in the subgraph $G[L_0\cup\cdots\cup L_{k-1}]$,
there is one or more cycles of level $i$ for every
$i\in\Set{0,\ldots,k-2}$, and zero or more cycles of level $k-1$.
In the subgraph $G[M_0\cup\cdots\cup M_{k-1}]$,
there is no ICT of level $0$, 
two or more ICT's of level $i$ for every $i\in\Set{1,\ldots,k-2}$,
and one or more ICT's of level $k-1$.
\end{definition}

\newcommand{\cycleE}[1]{\mathit{cycle\text{-}edges}(#1)}
\newcommand{\crossE}[1]{\mathit{cross\text{-}edges}(#1)}
\newcommand{\cycleV}[1]{\mathit{cycle\text{-}vertices}(#1)}
\newcommand{\crossV}[1]{\mathit{cross\text{-}vertices}(#1)}

We conclude by stating the relationship between the
standard notion of outerplanarity and the notion of
edge-outerplanarity used in this report.  If $G$ is a plane graph, let
$\OutPlan{V\!\!}{G}$ denote the smallest $k$ such that $G$ is
$k$-outerplanar (this is $k$-outerplanarity in the standard sense). 

\begin{proposition}
\label{prop:modified}
If $G$ is a $3$-regular plane graph, then
$\OutPlan{V\!\!}{G}$ and $\OutPlan{E}{G}$ are ``almost equal'', specifically:  
$\OutPlan{V\!\!}{G} \leqslant \OutPlan{E}{G} 
\leqslant 1+ \OutPlan{V\!\!}{G}$.
\end{proposition}

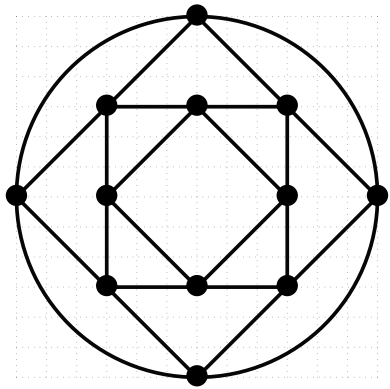
\begin{figure}[H] 
\vspace{-.2in}
\begin{centering}

\noindent
%
\begin{tikzpicture}[scale=.4] 
       \newcommand\EdgeOpacity{[line width=1.4,black,opacity=.98]};
       \draw [help lines, dotted] (0, 0) grid (12,12);
       \coordinate (A) at (0,0);
       \coordinate (B) at (6,0);
       \coordinate (C) at (12,0);
       \coordinate (D) at (12,6);
       \coordinate (E) at (12,12);
       \coordinate (F) at (6,12);
       \coordinate (G) at (0,12);
       \coordinate (H) at (0,6);
       \coordinate (I) at (3,3);
       \coordinate (J) at (6,3);       
       \coordinate (K) at (9,3);
       \coordinate (L) at (9,6);       
       \coordinate (M) at (9,9);
       \coordinate (N) at (6,9);       
       \coordinate (O) at (3,9);
       \coordinate (P) at (3,6);       

       \draw \EdgeOpacity (B) to[out=0,in=-90] (D) ;
       \draw \EdgeOpacity (D) to[out=90,in=0] (F) ;       
       \draw \EdgeOpacity (B) -- (K) ;              
       \draw \EdgeOpacity (D) -- (M) ;       
       \draw \EdgeOpacity (F) to[out=180,in=90] (H) ;
       \draw \EdgeOpacity (F) -- (O) ;       
       \draw \EdgeOpacity (H) to[out=-90,in=180] (B) ;
       \draw \EdgeOpacity (H) -- (I) ;
       \draw \EdgeOpacity (I) -- (J) ;
       \draw \EdgeOpacity (I) -- (B) ;       
       \draw \EdgeOpacity (J) -- (K) ;
       \draw \EdgeOpacity (J) -- (L) ;       
       \draw \EdgeOpacity (K) -- (D) ;
       \draw \EdgeOpacity (K) -- (L) ;       
       \draw \EdgeOpacity (L) -- (M) ;
       \draw \EdgeOpacity (L) -- (N) ;       
       \draw \EdgeOpacity (M) -- (F) ;
       \draw \EdgeOpacity (M) -- (N) ;       
       \draw \EdgeOpacity (N) -- (O) ;
       \draw \EdgeOpacity (N) -- (P) ;       
       \draw \EdgeOpacity (O) -- (H) ;
       \draw \EdgeOpacity (O) -- (P) ;              
       \draw \EdgeOpacity (P) -- (I) ;       
       \draw \EdgeOpacity (P) -- (J) ;

       \node at (B) {\huge $\bullet$}; 
       \node at (D) {\huge $\bullet$};
       \node at (F) {\huge $\bullet$};   
       \node at (H) {\huge $\bullet$};
       \node at (I) {\huge $\bullet$};
       \node at (J) {\huge $\bullet$};
       \node at (K) {\huge $\bullet$};   
       \node at (L) {\huge $\bullet$}; 
       \node at (M) {\huge $\bullet$};
       \node at (N) {\huge $\bullet$};
       \node at (O) {\huge $\bullet$};
       \node at (P) {\huge $\bullet$};   
\end{tikzpicture}

\end{centering}   
   \caption{A four-regular plane graph $G$, 
   with
   $\OutPlan{V\!\!}{G} = 2$ and $\OutPlan{E}{G} = 4$. \\ Contrast
   with Proposition~\ref{prop:modified}.
   }
\label{fig:four-regular-plane-graph}
\end{figure}

This proposition is not true for arbitrary plane graphs, even if they
are regular.  Consider, for example, the four-regular plane graph $G$
in Figure~\ref{fig:four-regular-plane-graph}, where
$\OutPlan{V\!\!}{G} = 2$ while $\OutPlan{E}{G} = 4$.

\begin{sketch}
  For a $3$-regular plane graph, the difference between
  $\OutPlan{V\!\!}{G}$ and $\OutPlan{E}{G}$ occurs in the last stage
  in the process of repeatedly removing (in the case of the standard
  definition) all vertices on the outer face and all their incident
  edges. The corresponding last stage in the modified definition may
  or may not delete all the edges; if it does not, then one extra
  stage is needed to delete all remaining edges.
\end{sketch}

\section{Algorithm $\KS$}
\label{sect:algorithm-KS}

We start by stating the main result of this section.

\begin{theorem}
\label{thm:algorithm-KS}
There is an algorithm, herein named $\KS$, which takes as input
a biconnected $3$-regular simple plane graph $G$
with $\OutPlan{E}{G} = k \geqslant 2$, and satisfies the following
properties:
\begin{enumerate}[itemsep=1pt,parsep=2pt,topsep=2pt,partopsep=0pt]   
\item $\KS$ terminates in linear time $\bigO{n}$, using $\bigO{n}$ space,
  where $n = \size{\vv(G)}$, and
   \item $\KS$ returns a binary reassembling $\B = \KS(G)$ of $G$ such
         that $\alpha(G,\B) \leqslant 2k$.
\end{enumerate}
Note, in particular, the value of $\alpha(G,\B)$ in the output $\B$ is
independent of $n$.
\end{theorem}
The rest of this section is devoted to the proofs of
Theorem~\ref{thm:algorithm-KS}, 
several lemmas, and several supporting examples. In
Corollary~\ref{cor:algorithm-KS}, we lift the biconnectedness
restriction. We begin with an informal description of algorithm
$\KS$. We use the terminology and notation introduced in
Section~\ref{sect:plane-and-planar}.

\subsection{Informal Description}
\label{sect:informal}

At the topmost level of algorithm $\KS$, there are two phases:
  \begin{description}
  \item[Pre-Processing Phase.] This phase 
    partitions the set $\ee(G)$ of edges into two non-empty disjoint sets,
    each partitioned into $k$ disjoint subsets (some of the latter possibly empty):
    \begin{itemize}
    \item the set $L_0\uplus\ldots\uplus L_{k-1}$ of \emph{cycle edges},
    \item the set $M_1\uplus\ldots\uplus M_{k-1}$ of \emph{ICT edges},
    \end{itemize}
    and carries out further classification of edges and vertices.
    We omit $M_0 = \varnothing$
    (part 3 in Proposition~\ref{prop:partition-for-3-regular}).
    \item[Processing Phase.] This is the actual algorithm at work,
      consisting of a sequence of repeated \emph{contractions}, which we
      divide into two kinds, \emph{collapses} and \emph{merges}:
       \begin{itemize}
       \item a \emph{collapse} contracts all the edges of an ICT
          and turns it into what we call a \emph{super vertex},
        \item a \emph{merge} contracts all the cycle
          edges that connect a \emph{super vertex} to what we call
          its \emph{clockwise neighbor}, thus producing a larger super vertex.
       \end{itemize}
        A \emph{super vertex} resulting from
        a collapse or a merge is not restricted to degree $=
        3$. With the introduction of super vertices,
        two vertices may be connected by more than one edge and self-loops
        may be introduced, \ie, with super vertices the graph becomes
        a \emph{multigraph}. (More on super vertices below.)
  \end{description}
  Although the algorithm's pseudocode and its Python implementation
  are sequential, it is better understood as carrying out its operations
  in parallel. Specifically, the algorithm starts by collapsing
  all eligible ICT's, \ie, ICT's satisfying certain
  conditions $\CC$ (Section~\ref{sect:contract-ICT-edges}),
  then it carries out all merge operations satisfying certain conditions
  $\DD$ (Section~\ref{sect:contract-cycle-edges}).
  The purpose of carrying out the merges according to $\DD$ is to make
  a second group of ICT's eligible for collapse according to
  $\CC$. After a second round of merges according to $\DD$, a third
  group of ICT's becomes eligible for collapse according to
  $\CC$. This process continues by following every round of collapses
  according to $\CC$ by a round of merges according to $\DD$, with the
  latter making a new group of ICT's eligible for collapse according
  to $\CC$ -- until the whole graph $G$ becomes a single super vertex.

  \subsection{Further Classification and Terminology}
  \label{sect:classification}
  
  The set $\vv(L_0\uplus\ldots\uplus L_{k-1})$ is the set
  of \emph{cycle vertices}, which is the same as the set of leaf
  vertices of the ICT's. (See Figure~\ref{fig:three-regular-plane-graph} 
  for an example.) We classify cycle vertices into two
  kinds, \emph{inward} and \emph{outward}. Let $v$ be a vertex on a
  cycle $X$ of level-$i$ for some $i\in\Set{0,\ldots,k-1}$, and let
  $\Set{\set{v\ w_1},\set{v\ w_2}, \set{v\ w_3}}$ be the three edges
  incident to $v$ with $\Set{\set{v\ w_1},\set{v\ w_2}}\subseteq X$
  and $\Set{\set{v\ w_3}}\subseteq T$ for some ICT $T$ incident to
  $X$:
  \begin{description}[itemsep=1pt,parsep=2pt,topsep=2pt,partopsep=0pt]
      \item[A.\ \ Inward Cycle Vertices:]
        If $T$ is of level-$(i+1)$, then $v$ is
        an inward vertex. 
      \item[B.\ \ Outward Cycle Vertices:]
        If $T$ is of level-$i$, then $v$ is
        an outward vertex. 
      \item[C.\ \ Super Vertices:] Each \emph{collapse} and each \emph{merge}
       produces what we call a \emph{super vertex}, which can be viewed as a `bag'
       containing two or more ordinary vertices as well as the edges
       connecting them.
     \end{description}
  To distinguish vertices in the initial set $\vv(G)$ from super
  vertices, we sometimes call the former \emph{ordinary vertices}. In
  contrast to ordinary vertices, always of degree $=3$, super vertices
  can have arbitrary degrees $\geqslant 2$. We use late lower-case
  Roman letters $u$, $v$, and $w$, to denote ordinary vertices; late
  lower-case Greek letters $\varphi$, $\chi$, and $\psi$, to denote
  super vertices; and middle lower-case Greek letters $\mu$, $\nu$,
  and $\xi$, to denote both ordinary and super vertices. If $\varphi$
  is a super vertex, then $\vv(\varphi)$ is the set of ordinary
  vertices contained in $\varphi$.
  
  If a super vertex $\varphi$ is not the final super vertex containing
  the entire input graph $G$, then $\varphi$ always straddles one or
  more cycles. If we say super vertex $\varphi$ \emph{straddles} cycle
  $X$, we mean $\vv(\varphi)\cap \vv(X)\neq \varphi$ and
  $\vv(\varphi)\not\supseteq \vv(X)$, \ie, one or more of the vertices
  of $X$ are inside $\varphi$ and one or more are outside
  $\varphi$. If $\mu$ is a super vertex or an ordinary cycle-vertex,
  its set of cycles is:
  \[
      \cycles{\mu}
      \ \triangleq\ \Set{\, X\;|\;\text{$X$ is a cycle and $\mu$ straddles $X$}\,} . 
  \]
  The \emph{innermost} cycle 
  of $\mu$ is defined by:
  \begin{alignat*}{6}
    & \inmost{\mu}\ \triangleq\ X\quad &&\text{where}\ \ 
         \level{X}\; = \max\, && \Set{\,\level{X'}\;| && \;X'\in\cycles{\mu}\,} .
  \end{alignat*}
  An ordinary cycle-vertex always straddles a single cycle, which is
  also its innermost cycle.  Later in this section,
  Lemma~\ref{lem:properties-of-super-vertices} implies that
  $\inmost{\mu}$ is uniquely defined, and that
  $\cycles{\mu}$ is a chain of nested cycles of consecutive levels,
  with lower-level cycles nesting higher-level cycles and with the unique
  $\inmost{\mu}$ having the highest level in that chain.

  Once produced, a super vertex is viewed as a single vertex for the
  purposes of the algorithm, but the collection of ordinary vertices
  included in the super vertex are recorded and part of the final
  output returned by the algorithm. Super vertices are not classified
  into `inward' and `outward', in contrast to ordinary cycle-vertices
  which are always inward or outward.
  \begin{description}
  \item[D.\ \ Two Invariants of the Algorithm:]
  During algorithm execution, cycle edges and ICT edges never change
  their designation: they remain `cycle edges' and `ICT edges',
  respectively, until they get included into super
  vertices and removed from further consideration.
  \end{description}
  For the notion
  of \emph{clockwise neighbor} used below, we view \emph{every cycle
  edge as directed clockwise and this direction remains unchanged
  throughout algorithm execution}. ICT edges do not have a direction.

  \begin{description}[itemsep=1pt,parsep=2pt,topsep=0pt,partopsep=0pt]
  \item[E.\ \ Another Invariant of the Algorithm:]
  The edges of an ICT $T$, but not necessarily its leaf
  vertices, remain unchanged until the collapse of $T$ places all
  of $\ee(T)$ in the same super vertex; some of the leaf vertices of
  $T$ may already be included in super vertices in earlier steps of
  the algorithm.  More precisely, there is an ICT $U$ in the initial
  $G$ such that either $T$ is $U$ or else, if there is an
  edge $e_1 = \set{v\;\varphi}$ in $T$ where $\varphi$ is a super vertex,
  then $\varphi$ is a leaf vertex of $T$ and there is an edge
  $e_0 = \set{v\;w}$ in the initial $U$ such that:
  \[
    \text{ $w$ is a leaf vertex of $U$,
    \ \ $w$ is included in $\varphi$,
    \ \ and $\cycles{w}\subseteq\cycles{\varphi}$, }
  \]
  and we can view $e_0$ and $e_1$ as the same edge.
  This implies that, even though the edges of $T$ remain unchanged
  until a collapse operation places them all in a super vertex $\psi$, the
  cycles straddled by a leaf vertex of $T$ may become a larger set of
  cycles during algorithm execution such that, after the collapse operation,
  some edges of those cycles become self-loops of $\psi$ or 
  incoming edges to $\psi$ or outgoing edges from $\psi$. Only cycle
  edges become self-loops (of super vertices), ICT edges do not become
  self-loops.
  \end{description}

  We use the terms `sibling' and `clockwise neighbor' to qualify two
  different relationships between cycle vertices, whether ordinary or super:
  \begin{description}[itemsep=1pt,parsep=2pt,topsep=0pt,partopsep=0pt]
     \item[F.\ \ Siblings:]
     We call \emph{siblings} the leaf vertices
     of an ICT $T$. If they all -- except possibly for one \emph{outward} 
     ordinary vertex $v$ -- have the same innermost cycle $X$ and can be
     listed consecutively as $X$ is traversed clockwise, \ie, without 
     encountering interleaved vertices of $X$ belonging to ICT's other than $T$,  
     then we call them the \emph{consecutive siblings} of $T$ and the exception 
     vertex $v$, if it exists, the \emph{root} of $T$.
     \item[G.\ \ Clockwise Neighbors:]
     Let $\mu$ and $\nu$ be cycle vertices, super or ordinary,
     and $X$ the innermost cycle of $\mu$. We say $\nu$ is the
     \emph{clockwise neighbor} of $\mu$ if there is an edge of $X$ in the
     clockwise direction from $\mu$ to $\nu$.
     \\ \textbf{Note}: We require that $X$ be the innermost cycle of $\mu$
     only, not $\nu$, and we do not disallow that $\mu=\nu$
     in which case $\mu$ is a clockwise neighbor of itself.
\end{description}

\subsection{Conditions $\CC$ for the Collapse Operation: How to Contract ICT Edges}
\label{sect:contract-ICT-edges}
  
We start applying the collapse operation whenever condition $({\CC}.0)$
is satisfied, and we apply it as long as conditions
$\Set{({\CC}.1),({\CC}.2)}$ are simultaneously satisfied. Condition $({\CC}.3)$
specifies how each ICT is collapsed.
\begin{enumerate}[itemsep=1pt,parsep=4pt,topsep=2pt,partopsep=0pt]
  \item[$({\CC}.0)$] No merge is possible according to conditions
    $\DD$ below.
  \item[$({\CC}.1)$]  Consider all the level-$(i+1)$ ICT's occurring between a
    level-$i$ cycle $X$ and zero or more level-$(i+1)$ cycles. These
    are the ICT's immediately enclosed in $X$ and outside
    level-$(i+1)$ cycles, if any: These ICT's are not eligible
    for collapse before all the ICT's that are incident to $X$
    from the outside -- except possibly for one such incident ICT --
    have been collapsed.
   \item[$({\CC}.2)$]
    Assuming condition $({\CC}.1)$ is met,
    an ICT $T$ is eligible for collapse when all its leaf vertices,
    except possibly for one \emph{outward} ordinary vertex $v$,
    are consecutive siblings on a cycle $X$. Each of the consecutive siblings
    on $X$ is: either a super vertex or an \emph{inward} ordinary vertex.
   \item[$({\CC}.3)$] The collapse of an ICT $T$ turns $T$ into
     a super vertex, by contracting all of $T$'s tree edges,
     and is carried out to minimize the $\alpha$-measure and in linear time.    
\end{enumerate}
\vspace{-.1in}
\Hide{ You canNOT omit condition $({\CC}.1)$ -- unless you add another
  condition $({\DD}.2)$. }
\paragraph{Remark.}\label{rem:about-typeA-typeB} 
Condition $({\CC}.1)$ is
not necessary for the algorithm to work correctly; we include it only
to force execution to proceed in an `outside-in' fashion, \ie, by
applying the collapse operation to the ICT's outside a cycle as much
as possible before applying it to the ICT's inside the same cycle.
For later reference, we distinguish the two cases of ``an ICT
$T$ is eligible for collapse'' in $({\CC}.2)$:
\begin{description}[itemsep=0pt,parsep=1pt,topsep=1pt,partopsep=0pt]
  \item[Type-$\bm{a}$:]
    The leaf vertices of $T$ are all consecutive siblings.
  \item[Type-$\bm{b}$:]
  The leaf vertices of $T$, except for one outward ordinary $v$,
  are consecutive siblings.
\end{description}
A type-$a$ ICT is a \emph{rootless} tree, where non-leaf vertices
(all ordinary) have each degree three and leaf vertices (some
ordinary, some super) have each degree one.  A type-$b$ ICT is a
\emph{rooted} tree, where the sole outward ordinary vertex is the
\emph{root}, non-leaf vertices (all ordinary) have each degree
three, and the root and the leaf vertices (some ordinary, some super)
have each degree one. The classification of ICT's into type-$a$ and type-$b$ 
applies only to ICT's \emph{eligible for collapse}, not to
ICT's \emph{not eligible for collapse} at any time during execution;
the classification is thus best understood
dynamically as the algorithm progresses. 
Examples~\ref{ex:type-a-vs-type-b} and~\ref{ex:type-a-vs-type-b-again}
illustrate the difference between type-$a$ and type-$b$.
\Hide{
Examples showing the difference between type-$a$
and type-$b$ are in Figures~\ref{fig:54-vertices-1},
\ref{fig:54-vertices-2}, and~\ref{fig:54-vertices-3}, and again in
Figures~\ref{fig:42-vertices-1} and~\ref{fig:42-vertices-2} (read the
captions).
}

\subsection{Conditions $\DD$ for the Merge Operation: How to Contract Cycle Edges}
\label{sect:contract-cycle-edges}

A merge is applied in one of two cases: (i) two super vertices or
(ii) one super vertex and one ordinary vertex. It is not
applied to two ordinary vertices. Let $\varphi$ be
the super vertex involved in a merge operation and $\mu$ the vertex,
ordinary or super, involved in the same operation.  The result of
applying a merge to $\varphi$ and $\mu$ is a new super vertex
obtained by contracting \emph{all the cycle edges} (one or more) connecting
$\varphi$ and $\mu$.

We start applying the merge operation from the moment condition $({\DD}.0)$
is satisfied, and we apply it repeatedly as long conditions
 $\Set{({\DD}.1),({\DD}.2)}$ are simultaneously satisfied. 
\begin{enumerate}[itemsep=1pt,parsep=3pt,topsep=2pt,partopsep=0pt]
\item[$({\DD}.0)$]
    No collapse is possible according to conditions $\CC$ above.
  \item[$({\DD}.1)$]
    $\varphi$ is a super vertex, $\mu$ is the clockwise neighbor of $\varphi$,
    and $\varphi$ is not a leaf vertex of an ICT.
  \item[$({\DD}.2)$]
    If $\mu$ is additionally an \emph{outward} ordinary vertex,
    then $\varphi$ is also the clockwise neighbor of $\mu$; \\ \ie,
    $\varphi$ and $\mu$ are distinct (because $\varphi$ is super
    and $\mu$ is ordinary) clockwise neighbors of each other. 
\end{enumerate}
Note the fact that $\varphi$ is not a leaf vertex in $({\DD}.1)$, which
implies that $\varphi$ and $\mu$ are not siblings, \ie, leaf vertices of the
same ICT. Several remarks are in order in relation to conditions
$\Set{({\DD}.1),({\DD}.2)}$, which also spell out the different special
cases subsumed by these two conditions:
\begin{enumerate}[itemsep=1pt,parsep=3pt,topsep=2pt,partopsep=0pt]
\item
  $\mu$ is an \emph{inward} ordinary vertex: Contracting
  the clockwise cycle edge from  $\varphi$ to $\mu$ produces a super vertex
  which is a leaf vertex, thus not satisfying condition $({\DD}.1)$
  and not eligible for an additional merge.
\item
  $\mu$ is an \emph{outward} ordinary vertex: Condition 
  $({\DD}.2)$ requires that $\varphi$ is the clockwise neighbor of $\mu$,
  thus implying that $\mu$ is the only ordinary vertex on its cycle.
\item
  $\mu$ is a super vertex: $\varphi$ is distinct from $\mu$ and $\varphi$ is
  not the clockwise neighbor of $\mu$.
\item
  $\mu$ is a super vertex: $\varphi$ is distinct from $\mu$ and $\varphi$ is
  the clockwise neighbor of $\mu$ (in this case $\varphi$ and $\mu$ are
  super vertices that are clockwise neighbors of each other).
\item
  $\mu$ is a super vertex: $\varphi$ is the \emph{same super vertex}
  as $\mu$, which is thus a clockwise neighbor of itself. \\[.66ex]
  This last special case is the only case when
  the merge operation does not create a new super vertex
  that contains a larger subset of ordinary vertices; its purpose is
  only to contract all the self-loops of $\varphi=\mu$.  
\end{enumerate}
Because of conditions $({\CC}.0)$ and $({\DD}.0)$, the algorithm
proceeds by alternating rounds of collapses and merges. Each of the
two kinds of rounds executes a maximum number of operations of its
kind. \emph{A round of collapses creates self-loops in general, the
  succeeding round of merges eliminates all resulting self-loops} (and
contracts other cycle edges in general).  Since all the vertices in
the initial input graph $G$ are ordinary, the algorithm starts with a
round of \emph{collapses} and stops with a round of \emph{merges}. We
later identify these alternating rounds by numbering them;
\textbf{round $\bm{k}$} will be a round of \emph{collapses} when
$k\geqslant 1$ is odd, and it will be a round of \emph{merges} when
$k\geqslant 2$ is even.

\textbf{Remark About Self-Loops:} The only case of an ICT collapse
that does not generate a self-loop is when the ICT is a single tree-edge
connecting two distinct cycles (of the same level $i$, or
of two consecutive levels $i$ and $i+1$, for some $i\geqslant 0$).
A merge never generates self-loops; it only eliminates them,
if it is according to case 5 of condition $({\DD}.1)$.

Let $G$ be the initial input graph and $G'$ a graph
obtained from $G$ by a sequence of collapse and merge rounds, the last
of which being a round of collapses. Let $e'$ be a self-loop in $G'$,
which is the `descendant' of a cycle edge $e$ in $G$, in the sense that the
two endpoints of $e'$ are a transformation of the two endpoints of
$e$. (More precisely, if $e'$ is the self loop
$\set{\varphi\ \varphi}$ then $e = \set{v\ w}$ where both of the
ordinary vertices $v$ and $w$ are included in the super vertex $\varphi$.)
If we apply a round of merges to $G'$ to obtain $G''$, then $e'$ along
with all other self-loops disappear in $G''$ and the cycles in $G''$
are therefore a subset of the cycles in $G'$.

\subsection{Examples}
\label{sect:examples}


Before we tackle the correctness of our algorithm $\KS$ in
Section~\ref{sect:pseudocode}, we include six examples illustrating
the progression of $\KS$.  For the graph $G$ in each example, the
constructed reassembling $\B$ consists of the super vertices produced
by $\KS$, in addition to the singletons $\Set{v}$ for all
$v\in\vv(G)$. We will thus have:
\[
  \alpha(G,\B) = \max\, \Set{\,\degr{}{\mu}\;|
    \;\mu\text{ is an ordinary vertex or a super vertex produced by $\KS$}\,}
\]
and $\alpha(G,\B) \leqslant 2\cdot\OutPlan{E}{G}$ in each example,
as predicted by Theorem~\ref{thm:algorithm-KS}.
The first example is very simple, and each successive example
exhibits a few more complications than the preceding one.

In all the examples, super vertices are shown enclosed in colored boundaries:
{\color{red}{\bf red}} if produced by a round of collapses,
{\color{green}{\bf green}} if produced by a round of merges.

\begin{figure}[H]
\vspace{-.12in}
\begin{center}
  \includegraphics[scale=.45]{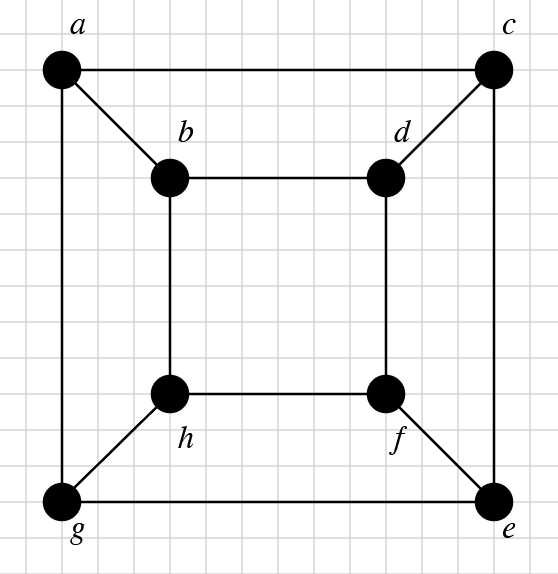}
  \hspace{2.5cm}
  \includegraphics[scale=.45]{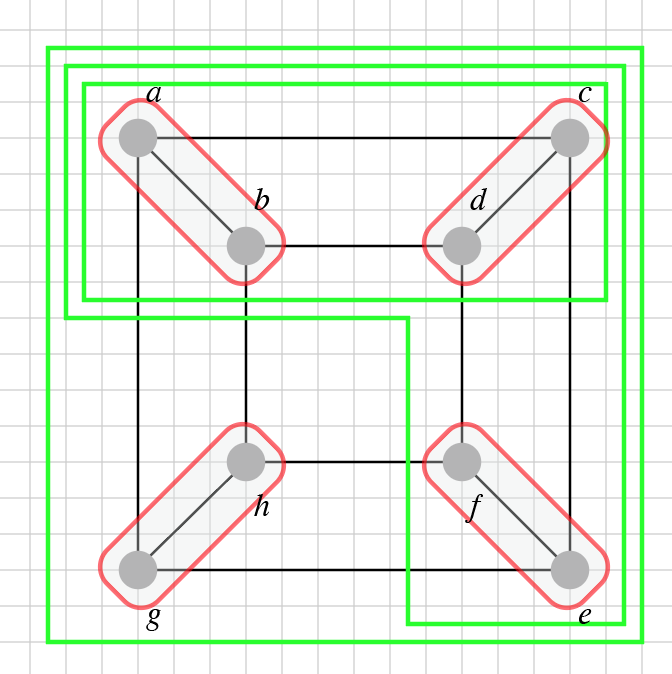}
\end{center}
\vspace{-.2in}
\caption{\label{fig:8-vertices}
  A $3$-regular plane graph (left)
  and the progression of algorithm $\KS$ on it (right). There are
  4 ICT's in this example, all single-edge, and all eligible for
  collapse of type-$b$.
  }
\end{figure}

\begin{example}
  \label{ex:8-vertices}
On the left in Figure~\ref{fig:8-vertices} is a $3$-regular plane 
graph (the ``cube''). It consists of two nested simple cycles, connected
by $4$ inter-cycle trees (ICT's); in this case, each ICT is a single edge.
On the right in Figure~\ref{fig:8-vertices}, we show the 
progression of our algorithm $\KS$.
The four innermost super vertices are obtained by the first round of 
\emph{collapses} (\textbf{round 1} of $\KS$) which contract only ICT's; 
these are enclosed in red boundaries. The ordinary vertices $\Set{a,b}$
are enclosed in one super vertex, the ordinary vertices $\Set{c,d}$
in a second super vertex, the ordinary vertices $\Set{e,f}$
in a third super vertex, and the ordinary vertices $\Set{g,h}$
in a fourth super vertex.

The following round of \emph{merges} (\textbf{round 2} of $\KS$)
contracts only cycle edges and 
produces three nested super vertices in succession. These are shown
on the right in Figure~\ref{fig:8-vertices} enclosed in green boundaries.
The merge of super vertex $\Set{a,b}$ with super vertex $\Set{c,d}$,
and then that of super vertex $\Set{a,b,c,d}$ with super vertex $\Set{e,f}$,
do not create self-loops; these are two \emph{merge} operations according to
\emph{case 3} of condition $({\DD}.1)$. One more \emph{merge}, according to
\emph{case 4} of condition $({\DD}.1)$, puts together super vertex
$\Set{a,b,c,d,e,f}$ and super vertex $\Set{g,h}$ to produce the final super
vertex, and again this last \emph{merge} does not create any self-loop.
\Hide{
The resulting reassembling $\B$ consists of all the super vertices that
are constructed during $\KS$'s execution, in addition to the singletons
$\Set{v}$ for all $v\in\vv(G)$. Hence, in this example,
$\alpha(G,\B) = \max \Set{\,\Bridges{}{X}\,|\,X\in\B} = 4$
which is the same as $2\cdot\OutPlan{E}{G} = 2\cdot 2= 4$, as predicted by
Theorem~\ref{thm:algorithm-KS}.}
\end{example}

\begin{figure}[H]
\vspace{-.12in}
\begin{center}
  \includegraphics[scale=.42]{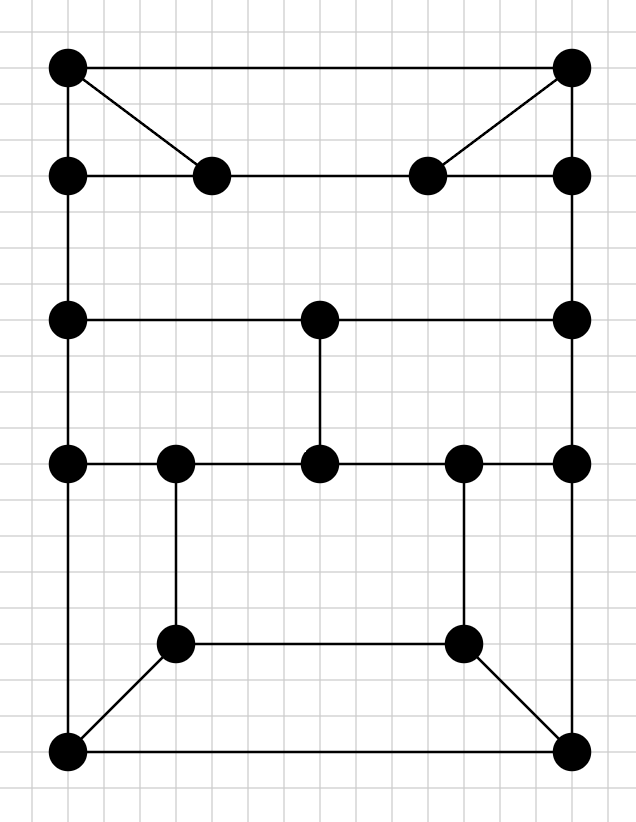}
  \hspace{2.5cm}
  \includegraphics[scale=.42]{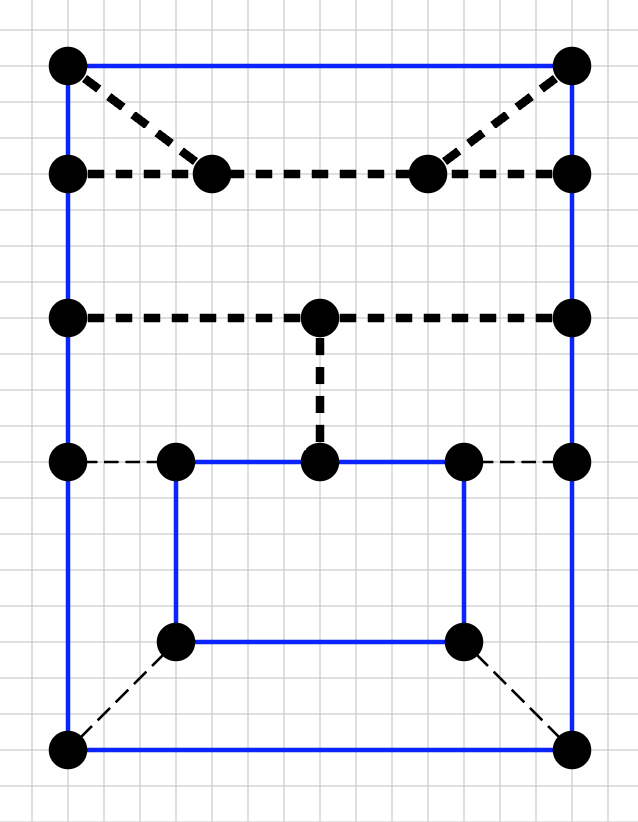}
\end{center}
\vspace{-.2in}
\caption{\label{fig:18-vertices}
  A $3$-regular plane graph (left figure)
  decomposed into $2$ cycles and $6$ ICT's (right figure),
  $4$ ICT's are single-edge (light dashed),
  $2$ multi-edge (bold dashed), only one not
  initially eligible for collapse (second from top). }
\end{figure}

\begin{figure}[H]
\vspace{-.2in}
\begin{center}
 \includegraphics[scale=.45]{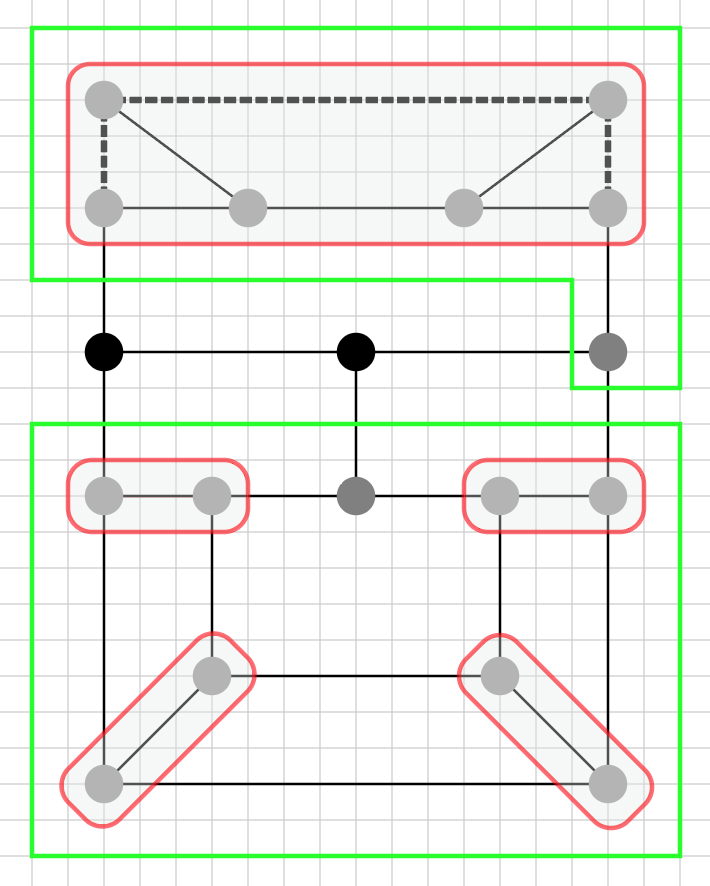}
 \hspace{1.5cm}
 \includegraphics[scale=.45]{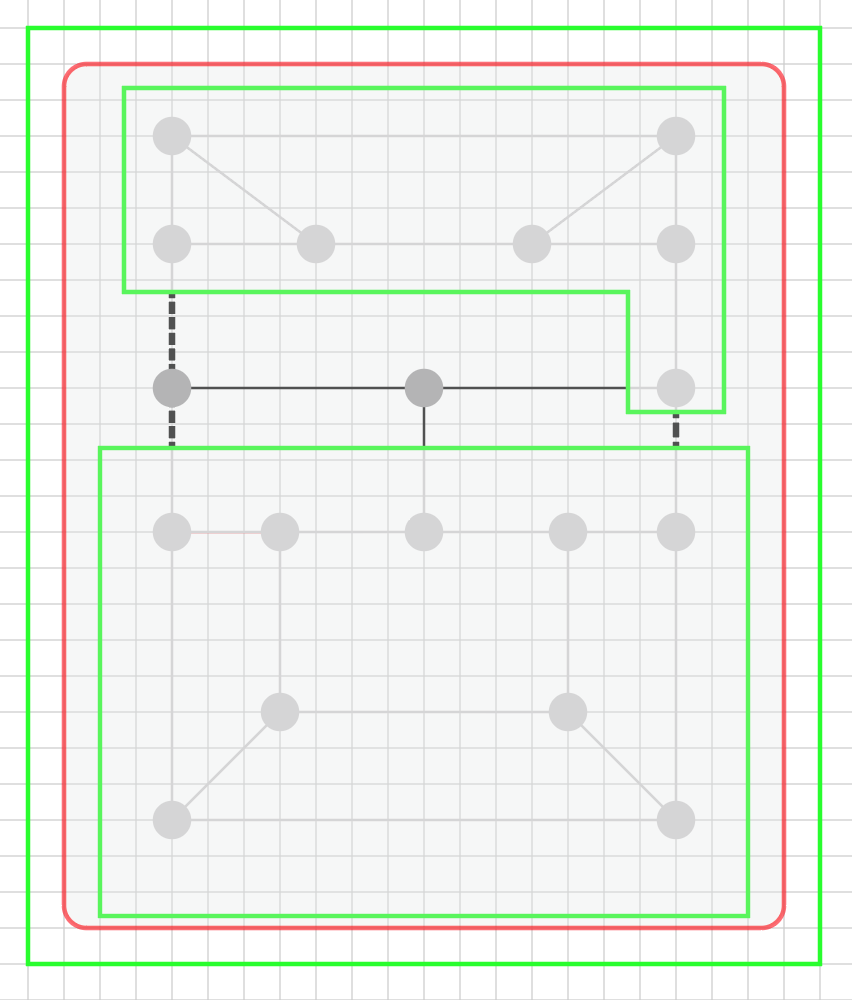}
\end{center}
\vspace{-.2in}
\caption{\label{fig:18-vertices-A} $\KS$'s progression
    on the graph in Figure~\ref{fig:18-vertices}.
    There is one ICT (left figure) with $2$ ordinary vertices
    (bold black) and $2$ super vertices (green), 
    eligible for collapse (type-$a$) and contracted in $\KS$'s
    \textbf{round~3} (right figure).
    $\KS$'s \textbf{round~4} contracts the self-loops (dashed edges
    in right figure) and produces the last super vertex (green). 
    }
\end{figure}

\begin{example}
  \label{ex:18-vertices}
In Figure~\ref{fig:18-vertices} we show a $3$-regular plane graph decomposed
into cycles and inter-cycle trees (ICT's).
In Figure~\ref{fig:18-vertices-A} we show the progression of our
algorithm $\KS$ on this graph. The five innermost super vertices
enclosed in red boundaries on the left of
Figure~\ref{fig:18-vertices-A} are obtained from a first round of
\emph{collapses} (\textbf{round 1} of $\KS$), followed by several
\emph{merges} (part of \textbf{round 2} of $\KS$) that contract the
self-loops of the five resulting super vertices. Actually, in this
case, only the top super vertex among these five has self-loops to be
contracted, shown as dashed edges on the left of
Figure~\ref{fig:18-vertices-A}. 

The remaining \emph{merges} of \textbf{round 2}
contract several cycle edges, producing the two super vertices
enclosed in green boundaries on the left of
Figure~\ref{fig:18-vertices-A}. The four lower super vertices
enclosed in red are put together to produce the lower super vertex in
green on the left of Figure~\ref{fig:18-vertices-A}; this is result of
three \emph{merge} operations according to \emph{case 3} of condition
$({\DD}.1)$, followed by one more \emph{merge} operation according
to \emph{case 2} of condition $({\DD}.1)$.

At this point, there remains only one ICT, with two ordinary vertices
and two super vertices. One more round of \emph{collapses}
(\textbf{round 3} of $\KS$) contracts this ICT into a super vertex
with three self loops (the three remaining
cycle edges that have not been yet contracted, shown as dashed edges
on the right of Figure~\ref{fig:18-vertices-A}), and the latter are
contracted by a final round of \emph{merges} (\textbf{round~4} of $\KS$)
which produce the final super vertex shown on the right of
Figure~\ref{fig:18-vertices-A} enclosed in the outermost green boundary.
\end{example}

\begin{figure}[H]
\vspace{-.12in}
\begin{center}
  \includegraphics[scale=.35]{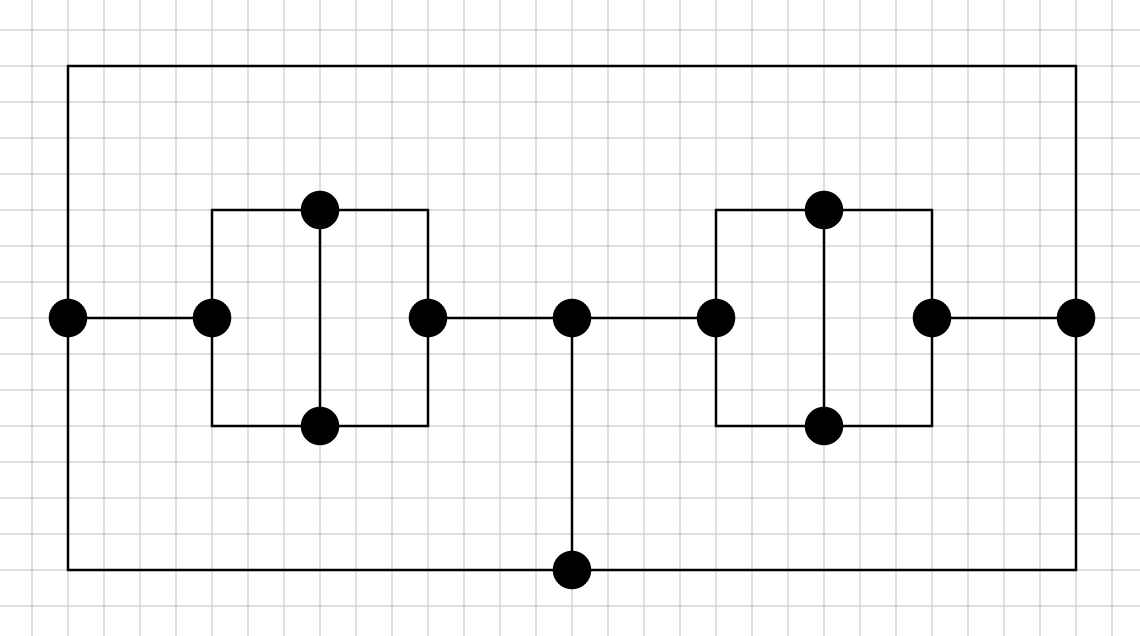}
  \hspace{1cm}
  \includegraphics[scale=.35]{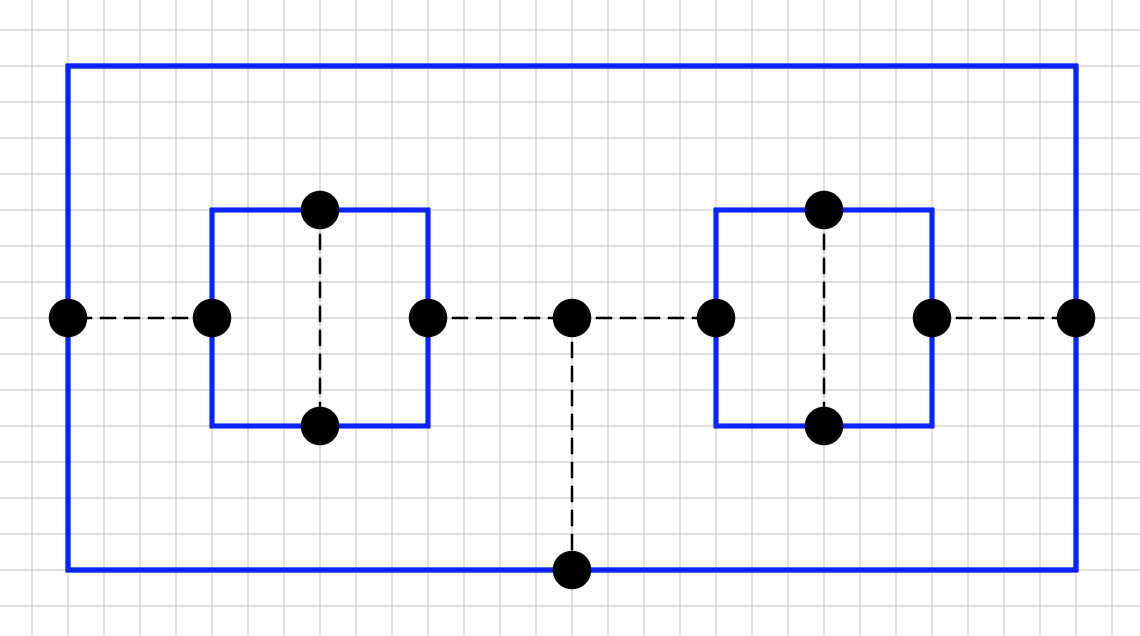}
\end{center}
\vspace{-.2in}
\caption{\label{fig:type-a-vs-type-b}
  A $3$-regular plane graph (left figure)
  decomposed into $3$ cycles and $5$ ICT's (right figure),
  $4$ ICT's are single-edge,
  one is three-edge. Two single-edge ICT's are initially eligible for
  collapse, the leftmost and the rightmost. 
  }
\end{figure}

\begin{figure}[H]
\vspace{-.12in}
\begin{center}
  \includegraphics[scale=.35]{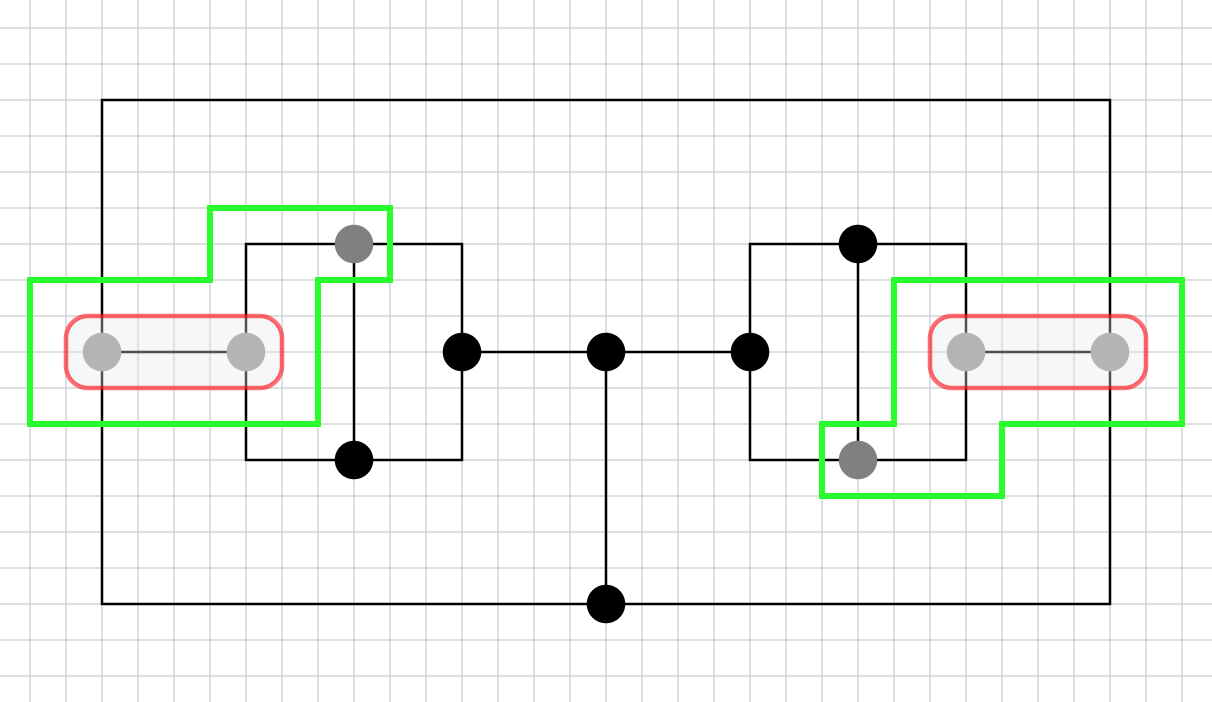}
  \hspace{0.5cm}
  \includegraphics[scale=.35]{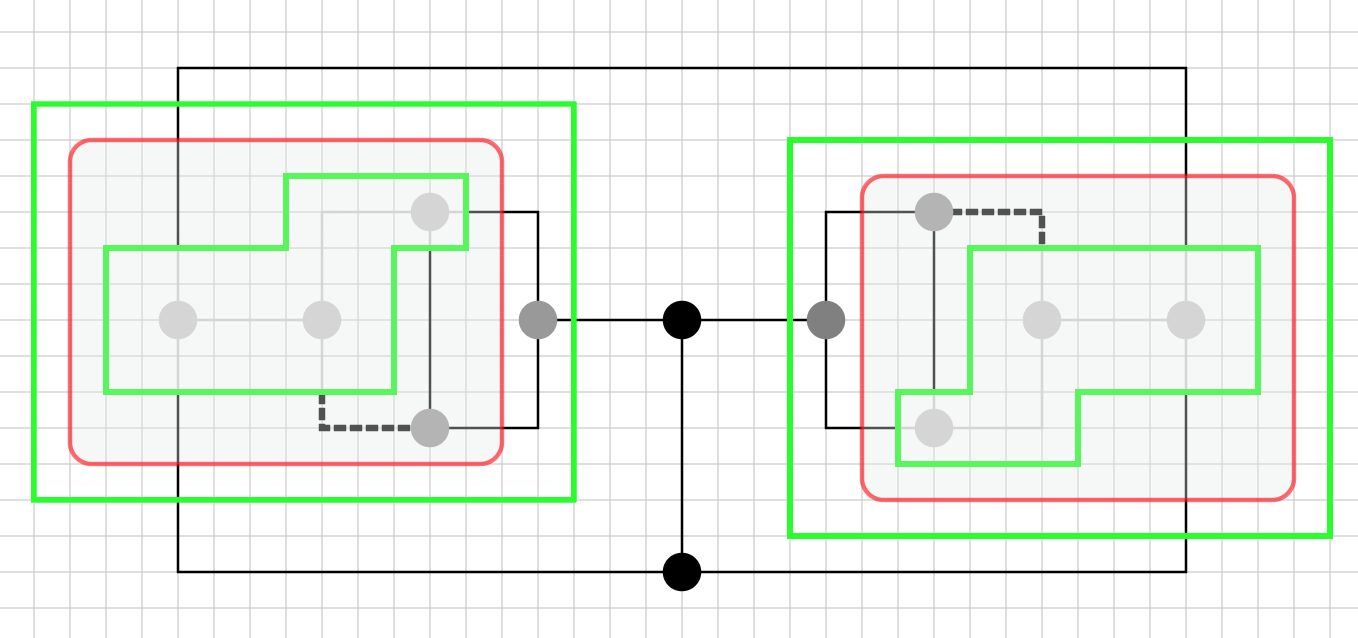}
  \includegraphics[scale=.35]{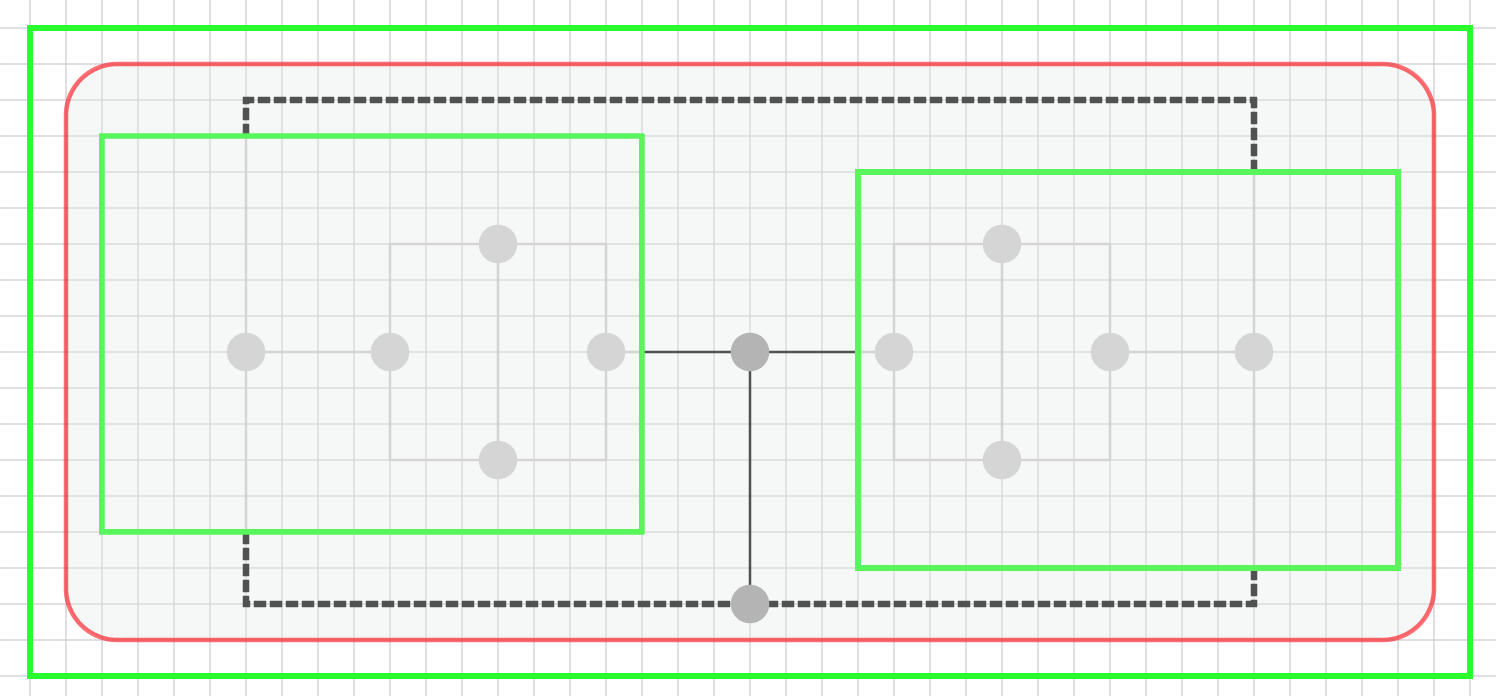}
\end{center}
\vspace{-.2in}
\caption{\label{fig:type-a-vs-type-b-A}
  $\KS$'s progression on the graph in Figure~\ref{fig:type-a-vs-type-b},
  after \textbf{rounds~1} and~\textbf{2} (top left),
  \textbf{rounds~3} and~\textbf{4} (top right),
  and \textbf{rounds~5} and~\textbf{6} (bottom). $\KS$'s
  \textbf{rounds~4} and \textbf{rounds~6} contract two and three
  self-loops (dashed edges).
  }
\end{figure}

\begin{example}
  \label{ex:type-a-vs-type-b}
The graph in Figure~\ref{fig:type-a-vs-type-b} (on the left) is
decomposed into $3$ cycles and $5$ ICT's (on the right).  The middle
three-edge ICT is not initially eligible for collapse, and becomes
eligible for collapse after \textbf{rounds~1, 2, 3} and~\textbf{4}
(middle figure in Figure~\ref{fig:type-a-vs-type-b-A}); and when it
becomes eligible for collapse, it is a type-$a$ ICT, because its three
leaf vertices (one ordinary and two super) are consecutive siblings on
the same cycle.
\end{example}

\begin{example}
  \label{ex:type-a-vs-type-b-again}
This is a variation on the graph in Example~\ref{ex:type-a-vs-type-b}:
All ICT's are single-edge, except for one which is three-edge, just like
in Example~\ref{ex:type-a-vs-type-b}. Whereas the single three-edge ICT
in Example~\ref{ex:type-a-vs-type-b} is type-$a$ when it becomes eligible
for collapse (after \textbf{round~4}), the single three-edge ICT
in this example is type-$b$ when it becomes eligible
for collapse (after \textbf{round~4}).
This illustrates how an ICT is classified as type-$a$ or type-$b$
dynamically, \ie, it is so classified at the time when it first becomes
eligible for collapse.
\end{example}

\begin{figure}[H]
\vspace{-.12in}
\begin{center}
  \includegraphics[scale=.3]{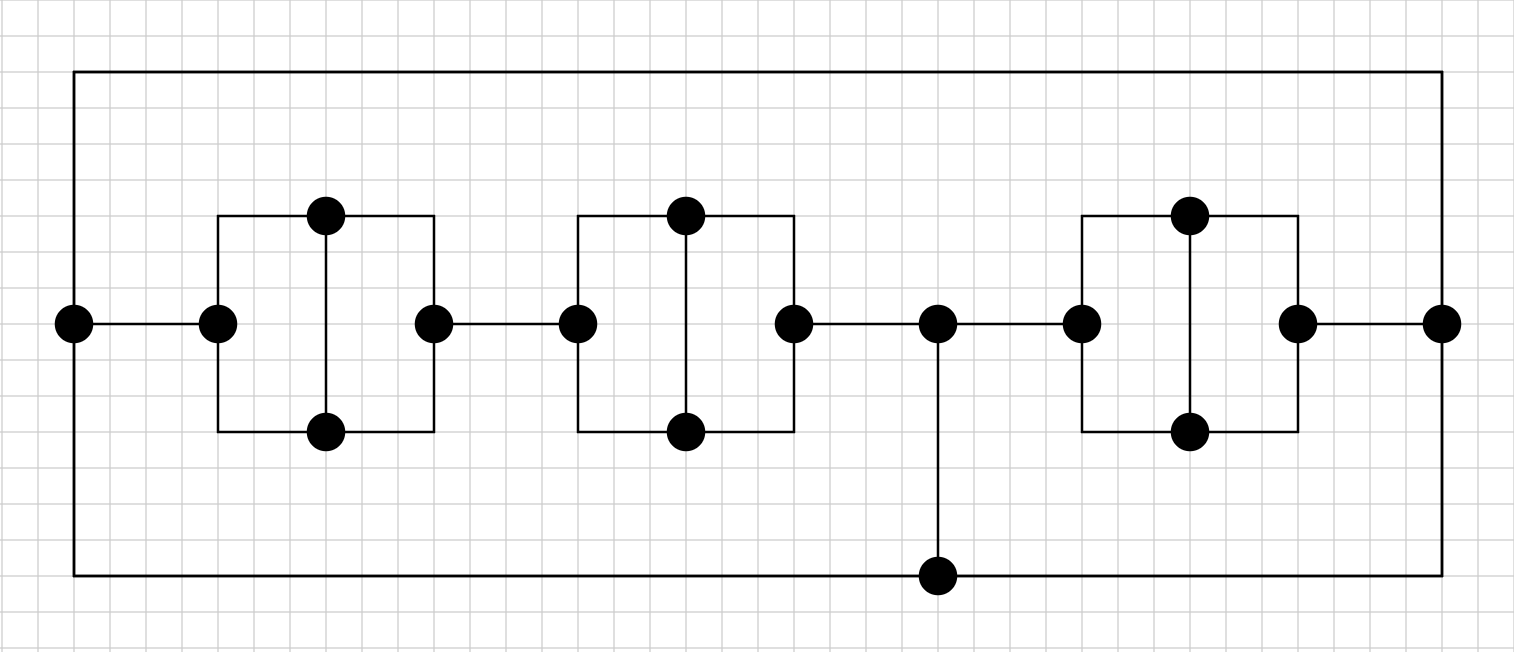}
  \hspace{0.5cm}
  \includegraphics[scale=.3]{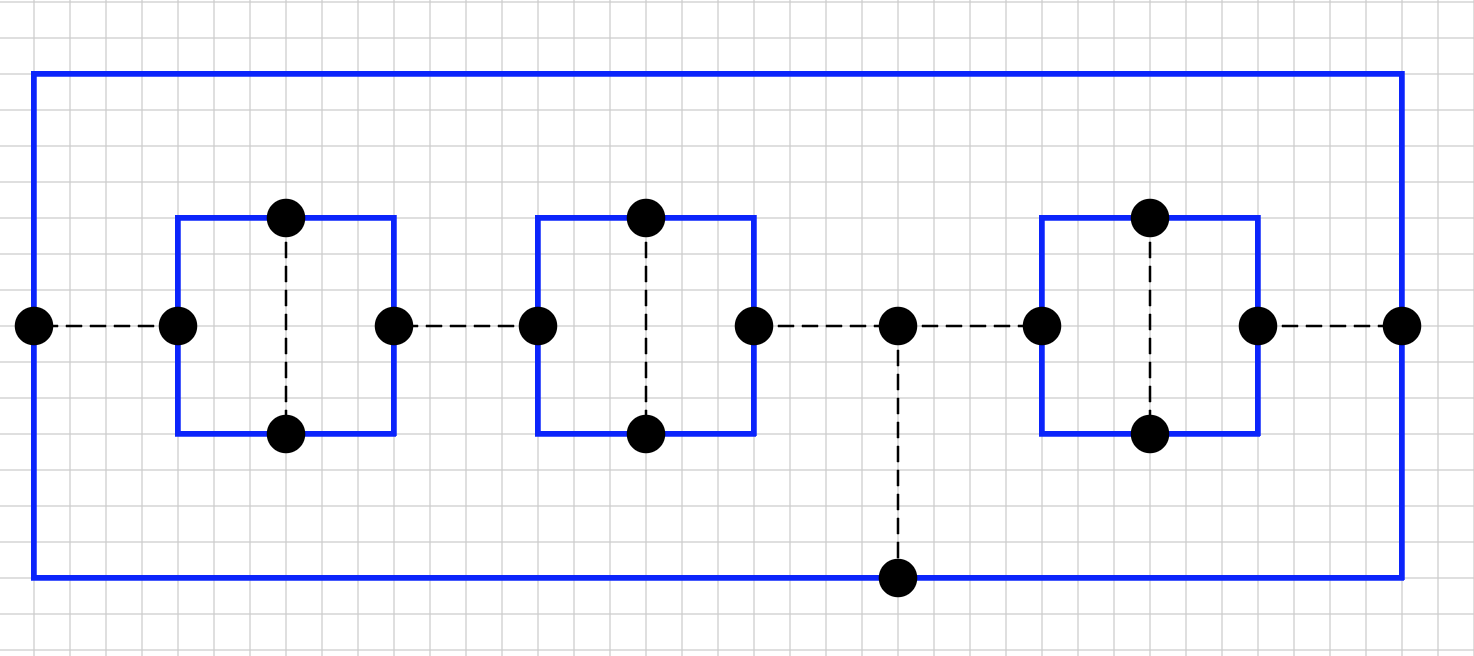}
\end{center}
\vspace{-.2in}
\caption{\label{fig:type-a-vs-type-b-again}
  A $3$-regular plane graph (left) decomposed into $4$ cycles and $7$
  ICT's (right).  One ICT is three-edge, $6$ ICT's are single-edge,
  two of the latter initially eligible for collapse, the leftmost and
  the rightmost, both type-$b$. }
\end{figure}

\begin{figure}[H]
\vspace{-.12in}
\begin{center}
  \includegraphics[scale=.4]{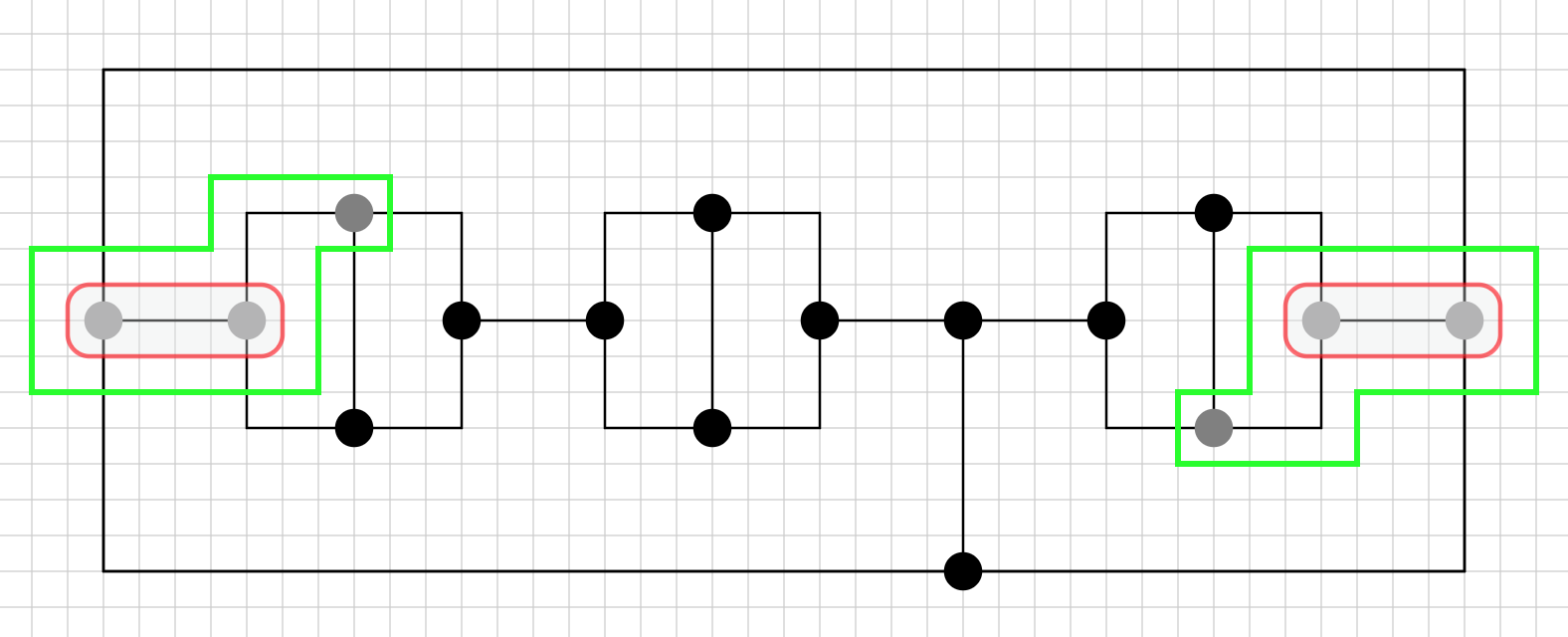}
  \includegraphics[scale=.4]{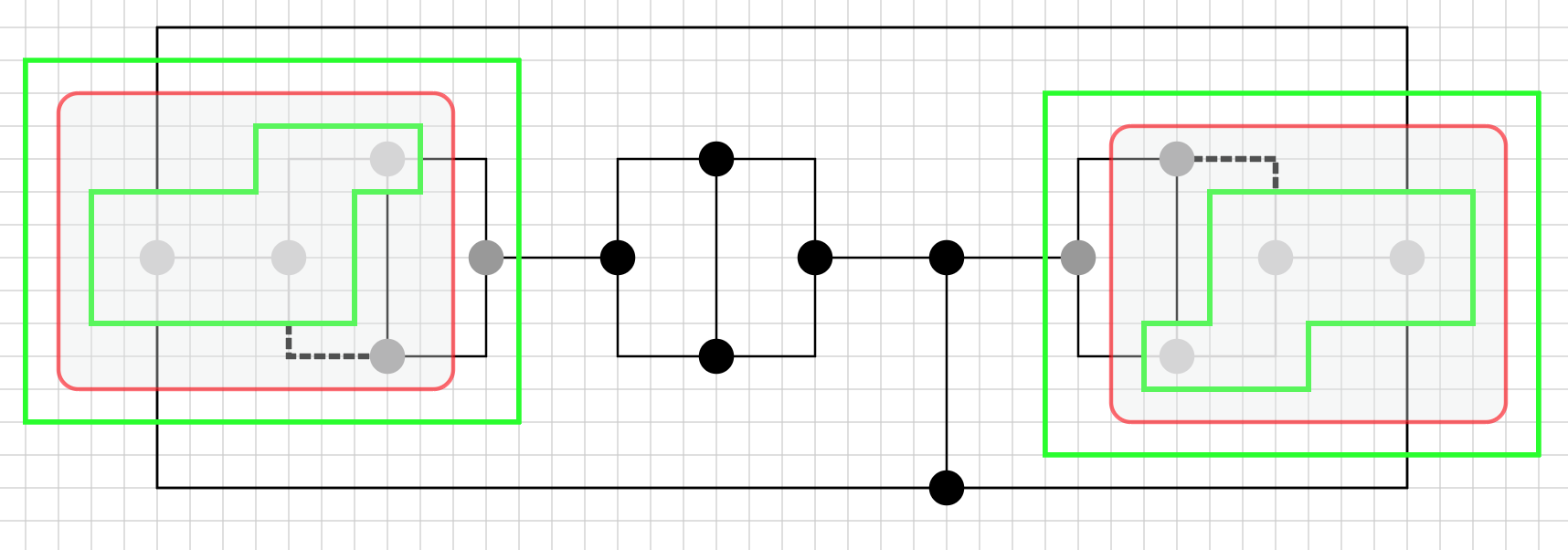}
  \includegraphics[scale=.4]{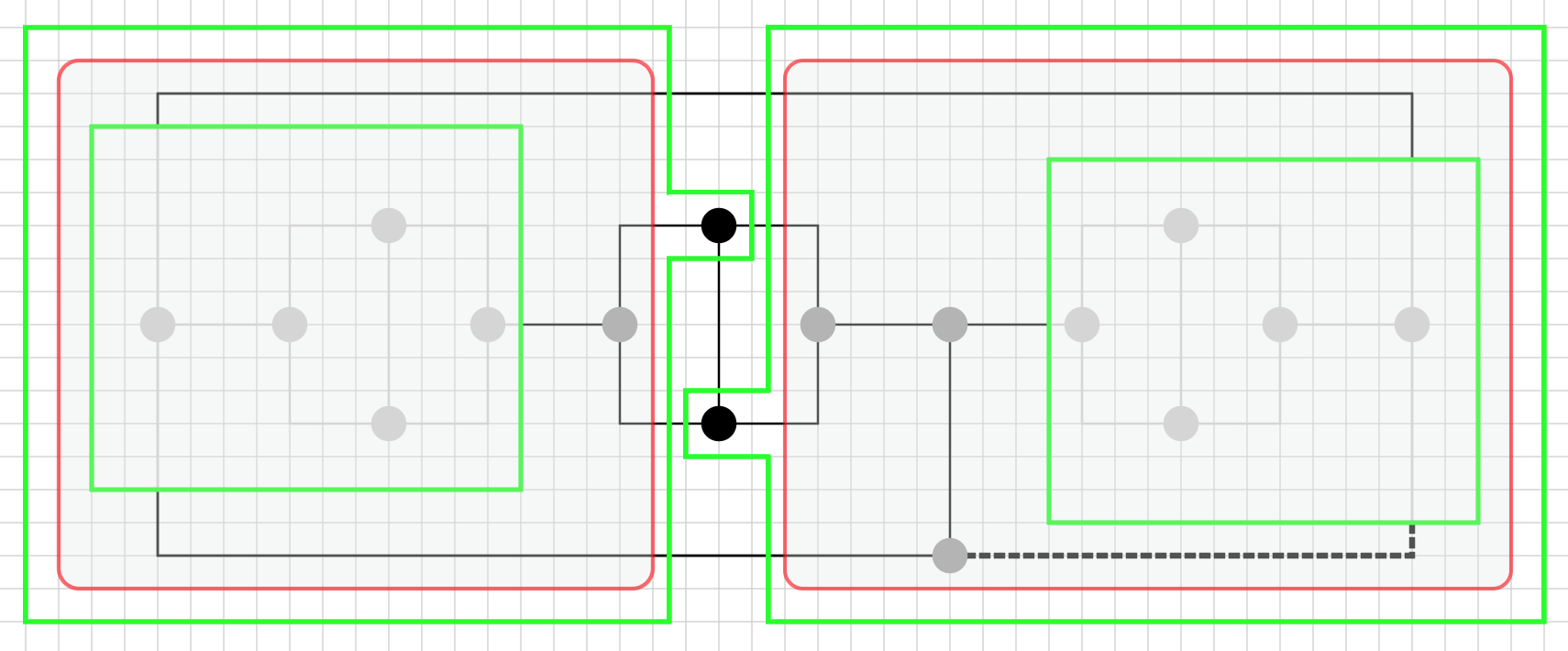}
  \includegraphics[scale=.4]{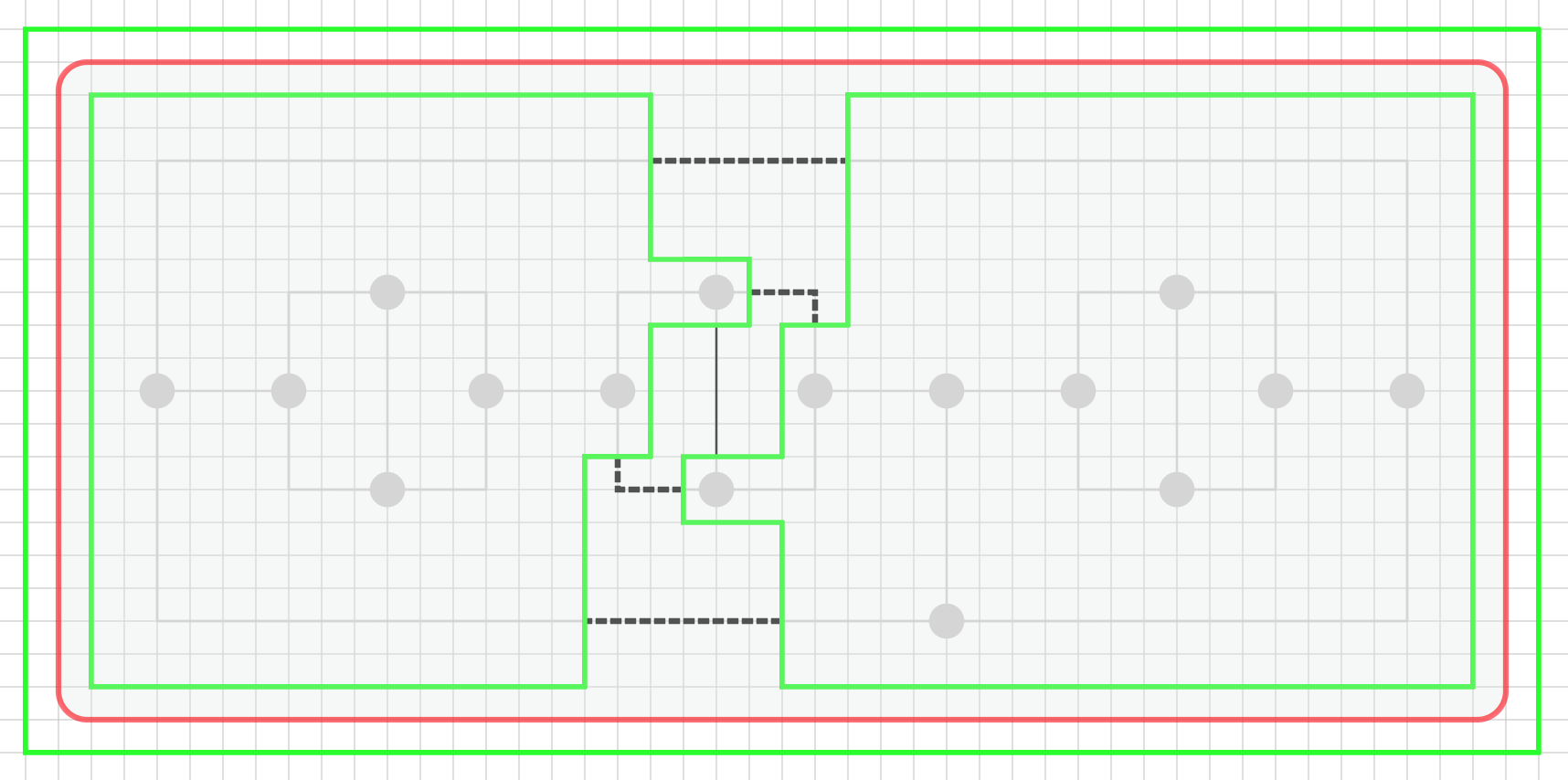}
\end{center}
\vspace{-.2in}
\caption{\label{fig:type-a-vs-type-b-again-A}
  $\KS$'s progression on the graph in Figure~\ref{fig:type-a-vs-type-b-again},
  after \textbf{rounds~1} and~\textbf{2} (top figure),
  \textbf{rounds~3} and~\textbf{4} (second figure from top),
  \textbf{rounds~5} and~\textbf{6} (third figure from top),
  and \textbf{rounds~7} and~\textbf{8} (bottom figure).
  $\KS$'s \textbf{rounds~4}, \textbf{rounds~6}, and \textbf{rounds~8},
  contract two, one, and four self-loops (dashed edges), respectively.
  }
\end{figure}

\begin{figure}[H]
\vspace{-.12in}
\begin{center}
  \includegraphics[scale=.45]{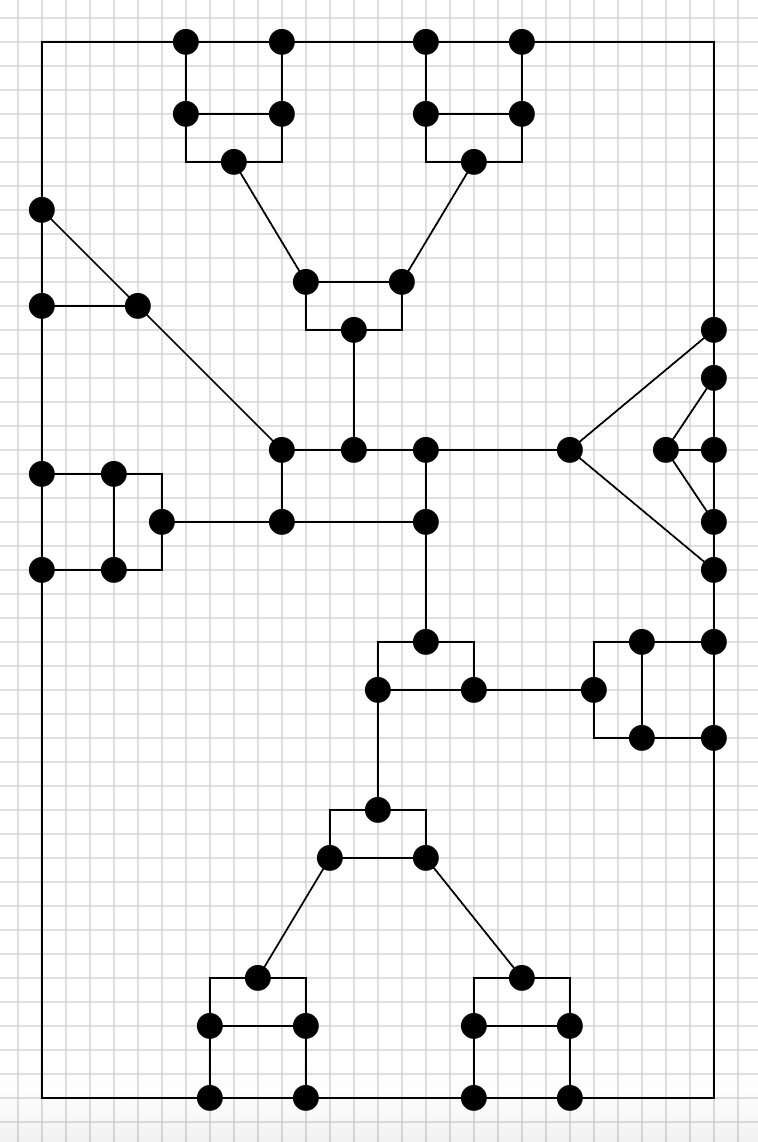}
  \hspace{2.0cm}
  \includegraphics[scale=.45]{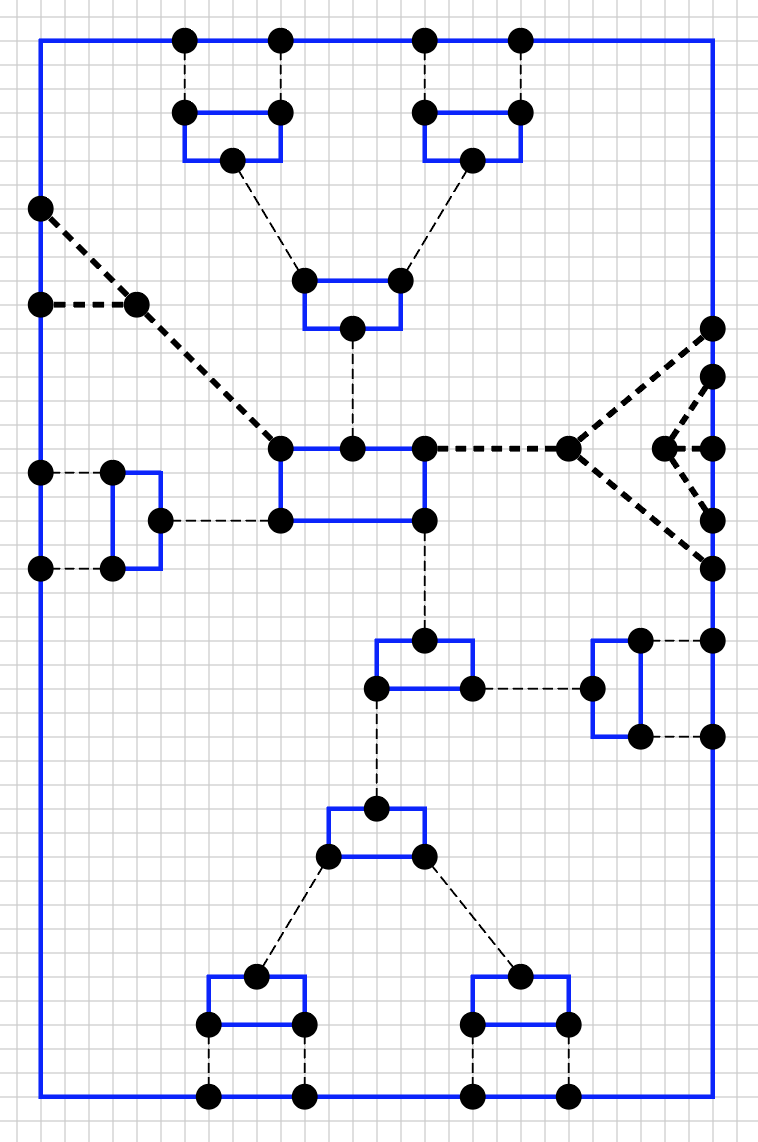}
\end{center}
\vspace{-.2in}
\caption{\label{fig:54-vertices-1}
  A $3$-regular plane graph (left)
  decomposed into 11 cycles and 24 ICT's (right), 21 
  ICT's are single-edge (light dashed edges) and 3 are multi-edge (bold
  dashed edges). Fourteen of the 24 initial ICT's are eligible for collapse;
  of these 14, one ICT is type-$a$ (nested multi-edge on the right),
  and 13 ICT's are type-$b$.
  }
\end{figure}

\begin{figure}[H]
\vspace{-.12in}
\begin{center}
  \includegraphics[scale=.45]{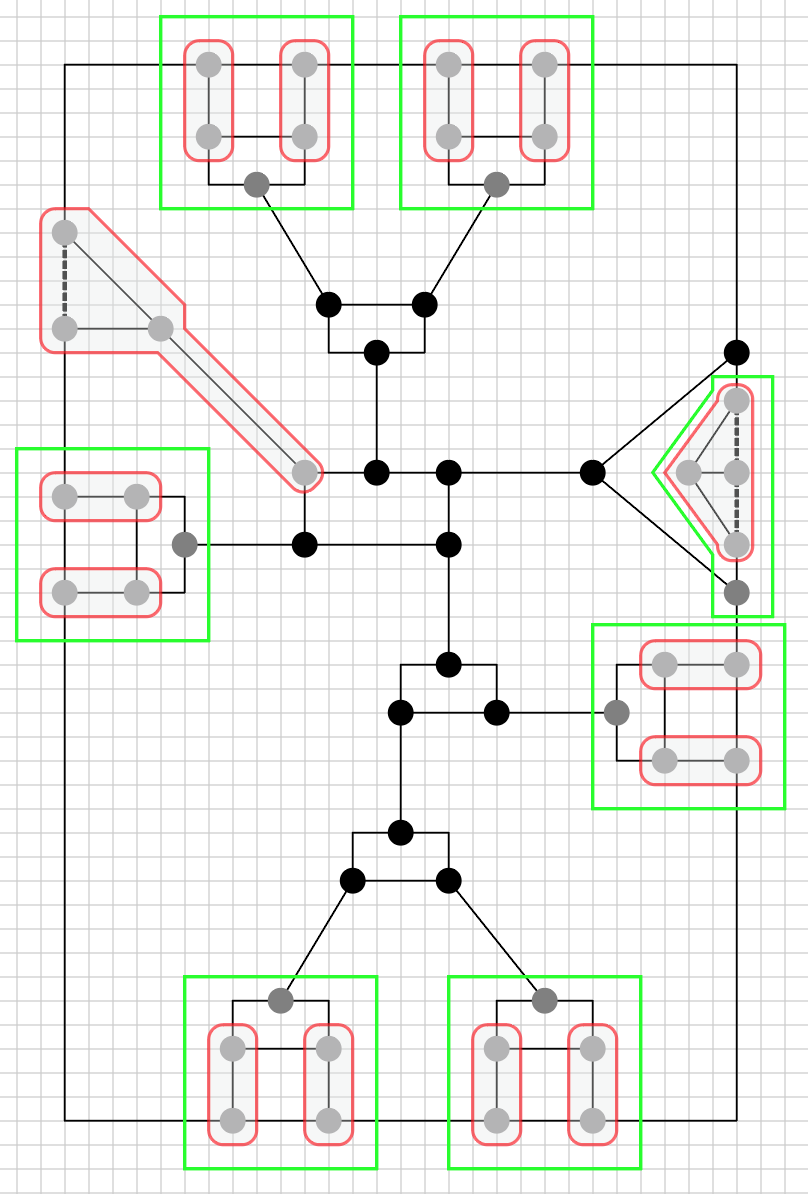}
  \hspace{0.95cm}
  \includegraphics[scale=.45]{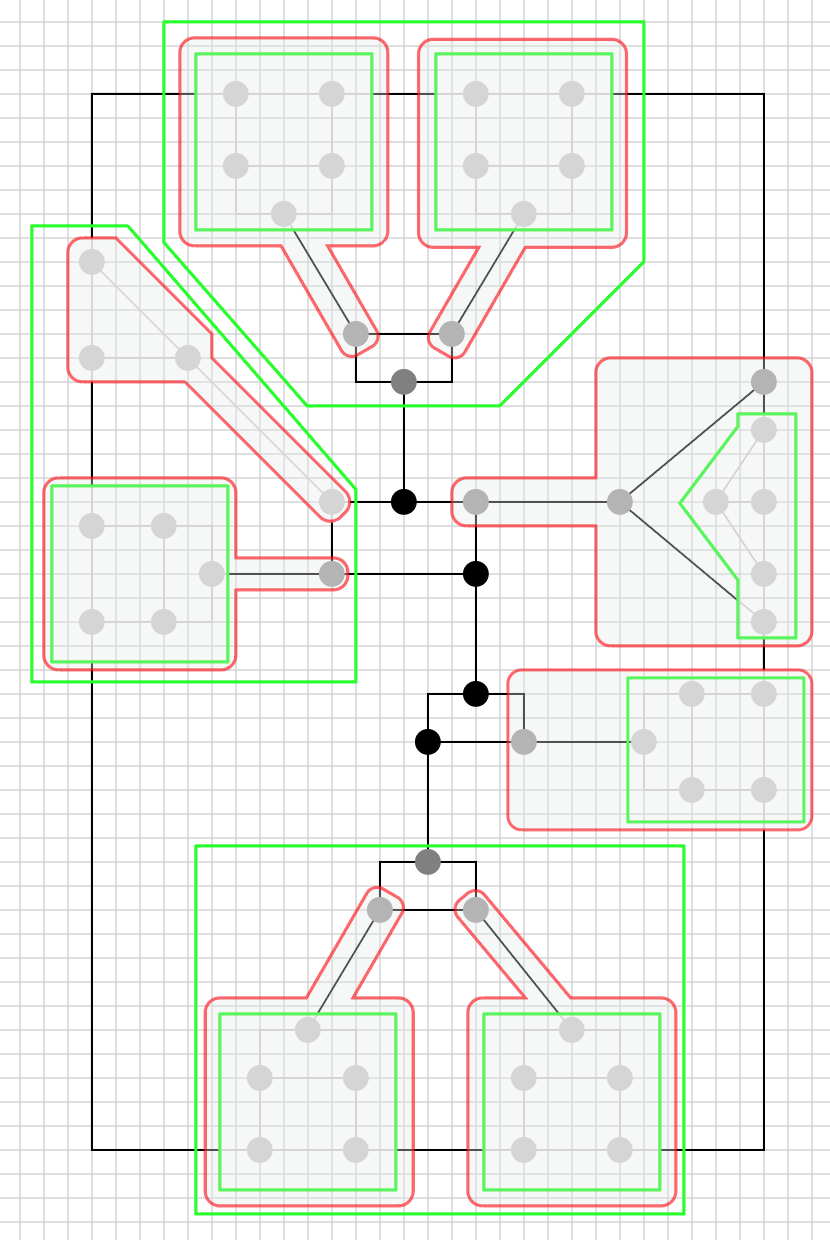}
\end{center}
\vspace{-.2in}
\caption{\label{fig:54-vertices-2}
  Progression of algorithm $\KS$ on the graph in Figure~\ref{fig:54-vertices-1}.
  All ICT's eligible for collapse on the left, 7 of them, are type-$b$.
  All ICT's eligible for collapse on the right, 2 of them, are type-$b$; one
  of these two involves the topmost super vertex, the other involves the
  bottommost super vertex.
  }
\end{figure}

\begin{figure}[H]
\vspace{-.12in}
\begin{center}
  \includegraphics[scale=.45]{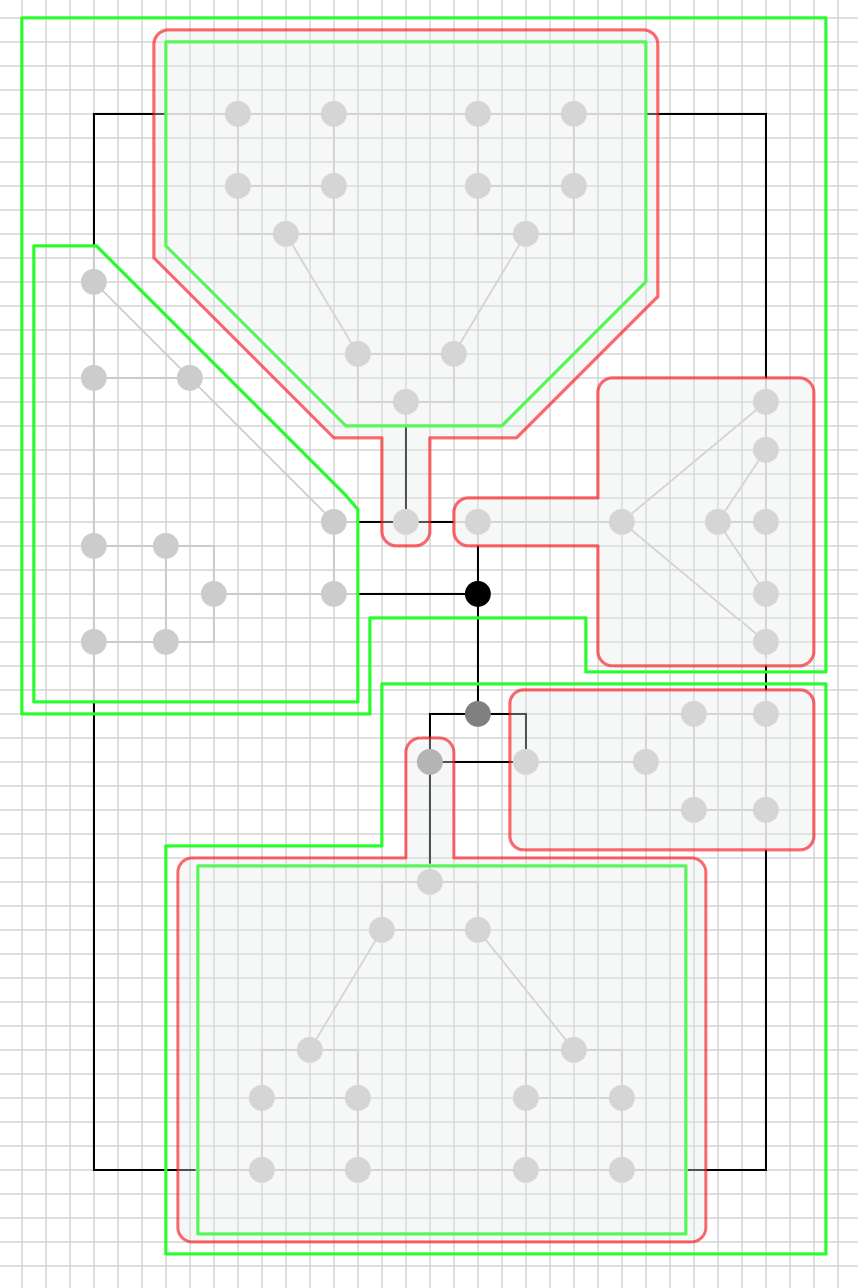}
  \hspace{0.2cm}
  \includegraphics[scale=.45]{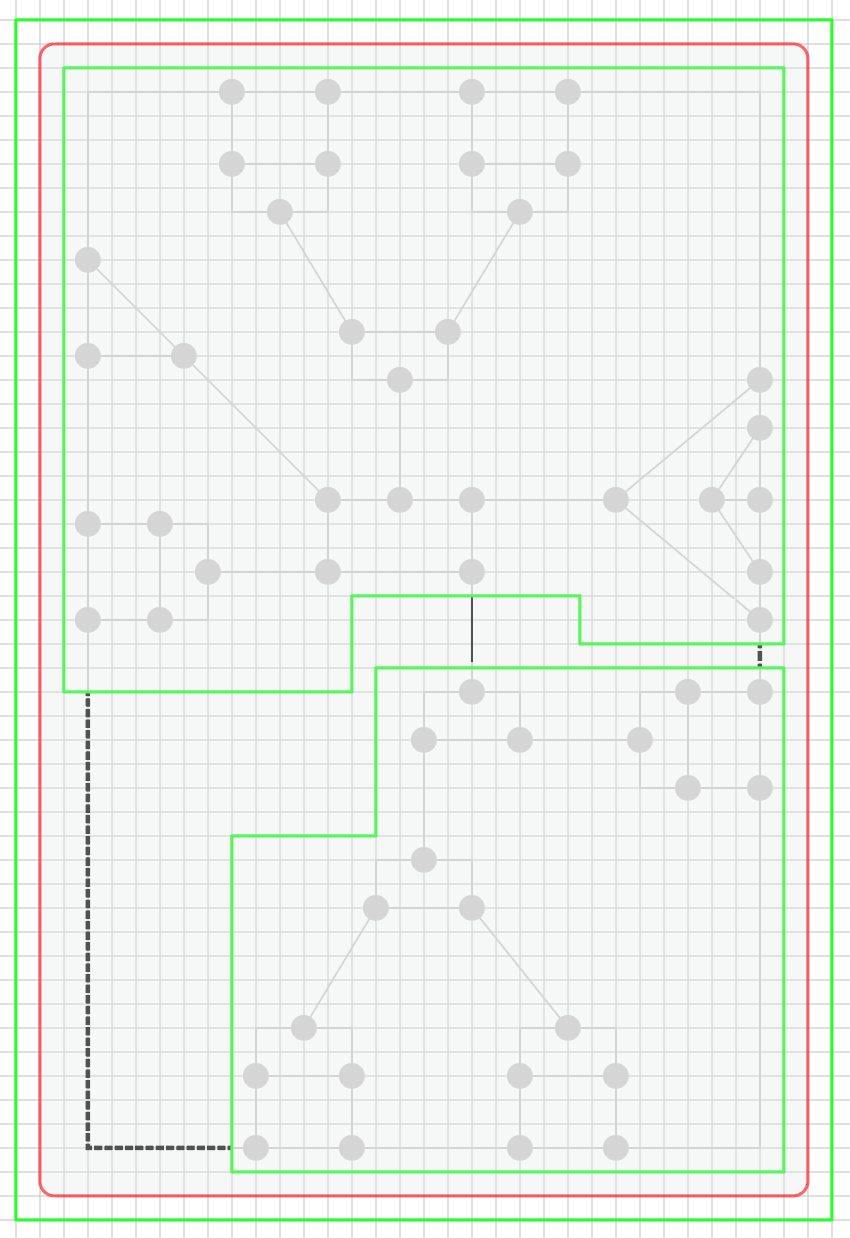}
\end{center}
\vspace{-.2in}
\caption{\label{fig:54-vertices-3}
  Progression of algorithm $\KS$ on the graph in Figures~\ref{fig:54-vertices-1}
  and~\ref{fig:54-vertices-2}. There is only one (single-edge) ICT on the left 
  which is type-$a$ and eligible for collapse. There are no ICT's on the right.
  }
\end{figure}

\begin{example}
  \label{ex:54-vertices}
Figure~\ref{fig:54-vertices-1} shows a $3$-regular plane 
graph with $54$ vertices, complicated enough to illustrate
different aspects of our algorithm $\KS$.
Figures~\ref{fig:54-vertices-2} and~\ref{fig:54-vertices-3}
show the progression of algorithm $\KS$ on this graph.

On the left in Figure~\ref{fig:54-vertices-2}, the 14 innermost super
vertices result from the first round of \emph{collapses} 
(\textbf{round 1} of $\KS$) which contract only ICT's;
these are enclosed in red boundaries. As a result of \textbf{round 1},
two self-loops are created, shown as dashed edges on the left in
Figure~\ref{fig:54-vertices-2}. The following round of
\emph{merges} (\textbf{round 2} of $\KS$) contracts the self-loops
and produces 7 new super vertices, enclosed in green boundaries
on the left in Figure~\ref{fig:54-vertices-2}. Six of these 7,
enclosed in square green boundaries, are each obtained in two steps:
first, by merging a super vertex with its clockwise neighbor, also a
super vertex, according to \emph{case 3} of $({\DD}.1)$; second,
by merging the resulting super vertex with its clockwise neighbor, an
outward ordinary vertex, according to \emph{case 2} of conditions
$\Set{({\DD}.1),({\DD}.2)}$. The remaining super vertex in green of those 7,
on the left in Figure~\ref{fig:54-vertices-2}, is obtained by merging a
super vertex with its clockwise neighbor, an inward ordinary vertex,
according to \emph{case 1} of $({\DD}.1)$.

Out of the 14 innermost super vertices in red on the left
in Figure~\ref{fig:54-vertices-2}, one is not involved in any
contractions of \textbf{round 2} (\ie, first round of \emph{merges});
conditions $\Set{({\DD}.1),({\DD}.2)}$ and their 5 special cases do
not apply to it.

The next round of \emph{collapses} (\textbf{round 3} of $\KS$) produces the
super vertices enclosed in red boundaries on the right in
Figure~\ref{fig:54-vertices-2}. There are 8 of these super vertices, 7
new and 1 from \textbf{round 1} that was not involved in any
contractions in the intermediate \textbf{round 2}. Also on the right in
Figure~\ref{fig:54-vertices-2}, three new super vertices in green
boundaries are shown, resulting from the following round of
\emph{merges} (\textbf{round 4} of $\KS$).

The result of the next round of \emph{collapses} and the following
round of \emph{merges}, \textbf{round 5} and \textbf{round 6} of $\KS$
respectively, is shown on the left in
Figure~\ref{fig:54-vertices-3}. At this point, all the ordinary
vertices in the initial graph are included in one of two disjoint
super vertices. The latter two super vertices are connected by two
cycle edges and one inter-cycle edge; these are the three remaining
uncontracted edges.

The contraction of that last inter-cycle edge is the result of one
more round of \emph{collapses}, \textbf{round 7} of $\KS$, which creates two
self-loops from the two remaining uncontracted cycle edges (shown
as dashed edges on the right in Figure~\ref{fig:54-vertices-3}). The latter
are contracted by one more round of \emph{merges}, \textbf{round 8} of $\KS$.
\end{example}

\begin{example}
  \label{ex:42-vertices}
  In Figure~\ref{fig:42-vertices-1} is a $3$-regular plane graph with $42$
  vertices, again complicated to exhibit additional aspects of our algorithm
  $\KS$. The top of Figure~\ref{fig:42-vertices-1} shows the initial input
  graph $G$, and the bottom shows $G$'s decomposition into $5$ cycles and $18$
  ICT's. In this case, it is easy to see that $\OutPlan{E}{G} = 4$.

  The progression of algorithm $\KS$ is shown in Figure~\ref{fig:42-vertices-2}.
  The top of the figure shows the result of \textbf{round 1} of \emph{collapses}
  ($7$ of them) followed by \textbf{round 2} of \emph{merges} (also
  $7$ of them). Super vertices
  resulting from \textbf{round 1} are enclosed in red boundaries, and 
  super vertices resulting from \textbf{round 2} are enclosed in green boundaries.
  There are only $6$ super vertices shown in green, not $7$, because one of them
  (at the north-west corner of the graph) is obtained by two \emph{merges}:
  the first according to \emph{case 3} of condition $({\DD}.1)$ and the
  second according to \emph{case 1} of condition $({\DD}.1)$.

  The middle of Figure~\ref{fig:42-vertices-2} shows the result of
  \textbf{round 3} of \emph{collapses} followed by \textbf{round 4}
  of \emph{merges}. As in the two previous rounds, super vertices
  resulting from \emph{collapses} are enclosed in red boundaries
  ($9$ such boundaries) and super vertices resulting from \emph{merges}
  are enclosed in green boundaries ($7$ such boundaries). Two of the
  $7$ super vertices in green are each the result of two consecutive
  \emph{merges}, one according to \emph{case 3} of condition $({\DD}.1)$
  followed by one according to \emph{case 1} of condition $({\DD}.1)$;
  these two super vertices in green are at the north-west corner of the graph
  in the middle of Figure~\ref{fig:42-vertices-2} and at the south-center in the
  middle of the same figure.

  One of the  \emph{merges} from \textbf{round 4} does not create a new
  super vertex; it only contracts self-loops, shown as dashed edges in the
  middle graph in Figure~\ref{fig:42-vertices-2}. A \emph{merge} operation
  that contracts self-loops is one according to \emph{case 5} of condition
  $({\DD}.1)$.

  At this point, there are only two ICT's that have not yet been contracted.
  Each of these two consists of a single ordinary vertex (shown as a boldface
  vertex in the middle graph in Figure~\ref{fig:42-vertices-2}) and
  three super vertices that share a same innermost cycle.

  The bottom of Figure~\ref{fig:42-vertices-2} shows the result of
  \textbf{round 5} of \emph{collapses} (two of them) followed by
  \textbf{round 6} of \emph{merges} (four of them). The two
  \emph{collapses} produce two super vertices, enclosed in red
  boundaries at the bottom of Figure~\ref{fig:42-vertices-2}, each
  with $5$ self-loops, shown as dashed edges. These $10$ self-loops
  are contracted by two \emph{merges} according to \emph{case 5} of
  condition $({\DD}.1)$. There are now $3$ super vertices, connected
  by a total of $5$ cycle edges that are yet to be contracted. A
  \emph{merge} according to \emph{case 3} of condition $({\DD}.1)$,
  followed by a \emph{merge} according to \emph{case 4} of condition
  $({\DD}.1)$, contract these $5$ cycle edges and produce the final
  and only super vertex.
\end{example}

\begin{custommargins}{-.70cm}{-.2cm} 
\begin{minipage}{1.1\textwidth}
\begin{figure}[H]
\begin{center}
  \includegraphics[scale=.35]{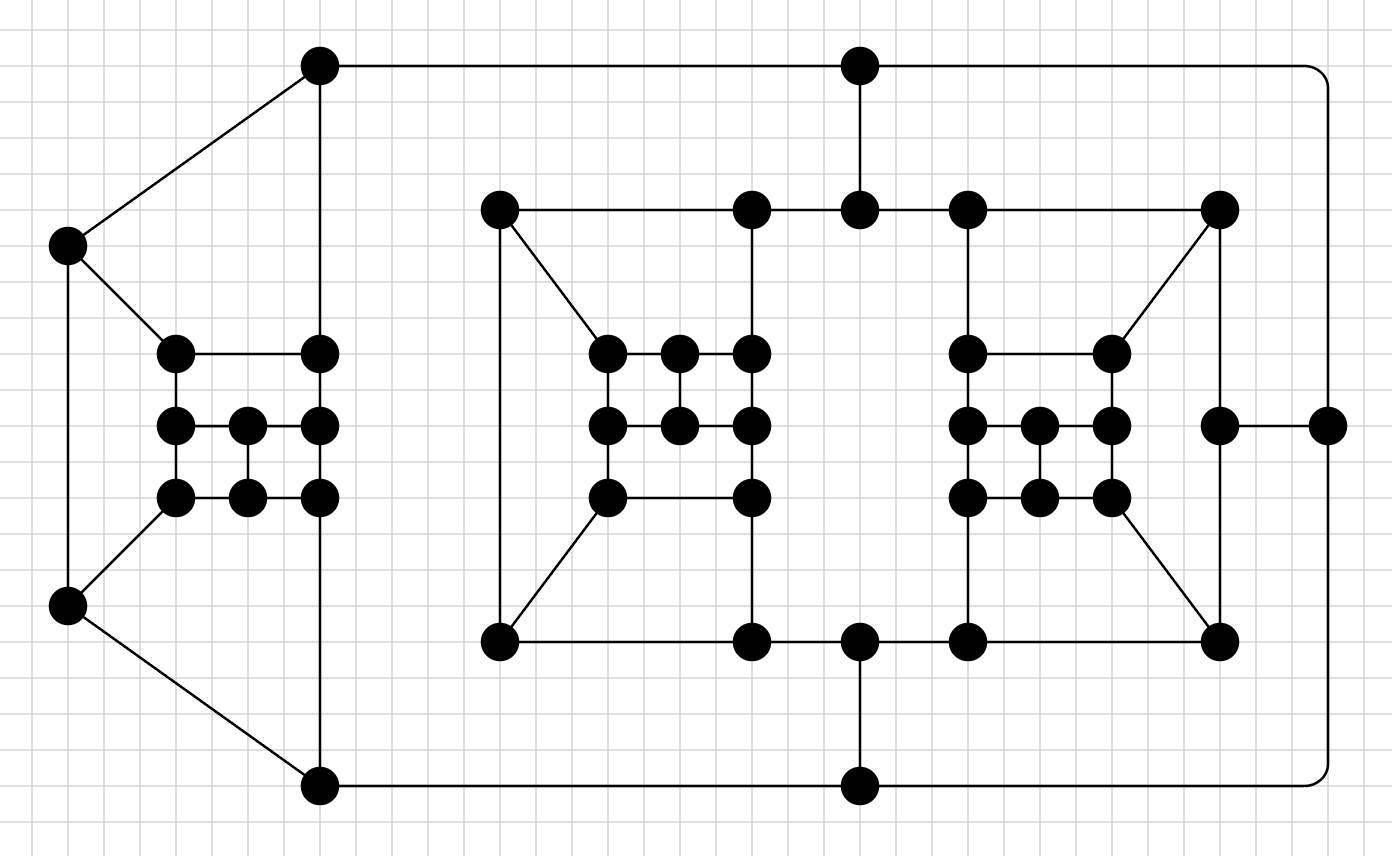} 
  \includegraphics[scale=.35]{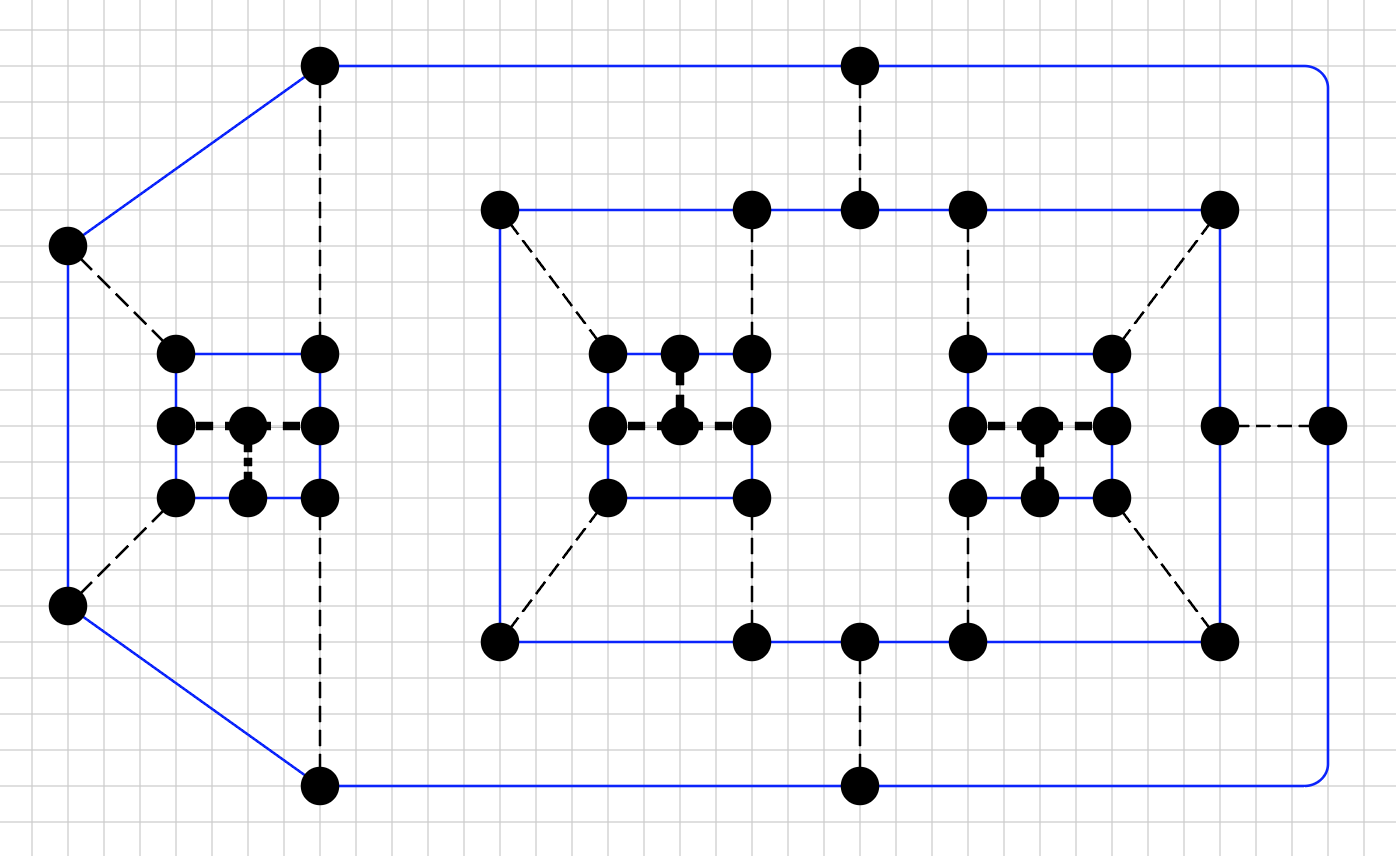}
\end{center}  
\caption{\label{fig:42-vertices-1}
  A $3$-regular plane graph (left)
  decomposed into 5 cycles and 18 ICT's (right), 15 
  ICT's are single-edge (light dashed) and 3 multi-edge (bold
  dashed). Seven of the 18 initial ICT's are eligible for collapse,
  all type-$b$.
  }
\end{figure}
\end{minipage}
\end{custommargins}

\begin{figure}[H]
\vspace{-.12in}
\begin{center}  
  \includegraphics[scale=.42]{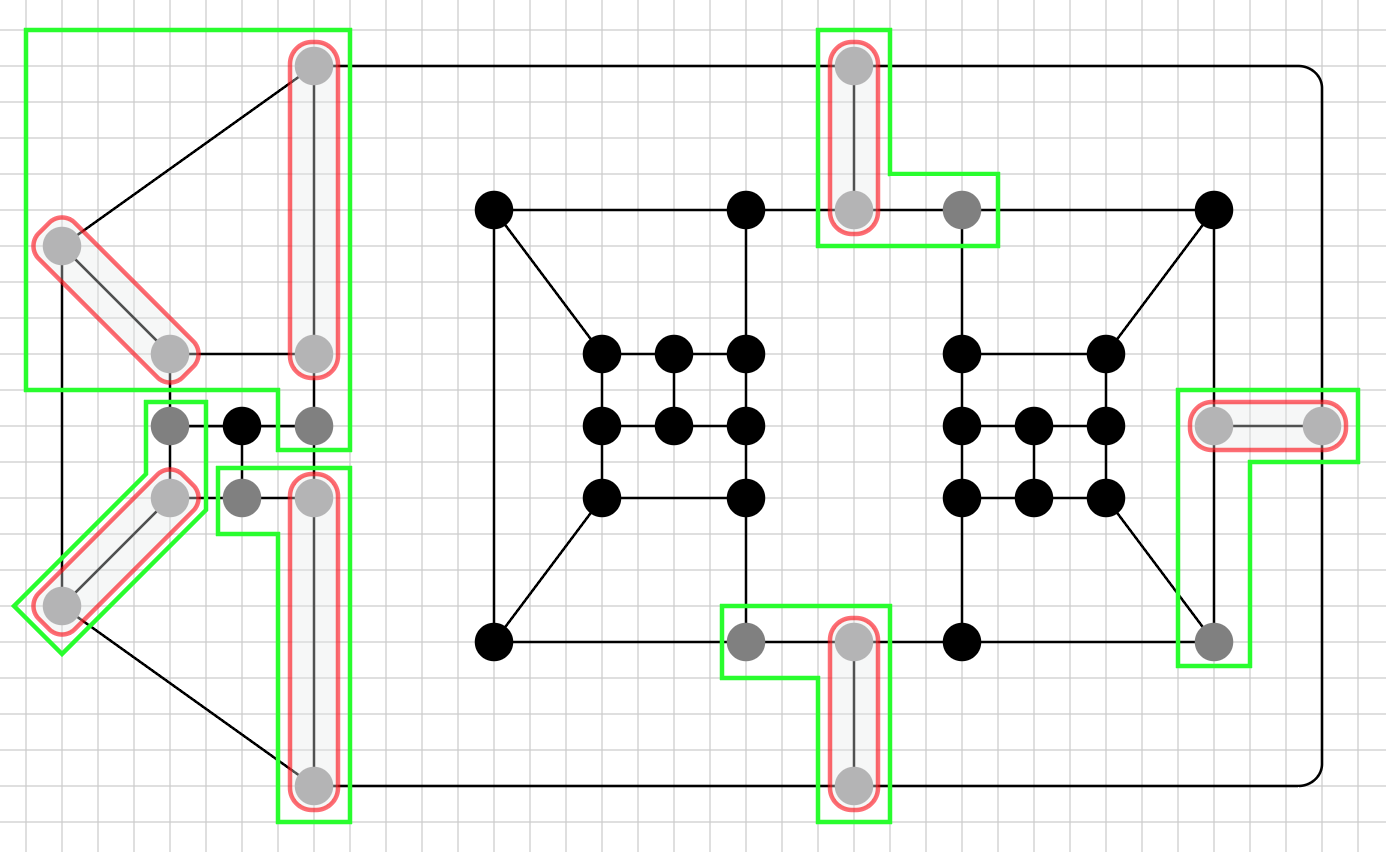} \\
  \vspace{.25cm}
  \includegraphics[scale=.42]{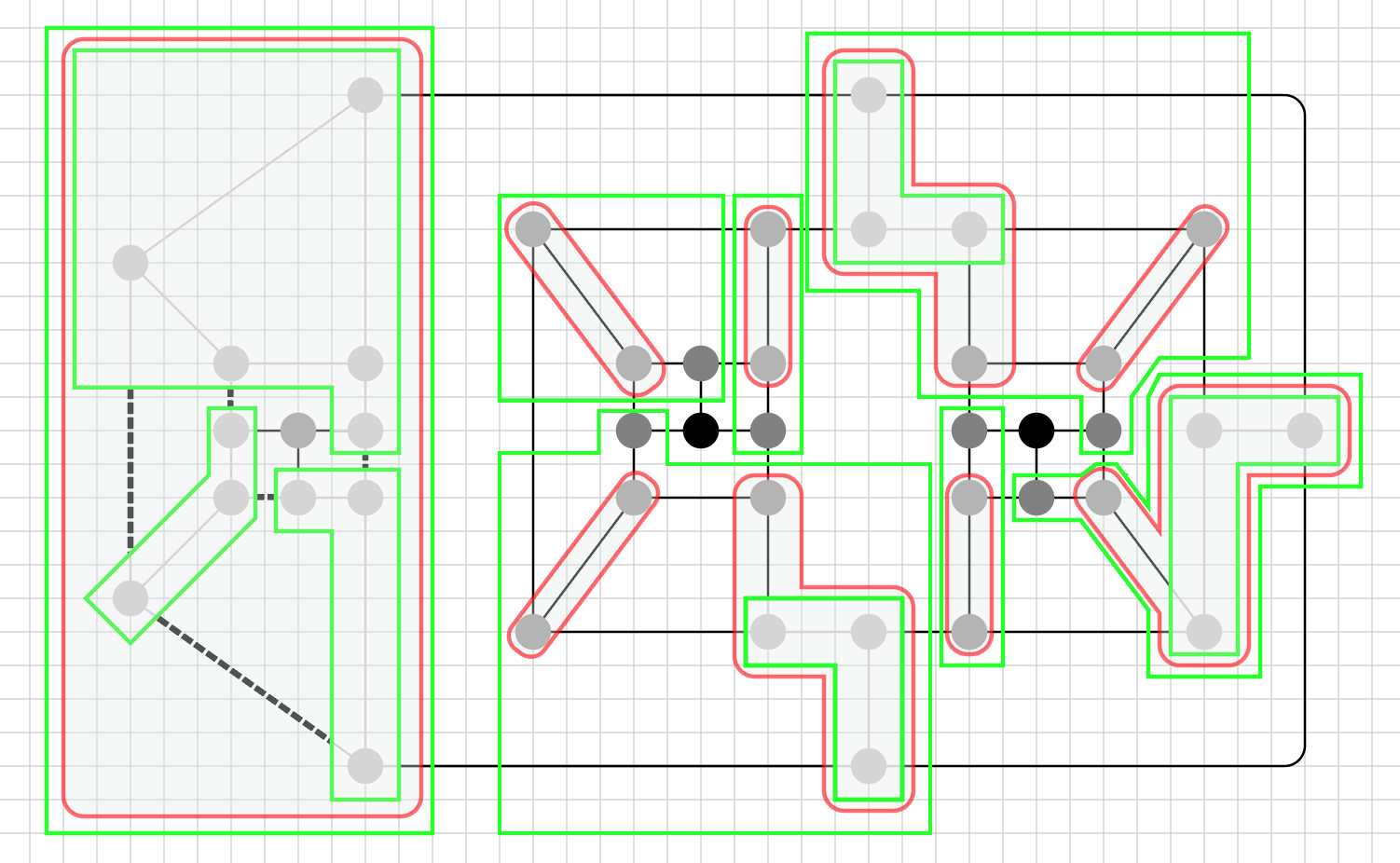} \\
  \vspace{.25cm}
  \includegraphics[scale=.42]{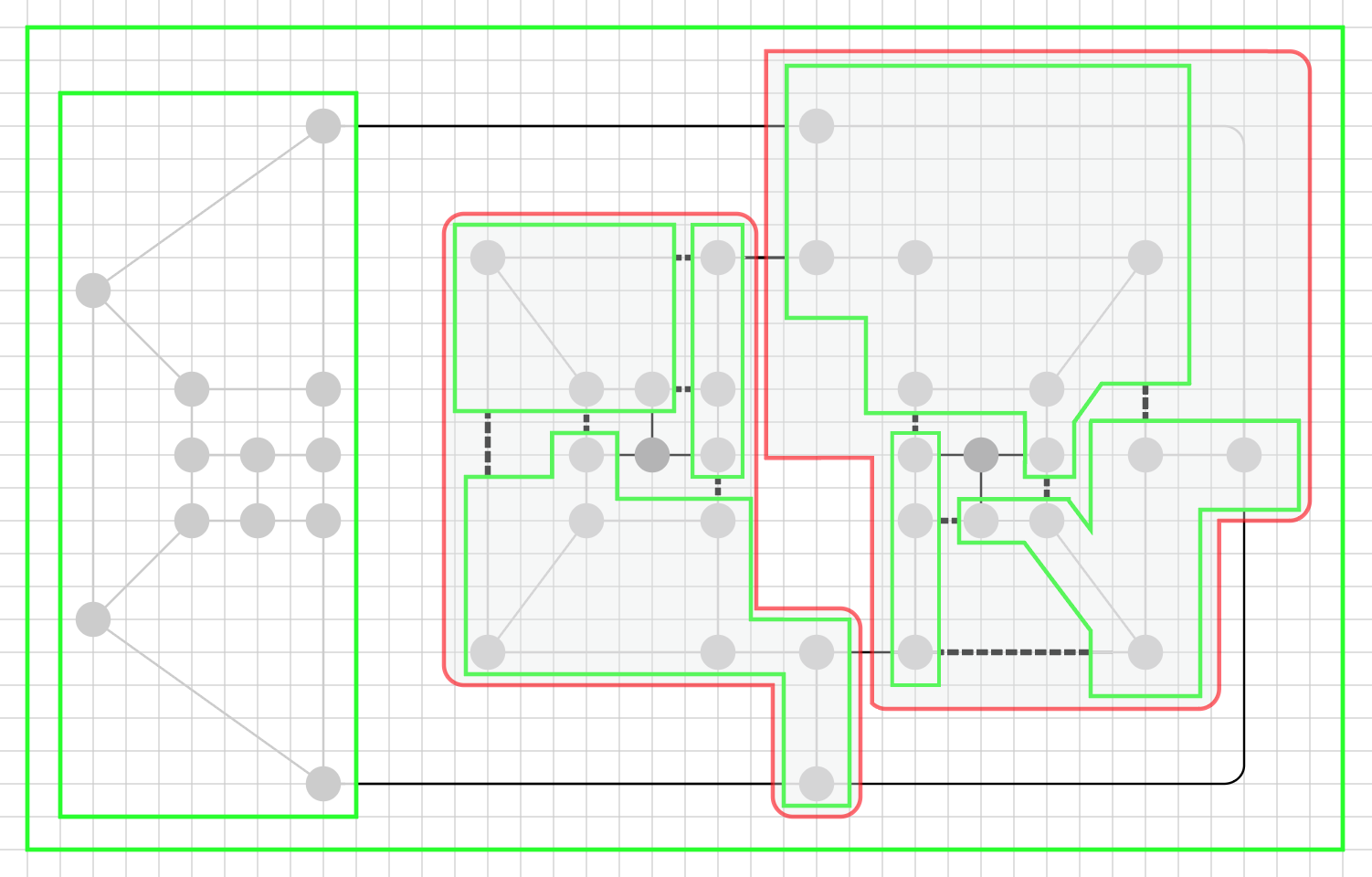}
\end{center}
\vspace{-.18in}
\caption{\label{fig:42-vertices-2}
  $\KS$'s progression on the graph in Figure~\ref{fig:42-vertices-1}.
  The top figure shows the graph after \textbf{rounds~1} and~\textbf{2},
  the middle figure after \textbf{rounds~3} and~\textbf{4},
  and the bottom figure after \textbf{rounds~5} and~\textbf{6}.
  In the top figure, there are 9 ICT's eligible for collapse; 8 of these are
  single-edge and type-$b$, and one (at the west) is multi-edge and type-$a$.
  In the middle figure, there are 2 ICT's eligible for collapse, both
  multi-edge and type-$a$. In the bottom figure, there are no ICT's
  eligible for collapse.
  }
\end{figure}


\subsection{Proof of Correctness and Complexity}
\label{sect:pseudocode}

\newcommand{\collapseA}{\textsc{Collapse}} 

We prove the correctness of our algorithm $\KS$ by referring to the
conditions $\CC$ for \emph{collapses}
(Section~\ref{sect:contract-ICT-edges}) and the conditions $\DD$ for
\emph{merges} (Section~\ref{sect:contract-cycle-edges}).  We purposely
avoid any explicit reference to the pseudocode in
Appendix~\ref{appendix:pseudocode} for two reasons: first, any such
reference would obscure the intuition underlying the proofs of
Lemmas~\ref{lem:successful}, \ref{lem:properties-of-super-vertices},
\ref{lem:algorithm-collapse}, and~\ref{lem:complexity}, as well as an
informal understanding of their correctness; second, the pseudocode in
Appendix~\ref{appendix:pseudocode} proposes one particular way (there
are others) of implementing $\KS$ and programming the conditions in
$\CC$ and $\DD$.

Let $G$ be a biconnected $3$-regular simple plane graph as in
Theorem~\ref{thm:algorithm-KS}. The execution of our algorithm
can be represented by a sequence of graphs (multigraphs after the first $G_0$):
%
\begin{itemize}
\item[$(\spadesuit)$]\quad
  $G_0 = G\quad \ G_1\quad \ G_2\quad \ G_3\quad \ \ldots\ \ \quad
  \ G_i\quad \ G_{i+1}\ \ \ldots\ \ \quad \ G_p$
\end{itemize}
%
where for every odd $i\geqslant 1$ (resp. even $i\geqslant 2$), graph
$G_i$ is obtained from the preceding $G_{i-1}$ by a maximum round of
\emph{collapses} (resp.  \emph{merges}). Thus, \textbf{round 1} is a
round of \emph{collapses}, \textbf{round 2} is a
round of \emph{merges}, \textbf{round~3} is a
round of \emph{collapses}, etc. This sequence is bound to terminate after a
finite number $p$ of rounds, because the collapse and merge operations
contract edges and there are finitely many edges. 
We say the sequence $(\spadesuit)$
\emph{terminates successfully} if the last graph $G_p$ is a single
super vertex with no edges.  We use Lemma~\ref{lem:successful} to
prove Part 1 of Theorem~\ref{thm:algorithm-KS}.

\Hide{
For the proof of Lemma~\ref{lem:successful}, keep in mind the
following, based on the preceding description of our algorithm: In the
course of algorithm execution, the edges of an ICT $T$, though not
necessarily the edges' endpoints that are leaf vertices, remain
unchanged until they all get contracted in a single \emph{collapse}
operation. The leaf vertices of $T$ that may have changed from
ordinary inward vertices to super vertices are the result of earlier
rounds of \emph{collapses} each followed by a round of \emph{merges}. Moreover, a
collapse operation contracts only ICT edges (\ie, inter-cycle tree
edges), and a merge operation contracts only cycle edges.
}

\begin{lemma}
  \label{lem:successful}
  The sequence $(\spadesuit)$ terminates successfully.
\end{lemma}

\begin{proof}
    We prove the lemma by contradiction.
  Assume the sequence $(\spadesuit)$ does not terminate successfully, \ie:
    \begin{enumerate}[itemsep=1pt,parsep=2pt,topsep=2pt,partopsep=0pt]   
    \item $\widetilde{G} = G_p$ is not a single super vertex, and
    \item no \emph{collapse} operation and no \emph{merge}
      operation can be applied to $\widetilde{G}$.
    \end{enumerate}
    For an easy case first, assume that there are no ICT edges in
    $\widetilde{G}$, because they have all been contracted in earlier
    rounds, and that all the edges left in $\widetilde{G}$ are cycle
    edges. An ICT $T$ before a \emph{collapse} operation, whether of type-$a$
    or type-$b$ (see Remark in Section~\ref{sect:contract-ICT-edges}),
    has at least one ordinary vertex; after $T$'s
    collapse, all of it vertices, ordinary or super, become merged
    into a single super vertex. Hence, all vertices in $\widetilde{G}$
    are now super vertices. Consider an innermost cycle $X$ in
    $\widetilde{G}$. Then $X = \inmost{\varphi}$ for some super vertex
    $\varphi$. If there is no super vertex $\psi\neq\varphi$ such that
    $X = \inmost{\psi}$, then $X$ is a self-loop of $\varphi$,
    otherwise we can choose $\psi$ as the clockwise neighbor of
    $\varphi$. In either case, a round of \emph{merges} can be applied to
    $\widetilde{G}$, contradicting our initial assumption.
      
    Assume next that there is an ICT $T$ which has not been collapsed
    in $\widetilde{G}$. Let all of $T$'s leaf vertices, except
    possibly for one outward ordinary $w$ in case $T$ is of type-$b$,
    be siblings on a cycle $X$, which are necessarily super vertices or
    inward ordinary vertices. Among all ICT $T'\neq T$ in
    $\widetilde{G}$ that have not been collapsed, select $T$ so that
    $\level{X}\leqslant \level{X'}$ where $X'$ is the cycle on which
    the leaf vertices of $T'$ are located. Hence, there are no
    ordinary \emph{outward} vertices on $X$, because any such ordinary outward
    vertex would be the root of a type-$b$ ICT $T'$ incident
    to $X$ from the outside and such that: (i) $T'$ has not been collapsed
    and (ii) $\level{X'} < \level{X}$, which would contradict the conditions
    for our selection of $T$.

    Moreover, all the super vertices on $X$ must be leaf vertices of
    ICT's enclosed in $X$, \ie, ICT's that are incident to
    $X$ from the inside of $X$. Indeed, if any of these super
    vertices, say $\varphi$, were not a leaf vertex of any such
    enclosed ICT, it would be possible to apply the \emph{merge} operation to
    contract the cycle edge connecting $\varphi$ with its clockwise
    neighbor on $X$ and $\widetilde{G}$ would not be the last graph in
    the sequence $(\spadesuit)$. Hence, it follows that:
    \begin{itemize}
    \item[(\$)]
      all the vertices on $X$ are either super vertices or ordinary
      inward vertices, and all of them are leaf vertices of ICT's that are
      enclosed in $X$.
    \end{itemize}
    If all the leaf vertices of the selected $T$ are consecutive siblings
    on $X$, it contradicts the assumption that no \emph{collapse} operation can
    be applied to $\widetilde{G}$. So, the leaf vertices of $T$ cannot
    be consecutive on $X$. But from this, together with fact (\$), it is
    easy to argue there must be another ICT $T_0$ whose leaf vertices
    are consecutive siblings on $X$, implying again that the \emph{collapse}
    operation can be applied to $\widetilde{G}$. We conclude that the
    sequence $(\spadesuit)$ always terminate successfully. 
\end{proof}

 We use Lemma~\ref{lem:properties-of-super-vertices} in the proof of
 Lemma~\ref{lem:algorithm-collapse}, and we use the latter to prove
 Part 2 of Theorem~\ref{thm:algorithm-KS}. In the lemmas to follow,
 $G$ is possibly a plane multigraph resulting from earlier
 applications of the \emph{collapse} and \emph{merge} operations.

\begin{lemma}
  \label{lem:properties-of-super-vertices}
  Let $T$ be an ICT in $G$ satisfying condition $({\CC}.2)$ all of
  whose leaf vertices, except possibly for one outward ordinary $w$,
  are consecutive siblings on a cycle $X$. Let
  $M = \Set{\mu_1,\ldots,\mu_p}$ be the consecutive siblings on $X$,
  each of which is a super vertex or an inward ordinary vertex.
  Assume none of the super vertices in $M$ have self-loops (as a
  result of preceding \emph{merge} operations). It then holds that:
  \begin{enumerate}[itemsep=1pt,parsep=2pt,topsep=2pt,partopsep=0pt]   
  \item For every super vertex $\varphi$ in $G$, if its clockwise neighbor is
    a super vertex $\psi\in M$, then $\varphi\in M$ ($\varphi$ is thus a leaf
    vertex of $T$); in which case
    $\inmost{\varphi} = \inmost{\psi} = X$ and there are no ordinary vertices
    on $X$ between $\varphi$ and $\psi$.%
    \footnote{If $\inmost{\varphi}=X$,
      it does not necessarily follow that $\varphi \in M$, unless
      $\varphi$'s clockwise neighbor is in $M$ -- as asserted here.}
  \item
    For every leaf vertex ${\mu}_i\in M$, the set $\cycles{{\mu}_i}$
    is a chain of nested cycles of consecutive levels, with $\level{X}$
    being the highest level and $X$ being its innermost cycle.
  \item
    For all leaf vertices ${\mu}_i, {\mu}_j\in M$, it holds
    that $\cycles{{\mu}_i}\subseteq \cycles{{\mu}_j}$
    or $\cycles{{\mu}_i}\supseteq \cycles{{\mu}_j}$.
  \end{enumerate}
\end{lemma}

\begin{proof}
  For Part1, let $\inmost{\varphi}=Y$. If $\psi$ is the clockwise
  neighbor of $\varphi$, then $Y\in\cycles{\psi}$. If $Y\neq X = \inmost{\psi}$,
  then $\varphi$ would be merged with $\psi$ in a preceding round of
  \emph{merge} operations. Hence, if $\varphi$ and $\psi$ are distinct, it
  must be that $Y = X$. And since $\psi$ is $\varphi$'s clockwise
  neighbor, there are no intervening ordinary vertices on $X$ between
  $\varphi$ and $\psi$.

  We can prove Part 2 by induction on the number $q\geqslant 1$ of
  rounds of applying the \emph{collapse} operation on the initial simple
  three-regular plane graph.  For $q = 1$, the desired conclusion is
  immediate, because all the leaf vertices of ICT's that are eligible
  for collapse in the initial graph are ordinary vertices, and each of
  these ordinary vertices straddles exactly one cycle. For the
  induction hypothesis, we assume the conclusion holds for an
  arbitrary $q\geqslant 1$ rounds of \emph{collapses} interleaved with $q$
  rounds of \emph{merges}, and we next prove the conclusion after the
  $(q+1)$-st round of \emph{collapses} is applied.  The $q$-th round of
  \emph{merges}, immediately preceding the $(q+1)$-st round of \emph{collapses},
  contracts all self-loops of the same vertex.

  Suppose the $(q+1)$-st round of \emph{collapses} results in the contraction of
  an ICT $T$ with leaf vertices $M = \Set{\mu_1,\ldots,\mu_p}$ or
  $M \cup \Set{w} = \Set{\mu_1,\ldots,\mu_p}\cup\Set{w}$,
  depending on whether $T$ is type-$a$ or type-$b$ (see Remark in
  Section~\ref{sect:contract-ICT-edges}), with $X$ being the
  innermost cycle of all the members of $M$ and $X'$ being the sole
  cycle of $w$. The contraction of such an ICT $T$ into a super vertex
  $\chi$ generally creates self-loops of $\chi$. The self-loops thus
  created, if any, are all edges of cycles in the chain of nested cycles
  formed by the members of
  $\bigcup\Set{\,\cycles{\mu_i}\;|\;\mu_i\in M\,}$. Suppose this chain
  of nested cycles is:
  \[
     X_1,\ \ldots\ ,\ X_r,\ X_{r+1},\ \ldots\ ,\ X_{r+s},\ X_{r+s+1},\ \ldots\ ,
     \ X_{r+s+t}
  \]
  with $X =  X_{r+s+t}$, which is listed in order of increasing consecutive
  levels and divided into three groups:%
  \footnote{In general, $0\leqslant r\leqslant 1$, while $s$ and $t$
    can be arbitrarily large integers $\geqslant 0$. Moreover,
    the same cycle can contribute more than one self-loops of $\chi$.
    This is provided by a finer analysis which we can ignore here.
  }
  \begin{itemize}[itemsep=1pt,parsep=2pt,topsep=2pt,partopsep=0pt]   
  \item $\Set{X_1, \ldots , X_r}$ do not contribute self-loops of $\chi$
    and are included in $\cycles{\chi}$,
  \item $\Set{X_{r+1}, \ldots , X_{r+s}}$ contribute self-loops of $\chi$
    and are included in $\cycles{\chi}$,
  \item $\Set{X_{r+s+1}, \ldots , X_{r+s+t}}$ contribute self-loops of $\chi$
    and are not included in $\cycles{\chi}$, because all their edges are
    turned into self-loops.
  \end{itemize}
  After applying the $(q+1)$-st round of \emph{merges}, the self-loops of
  $\chi$, all necessarily of levels $\geqslant\level{X'}$, are
  eliminated by contraction. Let $\chi'$ be $\chi$ after
  the elimination of all self-loops.
  If $T$ is a type-$a$, then $\cycles{\chi'} = \Set{X_1, \ldots ,
    X_{r+s}}$ and $\inmost{\chi'} = X_{r+s}$. If $T$ is a type-$b$, it
  is not difficult to check that $\level{X'} = 1+\level{X_{r+s}}$, with
  $\cycles{\chi'} = \Set{X_1, \ldots , X_{r+s}}\cup\Set{X'}$ so that
  also $\inmost{\chi'} = X'$.  This implies the conclusion of Part 2.
  
  For Part 3, if ${\mu}_i$ or ${\mu}_j$ is an ordinary vertex,
  then the conclusion is immediate, because $X$ is the sole cycle of all
  the ordinary vertices in $M$ and $X$ is also the innermost cycle of
  all the vertices in $M$. Suppose next that neither ${\mu}_i$ nor ${\mu}_j$
  are ordinary vertices and, by way of getting a contradiction,
  that $\cycles{{\mu}_i}\not\subseteq \cycles{{\mu}_j}$
  and $\cycles{{\mu}_i}\not\supseteq \cycles{{\mu}_j}$.
  Hence, there are cycles $Y_i\in\cycles{{\mu}_i}$ and $Y_j\in\cycles{{\mu}_j}$
  such that $Y_i\not\in\cycles{{\mu}_j}$ and $Y_j\not\in\cycles{{\mu}_i}$.
  But cycles $Y_i$ and $Y_j$ both enclose cycle $X$, and
  each of $\cycles{{\mu}_i}$ and $\cycles{{\mu}_j}$ is a set of nested
  cycles of consecutive levels. Since cycles $Y_i$ and $Y_j$ cannot
  cross each other, it must be that either $Y_i$ is nested in $Y_j$
  or $Y_j$ is nested in $Y_i$, \ie, $Y_i\in\cycles{{\mu}_j}$
  or $Y_j\in\cycles{{\mu}_i}$ -- but this is a contradiction.  
\end{proof}

\begin{lemma}
  \label{lem:algorithm-collapse}
  Let $T$ be an ICT in $G$ satisfying condition $({\CC}.2)$ all of
  whose leaf vertices, except possibly for one outward ordinary $w$,
  are consecutive siblings on a cycle $X$. Let $M =
  \Set{\mu_1,\ldots,\mu_p}$ be the consecutive siblings on $X$, with
  $p\geqslant 1$, each of which is a super vertex or an inward
  ordinary vertex.  Let $N=\Set{v_1,\ldots,v_q}$ be the non-leaf vertices
  of $T$, which are all ordinary of degree $3$, with $q\geqslant
  0$. Assume none of the super vertices in $M$ have self-loops (as a
  result of preceding \emph{merge} operations).

  It then holds that there is an algorithm $\collapseA$ which on input
  $T$ returns in linear time $\bigO{p+q}$ a reassembling of $T$, \ie,
  $\collapseA(T) = \B$ where $\B$ is a binary reassembling of the
  vertices in $M\cup N$ (if $T$ is type-$a$) or in $M\cup N\cup\Set{w}$
  (if $T$ is type-$b$) such that:%
  \footnote{Algorithm $\collapseA$ is the function \texttt{collapse\_tree}
    in the pseudocode in Appendix~\ref{appendix:pseudocode} and in the
    full Python implementation downloadable from the website
    \href{http://cs-people.bu.edu/bmsisson/}{Graph Reassembling}.}
  \begin{enumerate}[itemsep=1pt,parsep=3pt,topsep=2pt,partopsep=0pt]
  \item $\alpha(T,\B) \leqslant 3$.
  \item $\alpha(G,\B)
    \leqslant 1+ \max\,\SET{\,\degr{G}{\mu_i}\;\big|\;\mu_i\in M\,}$.
    \item If $\varphi$ is the super vertex resulting from contracting all
      the edges of $T$, and
      $\psi$ is the super vertex resulting from contracting all
      the self-loops of $\varphi$, then
    \[
      \degr{G}{\psi}\ \leqslant\
      \begin{cases}
        \max\,\SET{\,\degr{G}{\mu_i}\;\big|\;\mu_i\in M\,}
          & \text{if $T$ is type-$a$},
        \\[.69ex]
        2 + \max\,\SET{\,\degr{G}{\mu_i}\;\big|\;\mu_i\in M\,} \qquad
          & \text{if $T$ is type-$b$}.
      \end{cases}
    \]
  \end{enumerate}
\end{lemma}

\begin{proof}
  We first define a traversal of $T$, which includes all the tree
  edges of $T$ and excludes all the cycle edges connecting its leaf
  vertices $\Set{\mu_1,\ldots,\mu_p}$ or $\Set{\mu_1,\ldots,\mu_p}\cup\Set{w}$,
  depending on whether $T$ is type-$a$ or type-$b$, respectively.
  The traversal starts at any of the leaf vertices, say $\mu_1$, and
  moves along the edge, call it $e_1$, that connects $\mu_1$ to a
  non-leaf vertex, say $v_1$. From $v_1$ (and from every subsequent
  non-leaf vertex), the traversal continues recursively by visiting
  the left subtree (first), then the right subtree (second), and then
  finally edge $e_1$ in the reverse direction from $v_1$ to $\mu_1$
  (third).
  It is easy to check that this traversal visits the starting leaf
  vertex $\mu_1$ twice, every non-leaf vertex three times, and every
  other leaf vertex in $\Set{\mu_2,\ldots,\mu_p}$ or
  $\Set{\mu_2,\ldots,\mu_p}\cup\Set{w}$ once, and can be
  carried out in time $\bigO{p+q}$. As the traversal proceeds
  recursively, it is useful to think that every non-leaf vertex $v_j$
  with $1\leqslant j\leqslant q$, which is first reached by traversing
  an edge, say $e_j$, is the root of a binary tree whose right and
  left subtrees are the subtrees that can be aligned with edge $e_j$
  by a counterclockwise and a clockwise rotation around $v_j$,
  respectively.

  Algorithm $\collapseA$ carries out the traversal of $T$ just defined
  and simultaneously builds the vertex clusters of a binary
  reassembling $\B$ as in the statement of the lemma. It builds a
  singleton cluster $X_i \triangleq \Set{\mu_i}$ the first time it
  visits leaf vertex $\mu_i$, which is also the only time for
  $i\in\Set{2,\ldots,p}$.  If $T$ is a type-$b$ and $w$ exists, it
  also builds a singleton cluster $X_{p+1} \triangleq \Set{w}$ the
  first and only time it visits leaf vertex $w$.  For every non-leaf
  vertex $v_j$, algorithm $\collapseA$ builds a cluster $X_j$ where
  $1\leqslant j\leqslant q$ right after it visits $v_j$ the third and
  last time; the desired $X_j$ is defined as:
  \[
  X_{p+j}\ \triangleq\ \Set{v_j}\cup Y\cup Z
  \quad\text{or}\quad
  X_{p+1+j}\ \triangleq\ \Set{v_j}\cup Y\cup Z,
  \]
  depending on whether $T$ is type-$a$ or type-$b$,
  where $Y$ and $Z$ are the clusters of all the vertices (both leaf and
  non-leaf) of the right and left subtrees of $v_j$,
  respectively.

  Part 1 of the lemma now readily follows from the preceding
  analysis. Part 2 is an easy consequence of Part 1. In Part
  1, the degree of every leaf vertex is one because $T$ is considered
  in isolation from the rest of $G$; in
  Part~2, the degrees of the leaf vertices (also cycle vertices in $G$) are
  $\degr{G}{\mu_1},\ldots,\degr{G}{\mu_p}$ and $\degr{G}{w} = 3$.

  Part 3 is a straightforward consequence of
  Lemma~\ref{lem:properties-of-super-vertices}. If $T$ is
  type-$b$, let $\cycles{w} = X'$ which is necessarily $\neq X$,
  in which case $\level{X'} = 1+\level{X}$. Whether $T$ is type-$a$
  or type-$b$, it is easy to see that:
  \[
  \cycles{\psi}\ =
       \ \cycles{\varphi}
       \ -\ \Set{\;\text{all the self-loops of $\varphi$}\;}
  \]
  where $\cycles{\varphi} = \bigcup\Set{\cycles{{\mu}_i}|\mu_i\in M}$ or
  $\cycles{\varphi} = \Set{X'}\cup \bigcup\Set{\cycles{{\mu}_i}|\mu_i\in M}$,
  depending on whether $T$ is type-$a$ or type-$b$, respectively.
\end{proof}

For the next lemma,
review the \emph{pre-processing phase} and \emph{processing phase} of algorithm
$\KS$ in Section~\ref{sect:informal}.

\begin{lemma}
  \label{lem:complexity}
  The pre-processing phase and processing phase of $\KS$ on graph $G$
  are each carried out in $\bigOO{n}$ steps, each using $\bigOO{n}$
  space, where $n = \size{\vv(G)}$.
\end{lemma}

\begin{proof}
  The initial input graph $G$ is represented by an adjacency list,
  where each vertex $v$ is identified by a pair of numbers ($v$'s coordinates
  in the Cartesian plane) together with a list containing the three
  vertices to which $v$ is connected. This requires $\bigOO{n}$ space.
  The pre-processing phase decomposes the input graph $G$ into
  ICT's and cycles at each level of edge-outerplanarity,
  where every ICT can be identified by a postorder traversal (left, right, root);
  this whole pre-processing does not need to exceed $\bigOO{n}$ time
  and $\bigOO{n}$ space for its work.

  In the processing phase, every ICT $T$ is collapsed in time linear
  in the size $\size{T}$; this is the result of $\collapseA(T)$ in
  Lemma~\ref{lem:algorithm-collapse} which, according to its proof,
  also uses a postorder traversal (left, right, root).  All vertices
  (including all cycle vertices), whether ordinary or super, are each
  part of exactly one ICT, and every ICT is collapsed exactly
  once. Because any reassembling can be viewed as a tree containing
  $(2n-1)$ nodes (review definitions of reassembling
  trees in Section~\ref{sect:reassembling}), corresponding to $n$
  initial ordinary vertices plus $(n-1)$ super vertices produced in
  the course of $\KS$'s operation, the entire processing phase also takes
  $\bigOO{n}$ time and $\bigOO{n}$ space.
\end{proof}

\paragraph{Proof of Theorem~\ref{thm:algorithm-KS}.}

Part 1 of Theorem~\ref{thm:algorithm-KS} is an immediate consequence
of Lemma~\ref{lem:complexity}.  Lemma~\ref{lem:successful},
Lemma~\ref{lem:properties-of-super-vertices}, and
Lemma~\ref{lem:algorithm-collapse}, together imply Part 2 of
Theorem~\ref{thm:algorithm-KS}. More precisely, by
Lemma~\ref{lem:successful}, algorithm $\KS$ terminates when there is
only one super vertex $\varphi$ left to consider such that
$\vv(\varphi) = \vv(G)$. By Lemma~\ref{lem:properties-of-super-vertices},
if $\psi$ is one of the super vertices produced during $\KS$'s
execution at the end of a round of \emph{collapses} (and prior to the
following round of \emph{merges}) with $\cycles{\psi} =
\Set{X_1,\ldots,X_{\ell}}$, then the latter form a chain of nested
cycles of consecutive levels, which implies that $\ell\leqslant
\OutPlan{E}{G} = k$.  By Lemma~\ref{lem:algorithm-collapse}, each
additional cycle in $\cycles{\psi}$ contributes at most $2$ to
$\degr{}{\psi}$. Hence, for every super vertex $\chi$
produced at the end of a round of \emph{collapses}
during $\KS$'s execution, it holds that $\degr{}{\chi} \leqslant 2k$.
Hence, $\KS$ returns a reassembling $\B$ such that
$\alpha(G,\B) \leqslant 2k$.

\subsection{Lifting the Restriction of Biconnectedness}
\label{sect:lifting-restriction}

We do not give the pseudocode, nor do we implement, the algorithm whose
existence is asserted by the conclusion of the next corollary. We leave these
to the interested reader.

\begin{corollary}
  \label{cor:algorithm-KS}
  Identical to the statement of Theorem~\ref{thm:algorithm-KS}, except
  that $G$ is not required to be a biconnected graph.
\end{corollary}

\begin{figure}[H] 
\begin{custommargins}{0cm}{0cm} 
\begin{centering}

\noindent
%
\begin{tikzpicture}[scale=.22] 
       \newcommand\EdgeOpacity{[line width=1.4,black,opacity=.98]};
       \draw [help lines, dotted] (0, 0) grid (25,23);
       \coordinate (A4) at (0,0);
       \coordinate (A1) at (0,12);
       \coordinate (A2) at (12,12);
       \coordinate (A3) at (12,0);
       \coordinate (A5) at (17,12);
       \coordinate (A6) at (25,12);
       \coordinate (A7) at (25,3);
       \coordinate (A8) at (17,3);
       \coordinate (A9) at (15,23);
       \coordinate (A10) at (18,20);
       \coordinate (A11) at (15,17);
       \coordinate (A12) at (12,20);
       \coordinate (A13) at (15,15);
       \coordinate (B1) at (2,10);
       \coordinate (B2) at (10,10);
       \coordinate (B3) at (10,2);
       \coordinate (B4) at (2,2);
       \coordinate (B5) at (19,10);
       \coordinate (B6) at (23,10);
       \coordinate (B7) at (23,5);
       \coordinate (B8) at (19,5);
       \coordinate (B9) at (15,20);
       \coordinate (C1) at (4,8);
       \coordinate (C2) at (8,8);
       \coordinate (C3) at (8,4);
       \coordinate (C4) at (4,4);
       \coordinate (C5) at (21,10);
       \coordinate (C6) at (21,7);
       \coordinate (C7) at (19,7);
       \coordinate (D) at (6,6);
       \draw \EdgeOpacity (A1) -- (A2) ;
       \draw \EdgeOpacity (A2) -- (A3) ;
       \draw \EdgeOpacity (A3) -- (A4) ;
       \draw \EdgeOpacity (A4) -- (A1) ;
       \draw \EdgeOpacity (A5) -- (A6) ;
       \draw \EdgeOpacity (A6) -- (A7) ;
       \draw \EdgeOpacity (A7) -- (A8) ;
       \draw \EdgeOpacity (A8) -- (A5) ;
       \draw \EdgeOpacity (A9) -- (A10) ;
       \draw \EdgeOpacity (A10) -- (A11) ;
       \draw \EdgeOpacity (A11) -- (A12) ;
       \draw \EdgeOpacity (A12) -- (A9) ;
       \draw \EdgeOpacity (A2) -- (A13) ;
       \draw \EdgeOpacity (A5) -- (A13) ;
       \draw \EdgeOpacity (A11) -- (A13) ;
       \draw \EdgeOpacity (B1) -- (B2) ;
       \draw \EdgeOpacity (B2) -- (B3) ;
       \draw \EdgeOpacity (B3) -- (B4) ;
       \draw \EdgeOpacity (B4) -- (B1) ;
       \draw \EdgeOpacity (B5) -- (C5) ;
       \draw \EdgeOpacity (C5) -- (B6) ;
       \draw \EdgeOpacity (C7) -- (B5) ;              
       \draw \EdgeOpacity (B6) -- (B7) ;
       \draw \EdgeOpacity (B7) -- (B8) ;
       \draw \EdgeOpacity (B8) -- (C7) ;       
       \draw \EdgeOpacity (C1) -- (C2) ;
       \draw \EdgeOpacity (C2) -- (C3) ;
       \draw \EdgeOpacity (C3) -- (C4) ;
       \draw \EdgeOpacity (C4) -- (C1) ;
       \draw \EdgeOpacity (C5) -- (C6) ;
       \draw \EdgeOpacity (C6) -- (C7) ;            
       \draw \EdgeOpacity (A1) -- (B1) ;
       \draw \EdgeOpacity (A3) -- (B3) ;
       \draw \EdgeOpacity (A4) -- (B4) ;
       \draw \EdgeOpacity (A6) -- (B6) ;
       \draw \EdgeOpacity (A7) -- (B7) ;
       \draw \EdgeOpacity (A8) -- (B8) ;
       \draw \EdgeOpacity (A12) -- (B9) ;
       \draw \EdgeOpacity (A9) -- (B9) ;
       \draw \EdgeOpacity (A10) -- (B9) ;
       \draw \EdgeOpacity (B2) -- (C2) ;
       \draw \EdgeOpacity (B5) -- (C6) ;
       \draw \EdgeOpacity (C1) -- (D) ;
       \draw \EdgeOpacity (C3) -- (D) ;
       \draw \EdgeOpacity (C4) -- (D) ;       
       \node at (A1) {\huge $\bullet$};
       \node at (A2) {\huge $\bullet$};
       \node at (A3) {\huge $\bullet$};
       \node at (A4) {\huge $\bullet$};
       \node at (A5) {\huge $\bullet$};
       \node at (A6) {\huge $\bullet$};
       \node at (A7) {\huge $\bullet$};
       \node at (A8) {\huge $\bullet$};
       \node at (A9) {\huge $\bullet$};
       \node at (A10) {\huge $\bullet$};
       \node at (A11) {\huge $\bullet$};
       \node at (A12) {\huge $\bullet$};
       \node at (A13) {\huge $\bullet$};
       \node at (B1) {\huge $\bullet$};
       \node at (B2) {\huge $\bullet$};
       \node at (B3) {\huge $\bullet$};
       \node at (B4) {\huge $\bullet$};
       \node at (B5) {\huge $\bullet$};
       \node at (B6) {\huge $\bullet$};
       \node at (B7) {\huge $\bullet$};
       \node at (B8) {\huge $\bullet$};
       \node at (B9) {\huge $\bullet$};
       \node at (C1) {\huge $\bullet$};
       \node at (C2) {\huge $\bullet$};
       \node at (C3) {\huge $\bullet$};
       \node at (C4) {\huge $\bullet$};
       \node at (C5) {\huge $\bullet$};
       \node at (C6) {\huge $\bullet$};
       \node at (C7) {\huge $\bullet$};
       \node at (D) {\huge $\bullet$};       
 \end{tikzpicture}
\qquad\quad
\begin{tikzpicture}[scale=.27] 
       \newcommand\EdgeOpacity{[line width=1.4,black,opacity=.98]};
       \newcommand\EdgeOpacityA{[line width=1.4,dashed,black,opacity=.3]};       
       \draw [help lines, dotted] (0, 0) grid (25,23);
       \coordinate (A4) at (0,0);
       \coordinate (A1) at (0,12);
       \coordinate (A2) at (12,12);
       \coordinate (A3) at (12,0);
       \coordinate (A5) at (17,12);
       \coordinate (A6) at (25,12);
       \coordinate (A7) at (25,3);
       \coordinate (A8) at (17,3);
       \coordinate (A9) at (15,23);
       \coordinate (A10) at (18,20);
       \coordinate (A11) at (15,17);
       \coordinate (A12) at (12,20);
       \coordinate (A13) at (15,15);
       \coordinate (B1) at (2,10);
       \coordinate (B2) at (10,10);
       \coordinate (B3) at (10,2);
       \coordinate (B4) at (2,2);
       \coordinate (B5) at (19,10);
       \coordinate (B6) at (23,10);
       \coordinate (B7) at (23,5);
       \coordinate (B8) at (19,5);
       \coordinate (B9) at (15,20);
       \coordinate (C1) at (4,8);
       \coordinate (C2) at (8,8);
       \coordinate (C3) at (8,4);
       \coordinate (C4) at (4,4);
       \coordinate (C5) at (21,10);
       \coordinate (C6) at (21,7);
       \coordinate (C7) at (19,7);
       \coordinate (D) at (6,6);
       \draw  \EdgeOpacity (A1) -- (A2) ;
       \draw  \EdgeOpacity (A2) -- (A3) ;
       \draw  \EdgeOpacity (A3) -- (A4) ;
       \draw  \EdgeOpacity (A4) -- (A1) ;
       \draw \EdgeOpacity  (A5) -- (A6) ;
       \draw \EdgeOpacity  (A6) -- (A7) ;
       \draw \EdgeOpacity  (A7) -- (A8) ;
       \draw \EdgeOpacity  (A8) -- (A5) ;
       \draw \EdgeOpacity  (A9) -- (A10) ;
       \draw \EdgeOpacity  (A10) -- (A11) ;
       \draw \EdgeOpacity  (A11) -- (A12) ;
       \draw \EdgeOpacity  (A12) -- (A9) ;
       \draw \EdgeOpacityA (A2) -- (A13) ;
       \draw \EdgeOpacityA (A5) -- (A13) ;
       \draw \EdgeOpacityA (A11) -- (A13) ;
       \draw \EdgeOpacity (B1) -- (B2) ;
       \draw \EdgeOpacity (B2) -- (B3) ;
       \draw \EdgeOpacity (B3) -- (B4) ;
       \draw \EdgeOpacity (B4) -- (B1) ;
       \draw \EdgeOpacity (B5) -- (C5) ;
       \draw \EdgeOpacity (C5) -- (B6) ;
       \draw \EdgeOpacity (C7) -- (B5) ;              
       \draw \EdgeOpacity (B6) -- (B7) ;
       \draw \EdgeOpacity (B7) -- (B8) ;
       \draw \EdgeOpacity (B8) -- (C7) ;       
       \draw \EdgeOpacity (C1) -- (C2) ;
       \draw \EdgeOpacity (C2) -- (C3) ;
       \draw \EdgeOpacity (C3) -- (C4) ;
       \draw \EdgeOpacity (C4) -- (C1) ;
       \draw \EdgeOpacity (C5) -- (C6) ;
       \draw \EdgeOpacity (C6) -- (C7) ;            
       \draw \EdgeOpacity (A1) -- (B1) ;
       \draw \EdgeOpacity (A3) -- (B3) ;
       \draw \EdgeOpacity (A4) -- (B4) ;
       \draw \EdgeOpacity (A6) -- (B6) ;
       \draw \EdgeOpacity (A7) -- (B7) ;
       \draw \EdgeOpacity (A8) -- (B8) ;
       \draw \EdgeOpacity (A12) -- (B9) ;
       \draw \EdgeOpacity (A9) -- (B9) ;
       \draw \EdgeOpacity (A10) -- (B9) ;
       \draw \EdgeOpacityA (B2) -- (C2) ;
       \draw \EdgeOpacity (B5) -- (C6) ;
       \draw \EdgeOpacity (C1) -- (D) ;
       \draw \EdgeOpacity (C3) -- (D) ;
       \draw \EdgeOpacity (C4) -- (D) ;       
       \node at (A1) {\huge $\bullet$};
       \node at (A2) {\huge $\bullet$};
       \node at (A3) {\huge $\bullet$};
       \node at (A4) {\huge $\bullet$};
       \node at (A5) {\huge $\bullet$};
       \node at (A6) {\huge $\bullet$};
       \node at (A7) {\huge $\bullet$};
       \node at (A8) {\huge $\bullet$};
       \node at (A9) {\huge $\bullet$};
       \node at (A10) {\huge $\bullet$};
       \node at (A11) {\huge $\bullet$};
       \node at (A12) {\huge $\bullet$};
       \node at (A13) [opacity = .4] {\huge $\bullet$};
       \node at (B1)  {\huge $\bullet$};
       \node at (B2)  {\huge $\bullet$};
       \node at (B3)  {\huge $\bullet$};
       \node at (B4)  {\huge $\bullet$};
       \node at (B5)  {\huge $\bullet$};
       \node at (B6)  {\huge $\bullet$};
       \node at (B7)  {\huge $\bullet$};
       \node at (B8)  {\huge $\bullet$};
       \node at (B9)  {\huge $\bullet$};
       \node at (C1)  {\huge $\bullet$};
       \node at (C2)  {\huge $\bullet$};
       \node at (C3)  {\huge $\bullet$};
       \node at (C4)  {\huge $\bullet$};
       \node at (C5)  {\huge $\bullet$};
       \node at (C6)  {\huge $\bullet$};
       \node at (C7)  {\huge $\bullet$};
       \node at (D)   {\huge $\bullet$};
       \node at (12,22) {$G_1$}; 
       \node at (6,13) {$G_2$}; 
       \node at (21,13) {$G_3$}; 
       \node at (6,8.8) {$G_4$}; 
 \end{tikzpicture}

\end{centering}
\caption{The $3$-regular plane graph $G$ of
  Figure~\ref{fig:three-regular-plane-graph} is 
  on the left, its $4$ maximal biconnected components are
  on the right.
  }
\label{fig:three-regular-plane-graph-components}
\end{custommargins}
\end{figure}
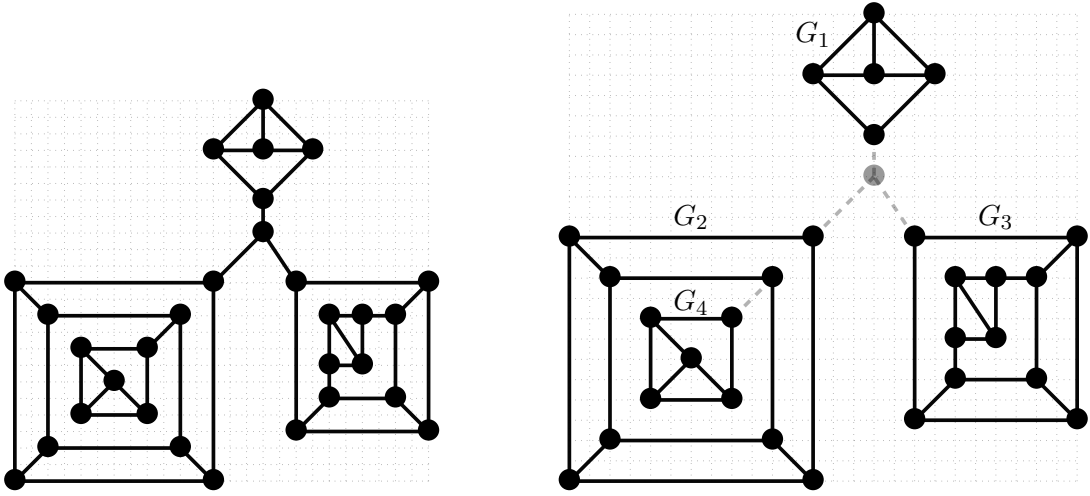

An example of a $3$-regular plane graph $G$ which is not biconnected is
shown in Figure~\ref{fig:three-regular-plane-graph}. It is reproduced
on the left in Figure~\ref{fig:three-regular-plane-graph-components}.
On the right of the latter figure, there
are $4$ biconnected components, shown as $\Set{G_1, G_2, G_3, G_4}$ such that
$\OutPlan{E}{G} = 4$  while $\OutPlan{E}{G_i} \leqslant 3$ for
$1\leqslant i\leqslant 4$. This is a general fact, implicit in the
proof of the corollary: the edge outerplanarity of every biconnected
component is bounded by the edge outerplanarity of the full graph.

\begin{sketch}
  Identifying the biconnected components of a simple graph is a
  classical result, and the operation can be carried out in linear
  time (see the original~\cite{hopcroft1973} or any of the standard
  textbooks discussing graph algorithms, and
  also~\cite{hochbaum1993}).  Let ${\KS}'$ be the algorithm to be
  defined in order to satisfy the conclusion of this corollary.

  Throughout this proof, \emph{biconnected} means \emph{maximal
    biconnected} and containing at least three vertices, ordinary or
  super. With no loss of generality, we can assume that the input
  graph $G$ is connected. 
  Initially, all vertices are ordinary, but as algorithm
  ${\KS}'$ is progressing, super vertices are created. A biconnected
  component has an \emph{outermost cycle} consisting of all the edges
  that form the boundary of the component's outerface. Initially, all
  vertices have degree $=3$, but as algorithm ${\KS}'$ starts
  executing, super vertices of degree $=1$ (and other super vertices
  of arbitrary degrees) are created. Let the initial $G$ have $p\geqslant 2$
  biconnected components, denoted:
  \[
     G_1,\ G_2,\ \ldots\ ,\ G_p .
  \]
  Each $G_i$ is such that
  $2 \leqslant \OutPlan{E}{G_i} \leqslant \OutPlan{E}{G}$.
     
  ${\KS}'$ calls algorithm $\KS$ $p$ times.  The
  reassembling of each $G_i$ is carried out separately, by applying
  $\KS$ to it, which is thus turned into a single super vertex.  The
  order in which $\Set{G_1,G_2,\ldots,G_p}$ are reassembled is not
  arbitrary: The next $G_i$ selected for reassembling is
  \emph{innermost} and with edge-boundary degree
  $\Bridges{}{\vv(G_i)} = 1$:
  \begin{itemize}[itemsep=0pt,parsep=2pt,topsep=2pt,partopsep=0pt]
  \item $G_i$ is \emph{innermost} if none of its faces contains
      another $G_j$ with $i\neq j$ and/or (super) vertices of degree
      $=1$; put differently, $G_i$ is
      \emph{innermost} if $\vv(G_i)$ is the set of all the vertices,
      ordinary or super, which are on, or enclosed in, the outermost cycle
      of $G_i$.
    \item Among the biconnected components $\Set{G_1,G_2,\ldots,G_p}$
      there is always one $G_i$ such that $\Bridges{}{\vv(G_i)} = 1$.    
  \end{itemize}
  Because the initial $G$ is $3$-regular, a biconnected component
  $G_i$ is connected to the rest of $G$ by a bridge $e = \set{\mu\,\nu}$
  both of whose endpoints $\Set{\mu,\nu}$ are articulation
  vertices. Suppose vertex $\nu\in\vv(G_i)$, so that
  $\mu\not\in\vv(G_i)$. (We denote the endpoints of $e$ 
  by the letters ``$\mu$'' and ``$\nu$'' because they may be super vertices
  as algorithm ${\KS}'$ progresses in its execution.)
  Suppose also that $G_i$ is innermost. Applying
  algorithm $\KS$ to $G_i$ produces a super vertex ${\varphi}$ of
  degree $=1$ containing exactly all the vertices in $\vv(G_i)$.
  The initial edge $e = \set{\mu\,\nu}$ is transformed into the edge
  $e' = \set{\mu\,\varphi}$, and contracting $e'$ produces a super
  vertex ${\varphi}'$ of degree $=2$ containing all the vertices in
  $\Set{\mu}\cup\vv(G_i)$. There are now two edges $\set{{\xi}_1\,{\varphi}'}$
  and $\set{{\xi}_2\,{\varphi}'}$ for some distinct vertices ${\xi}_1$ and
  ${\xi}_2$, which may be ordinary or super.

  Before proceeding to select the next innermost biconnected component
  $G_j$, algorithm ${\KS}'$ contracts one of the two edges,
  $\set{{\xi}_1\,{\varphi}'}$ or $\set{{\xi}_2\,{\varphi}'}$, to produce a new super
  vertex ${\varphi}''$ of degree $=3$.
\end{sketch}

\section{Conditions for the Optimality of Algorithm $\KS$}
\label{sect:optimality-of-KS}

We show that for a family of $3$-regular plane graphs with a
sufficiently high ``inter-cycle density'' (density of inter-cycle
trees), algorithm $\KS$ returns $\alpha$-optimal reassemblings
(Theorem~\ref{thm:optimality-of-KS}). For the same family of graphs
with a low ``inter-cyle density'', $\KS$ does not return
$\alpha$-optimal reassemblings (Proposition~\ref{thm:non-optimality-of-KS}).
What the informal expressions ``high density'' and ``low density'' mean
is made precise right after 
Theorem~\ref{thm:optimality-of-KS}.

\begin{figure}[H] 
\begin{custommargins}{-0.40cm}{-2.2cm}
\begin{center}
   \includegraphics[scale=.65]{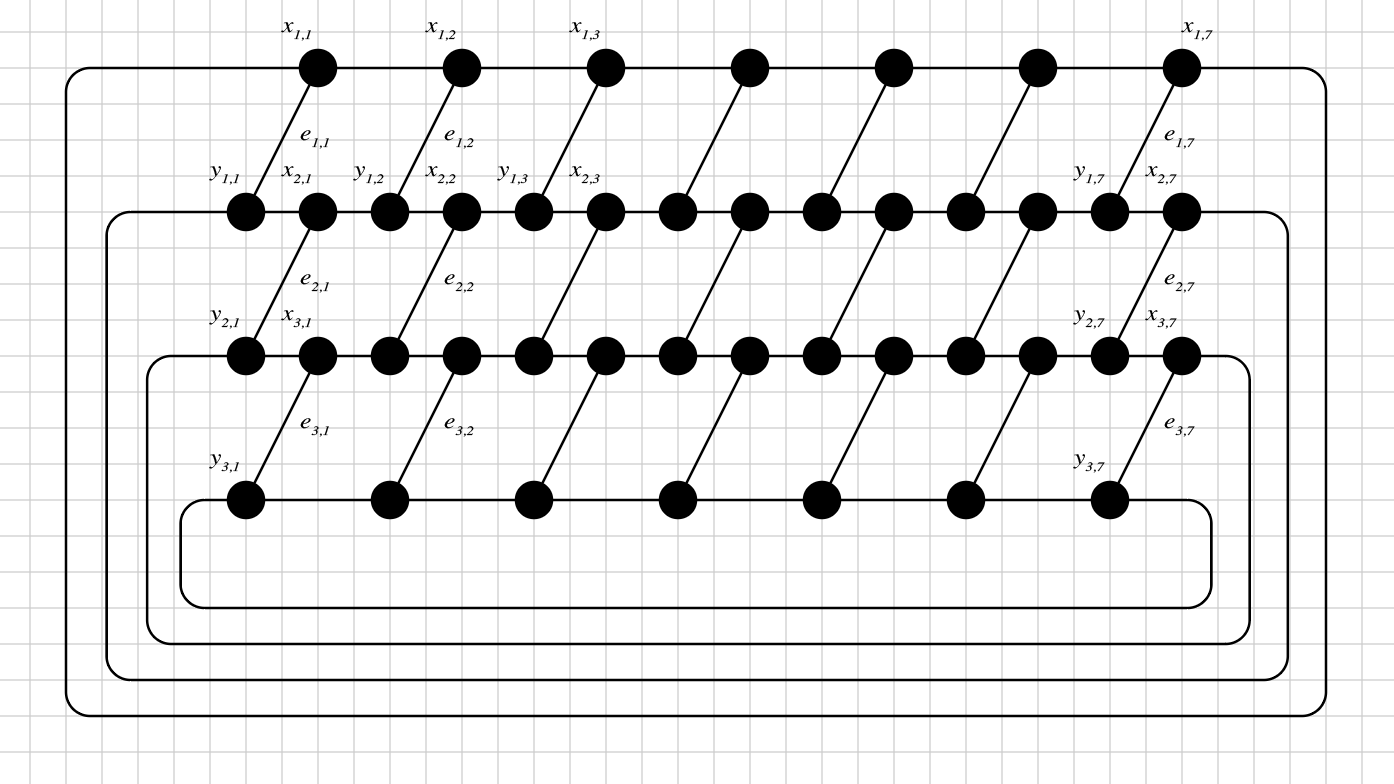}
\end{center}   
\vspace{-.25in}
   \caption{Graph $H_{f,k}$ when $k=4$ and $f(k) = 2\cdot k -1 = 7$,
   with $\OutPlan{E}{H_{f,k}} = 4$, showing
     the naming conventions on edges and vertices (on some of them only,
     not to clutter the figure).
   }
\label{fig:special-graph-E-outerplanarity-4}
\end{custommargins}
\end{figure}

Let $f$ be a monotonically increasing or constant function on the natural
numbers such that $f(x) \geqslant 3$ for all $x$.
We define an infinite family ${\HH}_{f}$ of $3$-regular plane graphs
parametrized with $f$. Each member of ${\HH}_{f}$ is assigned a
second parameter $k$, a natural number $\geqslant 2$:
\[
   {\HH}_{f}\ \triangleq\ \SET{\,H_{f,k}\;\big|\; \text{$k\geqslant 2$}\,}
\]
 such that $\OutPlan{E}{H_{f,k}} = k$.
 Until the proof of Lemma~\ref{lem:no-binary-reassembling}, we do not
 need to be specific about the function $f$, only $k$ needs to be
 operated on; until then, however, it is useful to keep in mind that
 we will choose $f$ so that $f(k) {\gg}\, k$.  The graph $H_{f,k}$ is
 shown in Figure~\ref{fig:special-graph-E-outerplanarity-4} when $k =
 4$ and $f(x) = 2x-1$, so that $f(k) = 7$.

 $H_{f,k}$ consists of $k$ concentric cycles
 $\Set{C_0,\ldots,C_{k-1}}$ such that cycle $C_{i}$ and $C_{i+1}$ are
 connected by $f(k)$ one-edge ICT's, 
 henceforth called ICE's in this section (ICE = inter-cycle edge).%
   \footnote{Note carefully that ``$C_{i}$'' here
   is unrelated to the standard notation ``$C_{i}$'' which
   refers to the cycle graph with $i$ vertices.}
 By our earlier conventions,
\[
 \level{C_0} = 0,\quad \level{C_1} = 1,\quad\ldots\quad,\quad\level{C_{k-1}}= k-1
\]
 For convenience, we use a double-indexing to denote the ICE's.
 The ICE $e_{i,j}$ is an edge whose first index $i$ denotes its level
 and its second index $j$ ranges over the set $\Set{1,\ldots, f(k)}$.
 Moreover, all the level-$i$ ICE's occur 
 in the following order: $e_{i,1}, e_{i,2}, \ldots, e_{i,f(k)}$ in a clockwise
 direction, for every $1\leqslant i\leqslant k-1$.

 For every $1\leqslant i\leqslant k-1$,
 we identify the two endpoints of ICE $e_{i,j}$ by the vertices
 $x_{i,j}$ and $y_{i,j}$, such that $x_{i,j}\in\vv(C_{i-1})$ and
 $y_{i,j}\in\vv(C_{i})$. The vertices and edges of cycle $C_0$ are
 therefore:
 \begin{alignat*}{5}
   & \vv(C_0)\ &&=\ &&\Set{\,x_{1,1},\;x_{1,2},\ \ldots\ ,\;x_{1,f(k)}}
   \\[1.2ex]
   & \ee(C_0)\ &&=
   \ &&\Set{\,\set{x_{1,1}\ \ x_{1,2}},\;\set{x_{1,2}\ \ x_{1,3}},
            \;\ldots\;,\;\set{x_{1,f(k)-1}\ \ x_{1,f(k)}},\;\set{x_{1,f(k)}\ \ x_{1,1}}\,}
 \end{alignat*}
 The vertices and edges of cycle $C_i$, for $1\leqslant i\leqslant k-2$, are:
 \begin{alignat*}{5}
   & \vv(C_i)\ &&=\ &&\Set{\,y_{i,1},\;x_{i+1,1},
       \;y_{i,2},\;x_{i+1,2},\ \ldots\ ,\;y_{i,f(k)},\;x_{i+1,f(k)}\,} 
   \\[1.2ex]
   & \ee(C_i)\ &&=
   \ &&\Set{\,\set{y_{i,1}\ \ x_{i+1,1}},\;\set{x_{i+1,1}\ \ y_{i,2}},
            \;\ldots\;,\;\set{y_{i,f(k)}\ \ x_{i+1,f(k)}},\;\set{x_{i+1,f(k)}\ \ y_{i,1}}\,}
 \end{alignat*}
 The vertices and edges of cycle $C_{k-1}$ are:
 \begin{alignat*}{5}
   & \vv(C_{k-1})\ &&=\ &&\Set{\,y_{k-1,1},
       \;y_{k-1,2},\ \ldots\ ,\;y_{k-1,f(k)}\,} 
   \\[1.2ex]
   & \ee(C_{k-1})\ &&=
   \ &&\Set{\,\set{y_{k-1,1}\ \ y_{k-1,2}},\;\set{y_{k-1,2}\ \ y_{k-1,3}},
      \;\ldots\;,\;\set{y_{k-1,f(k)-1}\ \ y_{k-1,f(k)}},\;\set{y_{k-1,f(k)}\ \ y_{k-1,1}}\,}
 \end{alignat*}

 Figure~\ref{fig:special-graph-E-outerplanarity-4}
 shows the graph $H_{f,k}$ and the naming conventions of edges and vertices,
 when $k=4$ and $f(k)=7$.
The sequence of definitions and lemmas, from 
Definition~\ref{def:clusters} to Lemma~\ref{lem:regular-clusters},
is to show that, if we want to maximize the size $\size{X}$ of
a vertex cluster $X\subseteq \vv(H_{f,k})$ whose edge-boundary degree
$\Bridges{}{X}$ is a fixed constant $c$ strictly less than $2k$,
then we can restrict attention
to clusters that we call \emph{strongly regular}
(Definition~\ref{def:regular-clusters}).

\begin{figure}[H] 
\begin{custommargins}{-0.40cm}{-2.2cm}
\begin{center}
   \includegraphics[scale=.45]{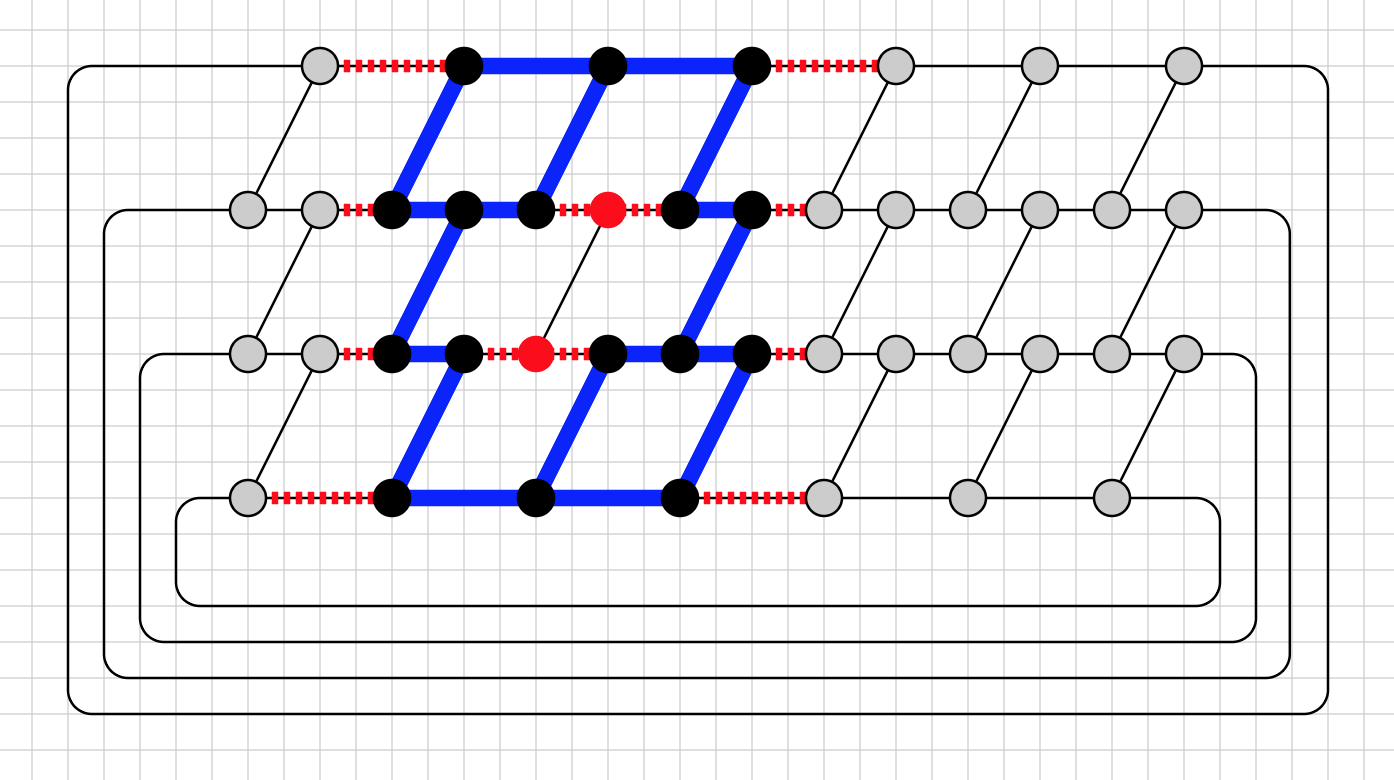}
\end{center}   
\vspace{-.25in}
   \caption{Example for Definition~\ref{def:clusters}: Graph $H_{f,k}$ is
   the same as in Figure~\ref{fig:special-graph-E-outerplanarity-4}. The black
   vertices form a cluster $X$ in $H_{f,k}$, the red vertices
   form a hole $Y$ in $X$. The union $Z = X\cup Y$ is a full cluster.}
\label{fig:clusters}
\end{custommargins}
\end{figure}

\begin{definition}{Clusters, Holes in Clusters, Full Clusters}
\label{def:clusters}
A non-empty subset $X\subseteq \vv(H_{f,k})$ is \emph{connected}
iff between any two vertices of $X$ there is a path. 
A \emph{cluster} in graph $H_{f,k}$
is a non-empty connected subset $X\subseteq \vv(H_{f,k})$.

Let $X, Y\subseteq \vv(H_{f,k})$ be clusters in graph $H_{f,k}$.
  We say $Y$ is a \emph{hole} in $X$, or that $X$ \emph{contains the
  hole} $Y$, iff all of the following conditions are satisfied:
\begin{enumerate}[itemsep=0pt,parsep=2pt,topsep=2pt,partopsep=0pt]
   \item
     $X\cap Y = \varnothing$, \ie, $X$ and $Y$ are disjoint.
   \item
     $\bridges{}{X,Y} \neq\varnothing$,
         \ie, there are edges with one endpoint in $X$
         and one endpoint in $Y$.
   \item
     $\bbridges{}{\vv(H_{f,k}) - (X\cup Y), Y} = \varnothing$,
     \ie, there is no edge with one endpoint in $\vv(H_{f,k}) - (X\cup Y)$
      and one endpoint in $Y$.         
\end{enumerate}
We say a cluster $X\subseteq \vv(H_{f,k})$ is \emph{full} iff $X$ contains no holes.
An example of a cluster $X$ with a hole $Y$
is shown in Figure~\ref{fig:clusters}, and a full cluster is $X\cup Y$.
\end{definition}

\begin{lemma}
\label{lem:full-clusters}
Let $X\subseteq \vv(H_{f,k})$ be a cluster in $H_{f,k}$. If $X$ contains holes,
then there is a cluster $Z\subseteq \vv(H_{f,k})$ without holes such that:%
    \footnote{
    A stronger conclusion in fact holds: The two
    inequalities ``$\geqslant$'' and ``$\leqslant$'' can be changed to strict
    inequalities ``$>$'' and ``$<$''. We do not need the stronger conclusion.}
\begin{enumerate}[itemsep=0pt,parsep=2pt,topsep=2pt,partopsep=0pt]
   \item $\Bridges{}{Z}  \leqslant \Bridges{}{X} \quad\text{and}
         \quad Z \supseteq X$.
   \item For every full cluster $Z' \supseteq X$, it holds that 
         $\size{Z}\leqslant\size{Z'}$.
\end{enumerate}
In words, we can minimally augment ${X}$ and eliminate all the holes in
it without increasing $\Bridges{}{X}$.
\end{lemma}
\begin{proof}
Full clusters exist, with $\vv(H_{f,k})$ being the largest full cluster.
The desired $Z$ in the lemma statement is the smallest full cluster such
that $Z\supseteq X$. It is straightforward to see that
$\Bridges{}{X} \geqslant \Bridges{}{Z}$, since each hole in $X$ delete
some vertices from $X$ and increases $\Bridges{}{X}$.
\end{proof}

The situation described in the statement of Lemma~\ref{lem:full-clusters}
is illustrated in Figure~\ref{fig:clusters}: 
For the full cluster $Z = X\cup Y$, we
have $8 = \Bridges{}{Z}  \leqslant \Bridges{}{X} = 12$.

\newcommand{\bound}[1]{\text{\em frontier}(#1)} 

\begin{figure}[H] 
\begin{custommargins}{-0.40cm}{-2.2cm}
\begin{center}
   \includegraphics[scale=.45]{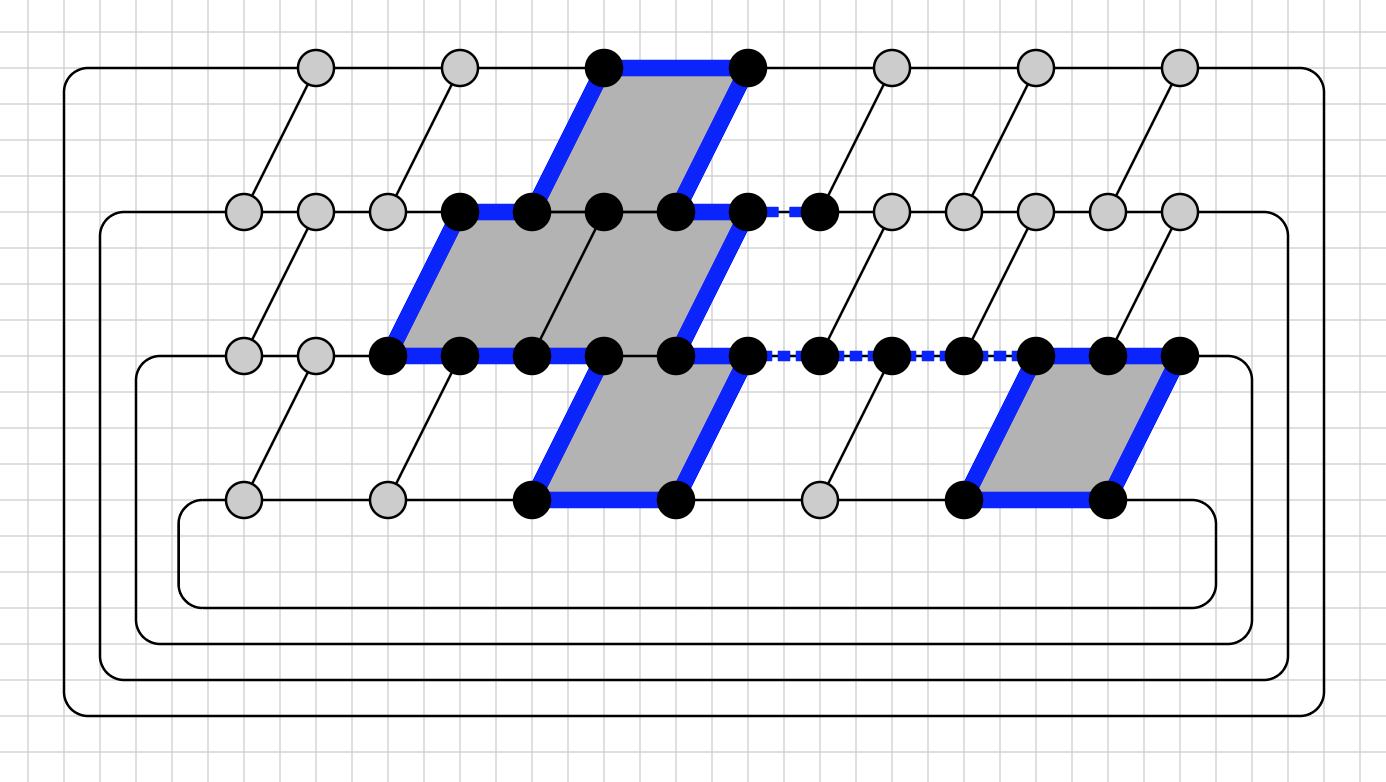}
\end{center}   
\vspace{-.25in}
   \caption{Example for Definitions~\ref{def:boundaries}
   and~\ref{def:cut-and-dangling}: Graph $H_{f,k}$ is
   the same as in Figure~\ref{fig:special-graph-E-outerplanarity-4}.
   The black vertices form a full cluster $X$ in $H_{f,k}$. The faces in
   light gray are \emph{inside} $X$; all the other faces are
   \emph{outside} $X$. The edges in boldface form $\bound{X}$; not all edges
   in $\bound{X}$ bound a face inside $X$ (the dashed boldface edges) . }
\label{fig:boundaries}
\end{custommargins}
\end{figure}

\begin{definition}{Frontiers of Full Clusters}
\label{def:boundaries}
\label{def:frontiers}
Other than its outermost face and its innermost face, every other face
of $H_{f,k}$ is bounded by $5$ or $6$ edges. Call the outermost and innermost
faces the \emph{large faces} of $H_{f,k}$ (there are only two of them),
and all the other faces the \emph{small faces} of $H_{f,k}$ (there are
$(k-1)\cdot f(k)$ of them).

Let $X\subseteq \vv(H_{f,k})$ be a full cluster in graph $H_{f,k}$. Every face
of $H_{f,k}$ is either inside or outside $X$. We say
a face of $H_{f,k}$ is \emph{inside} (resp. \emph{outside}) $X$ iff all
of the vertices (resp. one or more of the vertices) on the bounding edges
of the face are in $X$ (resp. not in $X$). The \emph{frontier} of
$X$ is a set of edges defined as follows:
\[
   \bound{X} \triangleq
   \ \Set{\,e\in\ee(H_{f,k})\;|
   \;\text{$e=\set{x\ y}$
     \ \ bounds a face outside $X$ and $\Set{x,y}\subseteq X$}\,}.
\]
Note that if $e\in\bound{X}$, it does not necessarily follow that
$e$ bounds a face inside $X$, even though the two endpoints of $e$
are in $X$; this is illustrated in Figure~\ref{fig:boundaries}.
And if $X$ is a face of $H_{f,k}$, large or small,
then $\bound{X}$ coincides with the boundary of $X$.
\end{definition}

\begin{lemma}
\label{lem:boundaries-of-full-clusters}
\label{lem:frontiers-of-full-clusters}
If $X\subseteq \vv(H_{f,k})$ be a full cluster in $H_{f,k}$, then
the set of vertices $\vv\big(\bound{X}\big)$ forms a cluster, \ie,
it is a non-empty connected subset of $\vv(H_{f,k})$.
\end{lemma}
It is worth noting that, unless the vertices of the outermost
cycle $C_0$ and/or the vertices of the innermost cycle $C_{k-1}$
are all in $X$, the vertices in $\vv\big(\bound{X}\big)$ form
a connected cactus.
\begin{proof}
Straightforward from the definition. Details omitted.
\end{proof}

\begin{figure}[H] 
\begin{custommargins}{-0.40cm}{-2.2cm}
\vspace{-.15in}
\begin{center}
   \includegraphics[scale=.45]{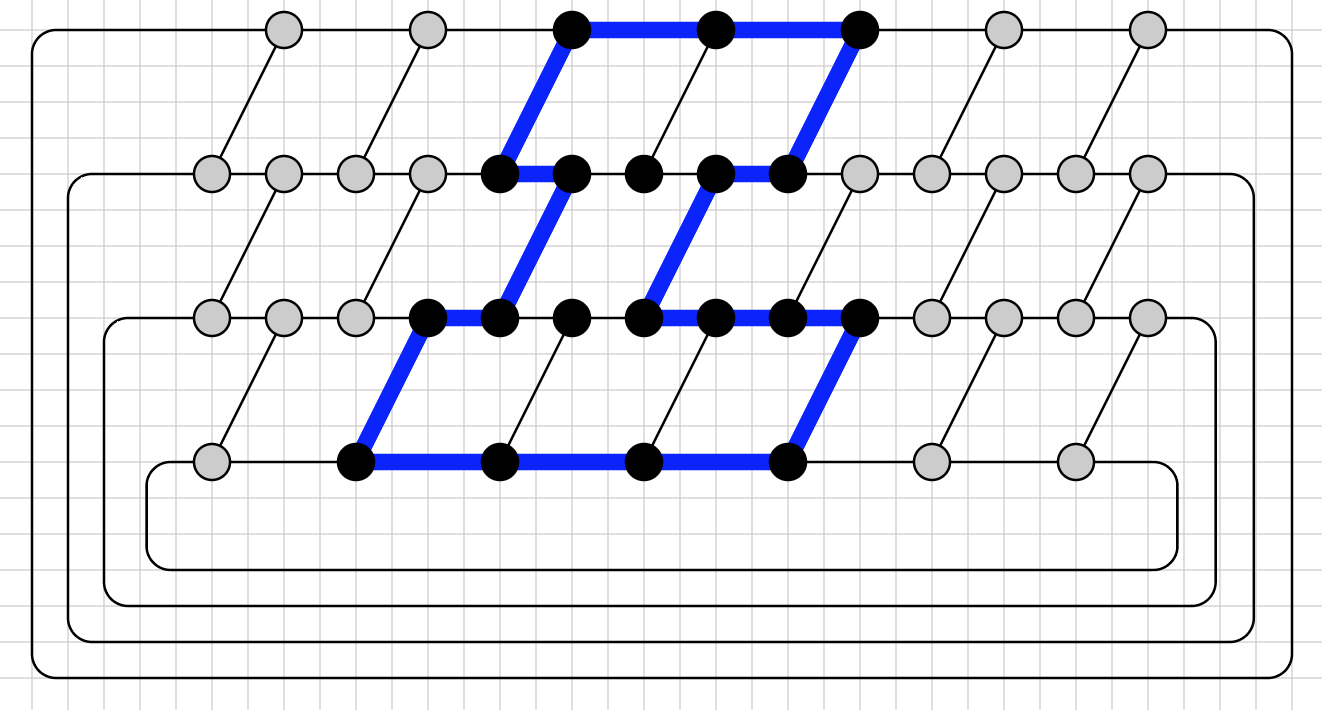}
\end{center}   
\vspace{-.25in}
   \caption{Example for Definition~\ref{def:cut-and-dangling}:
   Graph $H_{f,k}$ is
   the same as in Figure~\ref{fig:special-graph-E-outerplanarity-4}.
   The black vertices form a regular (full) cluster $X$ in $H_{f,k}$. There
   are no dangling edges and no cut edges in $\bound{X}$ . }
\label{fig:regular}
\end{custommargins}
\end{figure}

\begin{definition}{Cut Edges, Dangling Edges, and Regular Clusters}
\label{def:cut-and-dangling}
Let $X\subseteq \vv(H_{f,k})$ be a full cluster in graph $H_{f,k}$,
and consider $\bound{X}$.
If $e\in\bound{X}$ does not bound a face inside $X$,
then $e$ is one of two kinds:
\begin{itemize}[itemsep=0pt,parsep=2pt,topsep=2pt,partopsep=0pt]
   \item $e$ is a \emph{cut edge} of  $\bound{X}$,
   \item $e$ is a \emph{dangling edge} of $\bound{X}$. 
\end{itemize}
If the deletion of $e\in\bound{X}$ disconnects $\bound{X}$
into two components, then $e$ is a cut edge; otherwise, $e$
is a dangling edge. See Figure~\ref{fig:boundaries} for an illustration:
it shows one dangling edge and three cut edges.

If $X$ is a full cluster and $\bound{X}$ contains no cut edges and no
danling edges, then we say $X$ is a \emph{regular cluster}; this is
illustrated in Figure~\ref{fig:regular}. Observe
that a regular cluster is constructed from `piling on top of each
other' $5$-edge faces and $6$-edge faces.  The $5$-edge faces are
those adjacent to the outer face (bounded by $C_0$) and the innermost
face (bounded by $C_{k-1}$).
\end{definition}

\begin{lemma}
\label{lem:clean-boundaries}
\label{lem:clean-frontiers}
Let $X\subseteq \vv(H_{f,k})$ be a full cluster with $\size{X}\geqslant 3$.
If $\bound{X}$ contains cut edges and/or dangling edges, then there is a
regular cluster $Z\subseteq \vv(H_{f,k})$ such that:
\begin{enumerate}[itemsep=0pt,parsep=2pt,topsep=2pt,partopsep=0pt]
   \item $  \Bridges{}{Z} \leqslant \Bridges{}{X} \quad\text{and}
         \quad \size{Z}  \geqslant \size{X} $.
   \item For every regular cluster $Y$ such that $\size{Y}\geqslant\size{X}$, 
         it holds that $\size{Z}\leqslant\size{Y}$.
\end{enumerate}
In words, we can minimally augment the size $\size{X}$ and eliminate all cut edges
and dangling edges from $\bound{X}$ without incrasing $\Bridges{}{X}$.%
   \footnote{
   Note that we augment the size $\size{X}$ not $X$ itself, \ie, it is
   not necessarily that $Z\supseteq X$, but only that 
   $ \size{Z}  \geqslant \size{X} $.
   }
\end{lemma}

\begin{proof}
The proof is an exhaustive case analysis. 
We proceed repeatedly to eliminate cut edges and dangling edges,
one by one. There is no loss of generality in assuming that $X$
satisfies one of two conditions (or both):
\begin{enumerate}[itemsep=0pt,parsep=2pt,topsep=2pt,partopsep=0pt]
   \item
     $X\cap \vv(C_0) \neq\varnothing$ and $X\cap \vv(C_1) \neq\varnothing$,
   \item
     $X\cap \vv(C_{k-2}) \neq\varnothing$ and $X\cap \vv(C_{k-1}) \neq\varnothing$.
\end{enumerate}
In words, $X$ overlaps with the two outermost cycles (condition 1)
and/or the two innermost cycles of $H_{f,k}$ (condition 2).

Consider a particular $e_0\in\bound{X}$ which
is a cut edge or a dangling edge. Keep in mind that the two large
faces of $H_{f,k}$ do not have a common boundary, that every small
face of $H_{f,k}$ is bounded by $5$ or $6$ edges, and that every edge bounds
exactly two faces. 

Because $e_0\in\bound{X}$, there is a small face $F$ of $H_{f,k}$ outside
$X$ which is bounded by $e_0$.  Because $\bound{X}$ is connected, we
can choose the face $F$ to be bounded by at least two edges of
$\bound{X}$, say $e_1$ in addition to $e_0$. Let the set of edges
bounding the small face $F$ be $\Set{e_0,e_1,e_2,e_3,e_4}$ or
$\Set{e_0,e_1,e_2,e_3,e_4, e_5}$.

In all cases satisfying one of the two following conditions:
\begin{itemize}[itemsep=0pt,parsep=2pt,topsep=2pt,partopsep=0pt]
  \item[3.] $F$ is bounded by $5$ edges, with at least two of them
  $\Set{e_0,e_1}\subseteq\bound{X}$,
  \item[4.] $F$ is bounded by $6$ edges, with at least three of them 
  $\Set{e_0,e_1,e_2}\subseteq\bound{X}$,
\end{itemize}
it is straightforward to add all the vertices of $\vv(F)$ to $X$,
thus increasing $\size{X}$ and making $F$ a face inside $X$, without
increasing $\Bridges{}{X}$. All details of this straighforward case analysis
are omitted.

The remaining cases satisfy the following condition:
\begin{itemize}[itemsep=0pt,parsep=2pt,topsep=2pt,partopsep=0pt]
  \item[5.] $F$ is bounded by $6$ edges, with exactly two of them
  $\Set{e_0,e_1}\subseteq\bound{X}$.
\end{itemize}
Because $F$ is bounded by $6$ edges, $F$ is not adjacent to the outer
face (bounded by cycle $C_0$) nor to the innermost face
(bounded by cycle $C_{k-1}$).

We eliminate cut edges of $\bound{X}$ first, by starting from cut
edges of lowest level (those that are closest to cycle $C_0$), and we
then proceed inward until we reach cut edges of highest level (closest
to cycle $C_{k-1}$). In this order, it is easy to see that we only
need to eliminate cut edges that satisfy condition 3 of condition 4
above.

We are left with the case when there are only dangling edges and
condition $5$ is satisfied. Let therefore $e_0$ be a dangling
edge of $\bound{X}$, which implies that $F$ is a $6$-edge face outside $X$
which shares $e_1$ as a bounding edge with another face inside $X$.

Let $e_0 = \set{v_0\ v_1}$ and $e_1 = \set{v_1\ v_2}$. Let $Y = X - \Set{v_0}$,
so that $\Bridges{}{Y} = \Bridges{}{X} - 1$ and $\size{Y} = \size{X} - 1$.
Then $Y$ is a full cluster, such that $\bound{Y}$ contains no cut edges and
one dangling edge less than $\bound{X}$. The desired full cluster $Z$
in the conclusion of the lemma is obtained by appropriately adding two or
more vertices to $Y$ and by increasing $\Bridges{}{Y}$ by at most $1$.

Cluster $Y$ satisfies condition 1 or condition 2. Assume $Y$ satisfies
condition 1. This implies there is a face $F'$ with $5$ bounding
edges, say $\Set{e'_0,e'_1,e'_2, e'_3, e'_4}$, such that edge $e'_0$
is an ICE in $\bound{Y}$, edge $e'_1$ is an edge of $C_0$ and also in
$\bound{Y}$, edge $e'_2$ is an ICE which may or may not be in
$\bound{Y}$, and $\Set{e'_3, e'_4}$ are consecutive edges of cycle
$C_1$ which may or may not be in $\bound{Y}$. It is now easy to see
that we can add two (or more) vertices to $Y$ such that: (i)
$\Bridges{}{Y}$ is increased by at most $1$ and (ii) no cut edge and
dangling edge are added to $\bound{Y}$. The resulting cluster is the
desired $Z$.
\end{proof}

\begin{definition}{$\prec$-Sequences}
\label{def:prec-sequences}
  A \emph{maximal $\prec^{r}$-sequence} $S$ of ICE's in the graph $H_{f,k}$ is
  a sequence of the form:
 \[
     S\ \triangleq\ e_{1,j}\ \prec^{r}\ e_{2,j}\ \prec^{r}\ \cdots\ \prec^{r}\ e_{k-1,j}
 \]
 where $j\in\Set{1,\ldots,f(k)}$. A \emph{maximal $\prec^{\ell}$-sequence} $S$
 of ICE's in the graph $H_{f,k}$ is a sequence of the form:
 \[
     S\ \triangleq\ e_{1,j_1}\ \prec^{\ell}\ e_{2,j_2}\ \prec^{\ell}
         \ \cdots\ \prec^{\ell}\ e_{k-1,j_{k-1}}
 \]
 where $j_1\in\Set{1,\ldots,f(k)}$ and:
 \begin{alignat*}{8}
    & j_2\ &&=\ && 1 + \big( (j_1 - 2)\!\!\!\mod f(k)\big),
    \\
    & j_3  &&=  && 1 + \big((j_1 - 3)\!\!\!\mod f(k)\big),
    \\
    & && \cdots &&
    \\
    & j_{k-1} &&= && 1 + \big((j_1 - k +1)\!\!\!\mod f(k)\big) .
 \end{alignat*}
 A \emph{maximal $\prec$-sequence} $S$
 of ICE's in the graph $H_{f,k}$ is a sequence of the form:
 \[
     S\ \triangleq\ e_{1,j_1}\ \prec^{?}\ e_{2,j_2}\ \prec^{?}
         \ \cdots\ \prec^{?}\ e_{k-1,j_{k-1}}
 \]
 where every pair $e_{p,j_{p}} \prec^{?} e_{p+1,j_{p+1}}$ in the sequence,
 with $1\leqslant p\leqslant k-1$, is:
 \[
       \text{either}\quad
       e_{p,j_{p}}\ \prec^{r}\ e_{p+1,j_{p+1}}\quad
       \text{or}\quad
       e_{p,j_{p}}\ \prec^{\ell}\ e_{p+1,j_{p+1}} ,
 \]
 \ie, in a maximal $\prec$-sequence it does not matter whether
 $e_{p+1,j_{p+1}}$ is a clockwise successor or counter-clockwise successor of
 $e_{p,j_{p}}$, as we traverse the sequence from the outermost cycle to
 the innermost cycle.
 
 For graph $H_{f,k}$ with $k=4$ and $f(k) =8$, there are four possible
 shapes of maximal $\prec$-sequences; these are shown in boldface in
 Figure~\ref{fig:possible-shape-in-special-graph}, with interleaving
 cycle edges inserted in dashed boldface.

\begin{figure}[H] 
\begin{center}

\noindent
%
\begin{tikzpicture}[scale=.35] 
       \newcommand\EdgeOpacity{[line width=1.2,black,opacity=.98]};
       \draw [help lines, dotted] (0, 0) grid (36,18);
       \coordinate (A3-5)  at (3,5);
       \coordinate (A7-5)  at (7,5);
       \coordinate (A11-5) at (11,5);
       \coordinate (A15-5) at (15,5);
       \coordinate (A19-5) at (19,5);
       \coordinate (A23-5) at (23,5);
       \coordinate (A27-5) at (27,5);
       \coordinate (A31-5) at (31,5);
       \coordinate (B3-9)  at (3,9);
       \coordinate (B5-9)  at (5,9);       
       \coordinate (B7-9)  at (7,9);
       \coordinate (B9-9)  at (9,9);       
       \coordinate (B11-9) at (11,9);
       \coordinate (B13-9) at (13,9);       
       \coordinate (B15-9) at (15,9);
       \coordinate (B17-9) at (17,9);       
       \coordinate (B19-9) at (19,9);
       \coordinate (B21-9) at (21,9);       
       \coordinate (B23-9) at (23,9);
       \coordinate (B25-9) at (25,9);       
       \coordinate (B27-9) at (27,9);
       \coordinate (B29-9) at (29,9);       
       \coordinate (B31-9) at (31,9);
       \coordinate (B33-9) at (33,9);
       \coordinate (C3-13)  at (3,13);
       \coordinate (C5-13)  at (5,13);       
       \coordinate (C7-13)  at (7,13);
       \coordinate (C9-13)  at (9,13);       
       \coordinate (C11-13) at (11,13);
       \coordinate (C13-13) at (13,13);       
       \coordinate (C15-13) at (15,13);
       \coordinate (C17-13) at (17,13);       
       \coordinate (C19-13) at (19,13);
       \coordinate (C21-13) at (21,13);       
       \coordinate (C23-13) at (23,13);
       \coordinate (C25-13) at (25,13);       
       \coordinate (C27-13) at (27,13);
       \coordinate (C29-13) at (29,13);       
       \coordinate (C31-13) at (31,13);
       \coordinate (C33-13) at (33,13);
       \coordinate (D3-17)  at (3,17);
       \coordinate (D5-17)  at (5,17);       
       \coordinate (D7-17)  at (7,17);
       \coordinate (D9-17)  at (9,17);       
       \coordinate (D11-17) at (11,17);
       \coordinate (D13-17) at (13,17);       
       \coordinate (D15-17) at (15,17);
       \coordinate (D17-17) at (17,17);       
       \coordinate (D19-17) at (19,17);
       \coordinate (D21-17) at (21,17);       
       \coordinate (D23-17) at (23,17);
       \coordinate (D25-17) at (25,17);       
       \coordinate (D27-17) at (27,17);
       \coordinate (D29-17) at (29,17);       
       \coordinate (D31-17) at (31,17);
       \coordinate (D33-17) at (33,17);          

       \draw [line width=1.5mm, blue] (A3-5) -- (B5-9) ;
       \draw [line width=1.5mm, dashed, blue] (B3-9) -- (B5-9) ;
       \draw [line width=1.5mm, blue] (A7-5) -- (B9-9) ;
       \draw [line width=1.5mm, dashed, blue] (B9-9) -- (B11-9) ;       
       \draw \EdgeOpacity (A11-5) -- (B13-9) ;
       \draw [line width=1.5mm, blue] (A15-5) -- (B17-9) ;
       \draw [line width=1.5mm, dashed, blue] (B15-9) -- (B17-9) ;       
       \draw \EdgeOpacity (A19-5) -- (B21-9) ;
       \draw [line width=1.5mm, blue] (A23-5) -- (B25-9) ;
       \draw [line width=1.5mm, dashed, blue] (B25-9) -- (B27-9) ;              
       \draw \EdgeOpacity (A27-5) -- (B29-9) ;
       \draw \EdgeOpacity (A31-5) -- (B33-9) ;             
       \draw [line width=1.5mm, blue] (B3-9) -- (C5-13) ;
       \draw [line width=1.5mm, dashed, blue] (C3-13) -- (C5-13) ;
       \draw \EdgeOpacity (B7-9) -- (C9-13) ;
       \draw [line width=1.5mm, blue] (B11-9) -- (C13-13) ;
       \draw [line width=1.5mm, dashed, blue] (C11-13) -- (C13-13) ;       
       \draw [line width=1.5mm, blue] (B15-9) -- (C17-13) ;
       \draw [line width=1.5mm, dashed, blue] (C17-13) -- (C19-13) ;       
       \draw \EdgeOpacity (B19-9) -- (C21-13) ;
       \draw \EdgeOpacity (B23-9) -- (C25-13) ;
       \draw [line width=1.5mm, blue] (B27-9) -- (C29-13) ;
       \draw [line width=1.5mm, dashed, blue] (C29-13) -- (C31-13) ;           
       \draw \EdgeOpacity (B31-9) -- (C33-13) ;
       \draw [line width=1.5mm, blue] (C3-13) -- (D5-17) ;
       \draw \EdgeOpacity (C7-13) -- (D9-17) ;
       \draw [line width=1.5mm, blue] (C11-13) -- (D13-17) ;
       \draw \EdgeOpacity (C15-13) -- (D17-17) ;
       \draw [line width=1.5mm, blue] (C19-13) -- (D21-17) ;
       \draw \EdgeOpacity (C23-13) -- (D25-17) ;
       \draw \EdgeOpacity (C27-13) -- (D29-17) ;
       \draw [line width=1.5mm, blue] (C31-13) -- (D33-17) ;
       \draw[rounded corners,thick]
            (2,5) -- (34,5) -- (34,3) -- (2,3) -- cycle;
       \draw[rounded corners,thick]
            (1.5,9) -- (34.5,9) -- (34.5,1.8) -- (1.5,1.8) -- cycle;
       \draw[rounded corners,thick]
            (1,13) -- (35,13) -- (35,0.85) -- (1,0.85) -- cycle;
       \draw[rounded corners,thick]
            (.5,17) -- (35.5,17) -- (35.5,0) -- (.5,0) -- cycle;

       \node at (A3-5) {\Huge $\bullet$};  
       \node at (A7-5) {\Huge $\bullet$}; 
       \node at (A11-5) {\Huge $\bullet$}; 
       \node at (A15-5) {\Huge $\bullet$}; 
       \node at (A19-5) {\Huge $\bullet$}; 
       \node at (A23-5) {\Huge $\bullet$}; 
       \node at (A27-5) {\Huge $\bullet$}; 
       \node at (A31-5) {\Huge $\bullet$}; 
       \node at (B3-9) {\Huge $\bullet$};
       \node at (B5-9) {\Huge $\bullet$}; 
       \node at (B7-9) {\Huge $\bullet$}; 
       \node at (B9-9) {\Huge $\bullet$}; 
       \node at (B11-9) {\Huge $\bullet$}; 
       \node at (B13-9) {\Huge $\bullet$}; 
       \node at (B15-9) {\Huge $\bullet$}; 
       \node at (B17-9) {\Huge $\bullet$}; 
       \node at (B19-9) {\Huge $\bullet$};
       \node at (B21-9) {\Huge $\bullet$}; 
       \node at (B23-9) {\Huge $\bullet$}; 
       \node at (B25-9) {\Huge $\bullet$}; 
       \node at (B27-9) {\Huge $\bullet$}; 
       \node at (B29-9) {\Huge $\bullet$}; 
       \node at (B31-9) {\Huge $\bullet$}; 
       \node at (B33-9) {\Huge $\bullet$}; 
       \node at (C3-13) {\Huge $\bullet$};
       \node at (C5-13) {\Huge $\bullet$}; 
       \node at (C7-13) {\Huge $\bullet$}; 
       \node at (C9-13) {\Huge $\bullet$}; 
       \node at (C11-13) {\Huge $\bullet$}; 
       \node at (C13-13) {\Huge $\bullet$}; 
       \node at (C15-13) {\Huge $\bullet$}; 
       \node at (C17-13) {\Huge $\bullet$}; 
       \node at (C19-13) {\Huge $\bullet$};
       \node at (C21-13) {\Huge $\bullet$}; 
       \node at (C23-13) {\Huge $\bullet$}; 
       \node at (C25-13) {\Huge $\bullet$}; 
       \node at (C27-13) {\Huge $\bullet$}; 
       \node at (C29-13) {\Huge $\bullet$}; 
       \node at (C31-13) {\Huge $\bullet$}; 
       \node at (C33-13) {\Huge $\bullet$};
       \node at (D5-17) {\Huge $\bullet$}; 
       \node at (D9-17) {\Huge $\bullet$}; 
       \node at (D13-17) {\Huge $\bullet$}; 
       \node at (D17-17) {\Huge $\bullet$}; 
       \node at (D21-17) {\Huge $\bullet$}; 
       \node at (D25-17) {\Huge $\bullet$}; 
       \node at (D29-17) {\Huge $\bullet$}; 
       \node at (D33-17) {\Huge $\bullet$};        

   \end{tikzpicture} 
%

   \caption{Four possible shapes of maximal $\prec$-sequences 
     in graph $H_{f,k}$ when $k=4$ and $f(k) = 2\cdot k = 8$ (slightly
     different from the graph in
     Figure~\ref{fig:special-graph-E-outerplanarity-4}).
     The leftmost above is a maximal $\prec^{r}$-sequence, the rightmost is a
     maximal $\prec^{\ell}$-sequence, and the two in the middle are
     neither maximal $\prec^{r}$ nor maximal $\prec^{\ell}$.
   }
\label{fig:possible-shape-in-special-graph}
\end{center}
\end{figure}
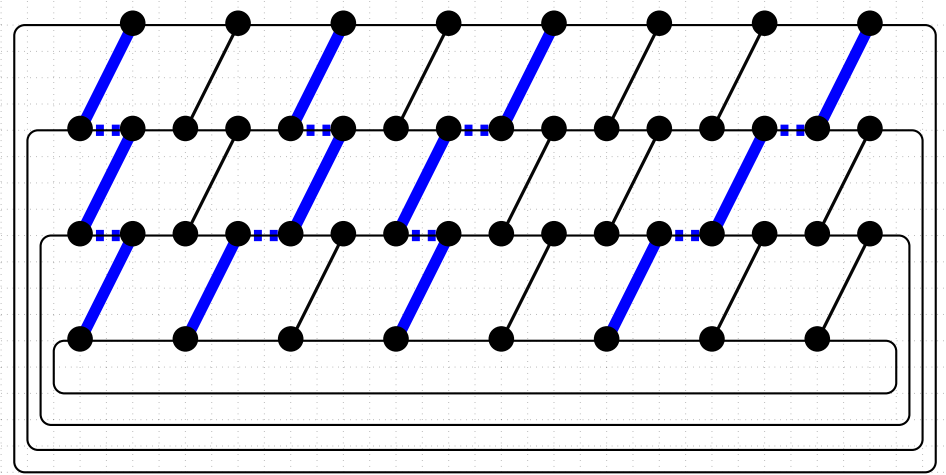
 
 A \emph{$\prec^{r}$-sequence} is a subsequence of a
 maximal $\prec^{r}$-sequence, \ie, the former is obtained 
 by omitting a prefix and/or a suffix from the latter.

 Similarly, a \emph{$\prec^{\ell}$-sequence} is a subsequence of a
 maximal $\prec^{\ell}$-sequence, and a \emph{$\prec$-sequence} is a
 subsequence of a maximal $\prec$-sequence.
 \end{definition}

\Hide{
\begin{definition}{Cycle Sequences}
\label{def:cycle-sequences}
A \emph{maximal $\bowtie$-sequence} $S$ of ICE's
is a sequence of consecutive ICE's which are all of the same level,
which we write as:
\[
  S\ \triangleq\ e_{i,j_1}\ \bowtie\ e_{i,j_2}\ \bowtie
         \ \cdots\ \bowtie\ e_{i,j_{2k}}
\]
where $i\in\Set{1,\ldots,k-1}$ and $j_1\in\Set{1,\ldots,2k}$ with:
 \[
     j_2 = 1 + \big( j_1 \!\!\!\mod 2k\big),\ \ 
     j_3 = 1 + \big((j_1 +1)\!\!\!\mod 2k\big),
     \ \ \ldots\ \ ,\ \ j_{2k} = 1 + \big((j_1 + 2k - 2)\!\!\!\mod 2k\big) .
 \]
A \emph{$\bowtie$-sequence of ICE's} is a subsequence of a maximal $\bowtie$-sequence
of ICE's. In Figure~\ref{fig:cycle-sequence}, we show a maximal 
$\bowtie$-sequence of ICE's in graph $H_4$ of level $i=2$.
\end{definition}
}

\begin{figure}[H] 
\begin{custommargins}{-0.40cm}{-2.2cm}
\begin{center}
   \includegraphics[scale=.45]{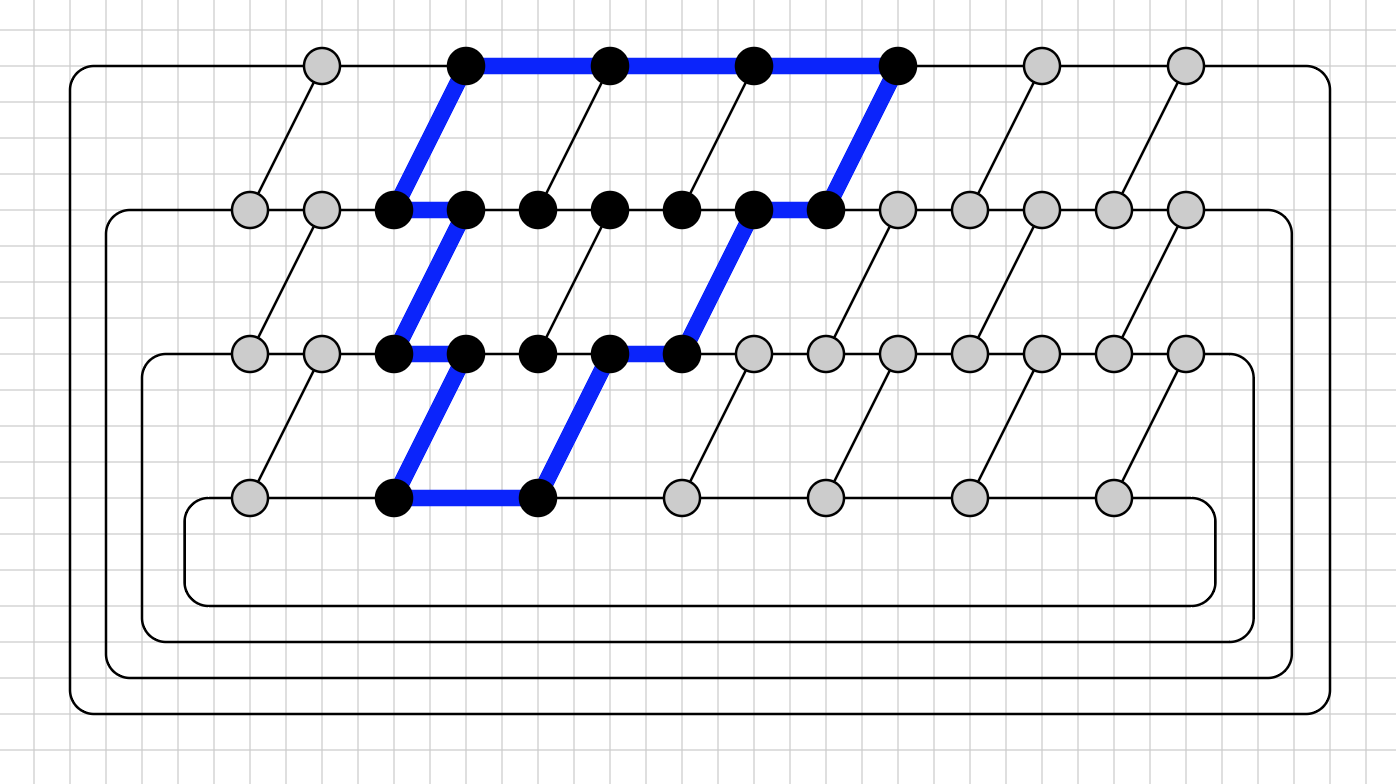}
\end{center}   
\vspace{-.25in}
   \caption{Example for Definition~\ref{def:strongly-regular-clusters}:
   Graph $H_{f,k}$ is
   the same as in Figure~\ref{fig:special-graph-E-outerplanarity-4}.
   The black vertices form a \emph{strongly regular cluster} bounded by
   the outermost cycle $C_{0}$ . }
\label{fig:strongly-regular-clusters-1} 
\end{custommargins}
\end{figure}

\begin{figure}[H] 
\begin{custommargins}{-0.40cm}{-2.2cm}
\begin{center}
   \includegraphics[scale=.45]{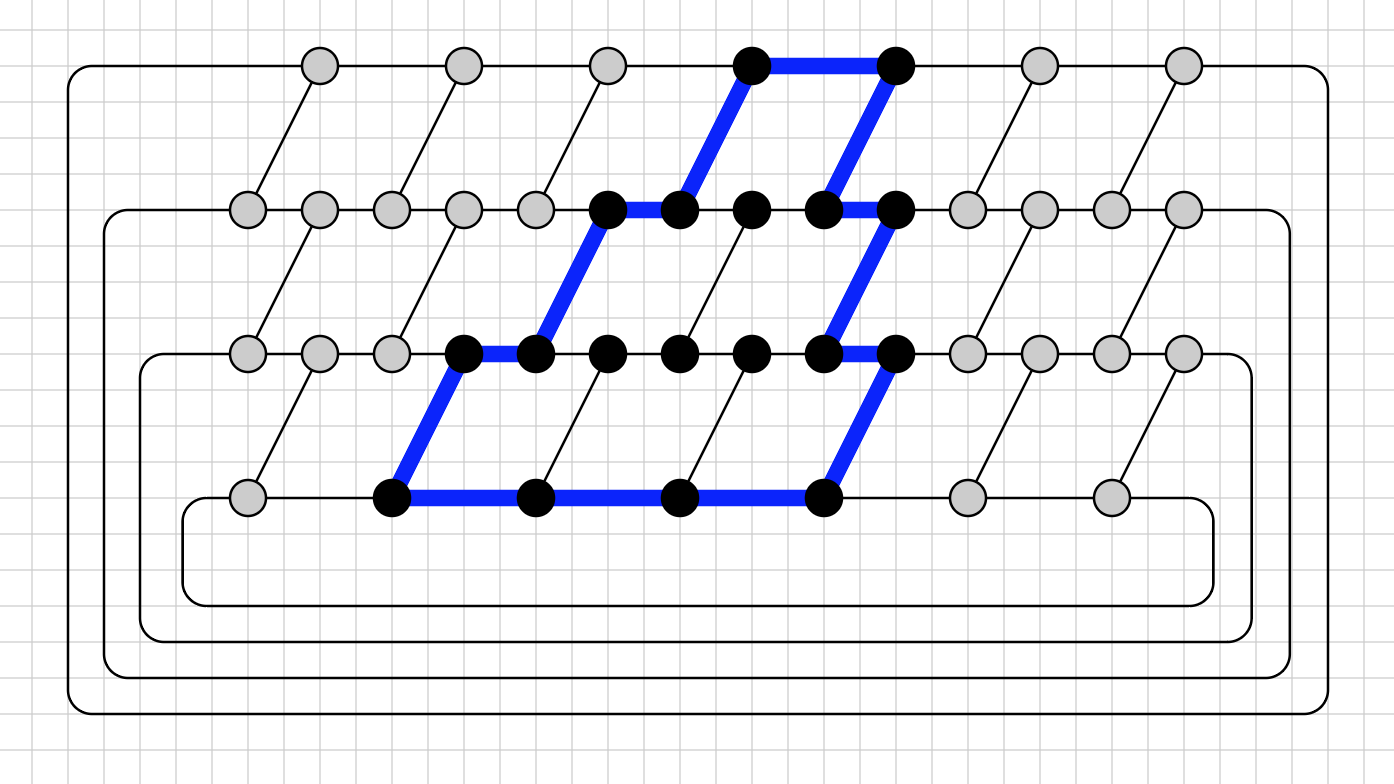}
\end{center}   
\vspace{-.25in}
   \caption{Example for Definition~\ref{def:strongly-regular-clusters}:
   Graph $H_{f,k}$ is
   the same as in Figure~\ref{fig:special-graph-E-outerplanarity-4}.
   The black vertices form a \emph{strongly regular cluster} bounded by
   the innermost cycle, which is here $C_{3}$ . }
\label{fig:strongly-regular-clusters-2} 
\end{custommargins}
\end{figure}

\begin{definition}{Strongly Regular Clusters}
\label{def:regular-clusters}
\label{def:strongly-regular-clusters}
Let $X\subseteq \vv(H_{f,k})$ be a regular cluster with $\size{X}\geqslant 3$.
We say $X$ is a \emph{strongly regular cluster bounded by the outermost
cycle $C_0$} iff $\bound{X}$ can be partitioned into $4$ disjoint subsets:
\begin{itemize}[itemsep=0pt,parsep=2pt,topsep=2pt,partopsep=0pt]
  \item a subsequence of $n$ consecutive edges in $C_0$,
  \item a prefix of $p$ ICE's in a maximal $\prec^{r}$-sequence together
        with their $(p-1)$ intermediate cycle edges,
  \item a subsequence of $q$ consecutive edges in $C_p$,
  \item a prefix of $p$ ICE's in a maximal $\prec^{\ell}$-sequence together with
        their $(p-1)$ intermediate cycle edges,
\end{itemize}
for some $n,p,q\geqslant 1$. Symmetrically, 
we say $X$ is a \emph{strongly regular cluster bounded by the innermost
cycle $C_{k-1}$} iff $\bound{X}$ can be partitioned into $4$ disjoint subsets:
\begin{itemize}[itemsep=0pt,parsep=2pt,topsep=2pt,partopsep=0pt]
  \item a subsequence of $n$ consecutive edges in $C_{k-1}$,
  \item a suffix of $p$ ICE's in a maximal $\prec^{r}$-sequence together
        with their $(p-1)$ intermediate cycle edges,
  \item a subsequence of $q$ consecutive edges in $C_{k-1-p}$,
  \item a suffix of $p$ ICE's in a maximal $\prec^{\ell}$-sequence together
        with their $(p-1)$ intermediate cycle edges,
\end{itemize}
for some $n,p,q\geqslant 1$. We call $n$, $p$, and $q$ the \emph{parameters}
of the strongly regular cluster $X$, the values of which are not
totally arbitrary, as stated in the next lemma; $p$ is the \emph{height}
of $X$, and $n$ and $q$ its \emph{bases}.
Figure~\ref{fig:strongly-regular-clusters-1} shows an example of 
a \emph{strongly regular cluster bounded by the outermost
cycle $C_{0}$} and Figure~\ref{fig:strongly-regular-clusters-2}
shows an example of a \emph{strongly regular cluster bounded by the innermost
cycle $C_{k-1}$}.
\end{definition}

\begin{lemma}
\label{lem:restrictions-on-regular-clusters}
If $X$ is a strongly regular cluster in $H_{f,k}$ with parameters  $n$, $p$, and $q$,
then:
\begin{enumerate}[itemsep=0pt,parsep=4pt,topsep=2pt,partopsep=0pt]
   \item $1\leqslant n\leqslant f(k) - 1$ and $1\leqslant p\leqslant k-1$.
   \item If $p < k-1$, then $q = 2i$ for some $1\leqslant i \leqslant f(k) - p$, \\
         $p + (q/2) = n + 1$, \ \ $\Bridges{}{X} = 2(p+1) + (q/2)$,
         \ and\ $\size{X} = p^2+pq+p+ (q/2)$.
   \item If $p = k-1$, then $q = i$ for some $1\leqslant i \leqslant f(k)-p$, \\
         $p + q = n + 1$, \ \ $\Bridges{}{X} = 2k$,\ and
         \ $\size{X} = p^2+2pq+p$.
\end{enumerate}
\Hide{
Since the total number of small faces in $H_{f,k}$ is $2 k(k-1)$, the 
largest possible regular cluster includes at most $3/4$ of the small
faces of $H_{f,k}$.
}
\end{lemma}

\begin{proof} 
Let $X$ be a strongly regular cluster bounded by the
outermost cycle $C_0$. (The argument applies again symmetrically
if $X$ is bounded by the innermost cycle $C_{k-1}$.)

For part 1, note that $n$ is restricted to be at most $f(k)-1$, it
cannot be $f(k)$, because Definition~\ref{def:regular-clusters} requires
that the two bounding $\prec$-sequences be disjoint.  Parts 2 and 3
follow from straightforward calculations (all details omitted).
\end{proof}

\begin{lemma}
\label{lem:very-clean-boundaries}
\label{lem:very-clean-frontiers}
Let $X\subseteq \vv(H_{f,k})$ be a regular cluster with $f(k)\gg k$.
Let $N = \size{\vv(H_{f,k})}$, the total number of vertices in $H_{f,k}$.
If $\size{X} < N/2$ and $\Bridges{}{X} \leqslant 2k-1$,
then there is a strongly regular cluster $Z\subseteq \vv(H_{f,k})$ such that:%
   \footnote{We do not need the restriction $\size{X}\geqslant 3$ here,
   which is part of the hypothesis in Lemma~\ref{lem:clean-boundaries},
   because if there are no cut edges and no dangling edges, it follows
   that $\size{X}\geqslant 3$ -- in fact, that $\size{X}\geqslant 5$.}
\[
   \Bridges{}{Z} \leqslant \Bridges{}{X} \quad\text{and}
   \quad \size{X} \leqslant \size{Z} < N/2.
\]
\end{lemma}

\begin{proof}
There is no loss of generality in assuming that $X$ satisfies
condition 1 or condition 2 in the proof of Lemma~\ref{lem:clean-boundaries}.
Since $\Bridges{}{X}\leqslant 2k - 1$, it is only one of these two conditions
that $X$ can satisfy, not both. We assume that $X$ satisfies condition 2 and
thus view that $X$ has one or more consecutive $5$-edge faces at the
bottom (all adjacent to the innermost face), on top of which there are
$6$-edge faces, piled upward no higher than cycle $C_1$ in order not to
violate the restriction $\Bridges{}{X}\leqslant 2k - 1$. Moreover, because
$\size{X} < N/2$ and $\Bridges{}{X}\leqslant 2k - 1$,
not all the vertices of the innermost cycle
$C_{k-1}$ are in $X$, and because $X$ is full, there are paths connecting
vertices in $C_0$ and $C_{k-1}$ whose vertices are all not in $X$.

Let $h\geqslant 1$
be the height of $X$, which is necessarily $\leqslant k - 2$, otherwise
we would have $\Bridges{}{X}\geqslant 2k$ contradicting the hypothesis.
There are two kinds of edges connecting $X$ to $\big(\vv(H_{f,k})-X\big)$
contributing to the total in $\Bridges{}{X}$: ICE's and cycle edges.
Consider all the ICE's in the set ${\bridges{}{X}}$ at height $h$, say, these
are $e_1, e_2, \ldots, e_r$ for some $r\geqslant 1$. These ICE's are
not necessarily consecutive, as there may be ``dips'' along $\bound{X}$.

The construction of the desired cluster $Z$ in the lemma conclusion
proceeds in two stages. First, we remove all the ``dips'' between the top-level
ICE's $e_1, e_2, \ldots, e_r$ in ${\bridges{}{X}}$, to obtain an intermediate
full cluster $X'$ of the same height $h$ and where all the top-level 
ICE's $e'_1, e'_2, \ldots, e'_s$ in ${\bridges{}{X'}}$ are now consecutive
and such that $\Bridges{}{X} \geqslant \Bridges{}{X'}$ and
$\size{X} \leqslant \size{X'}$. By a `top-level ICE' in ${\bridges{}{X}}$
we mean an ICE which is closest to the outermost cycle $C_0$. 
Note that
\[
   2k - 1\ \geqslant\ \Bridges{}{X}\ \geqslant\ \Bridges{}{X'}
   \ \geqslant 2(h+1) + s 
\]
In the last term, $2(h+1)$ is the number of cycle edges in ${\bridges{}{X'}}$
and $s$ is the number of ICE's in ${\bridges{}{X'}}$ at height $h$. In general,
there may be other ICE's in ${\bridges{}{X'}}$ at heights lower than $h$.

Second, we obtain the desired regular cluster $Z$ by setting its parameters
$p$ and $q$, as specified in Definition~\ref{def:regular-clusters}, and by
Lemma~\ref{lem:restrictions-on-regular-clusters}, we do not need to also set
its parameter $n$. We set $p= h$ and $q = 2s$. By 
part 2 of Lemma~\ref{lem:restrictions-on-regular-clusters}, we have
$\Bridges{}{Z} = 2(h+1) + s$ and also $\size{Z} \leqslant \size{X'}$.
\end{proof}

\begin{lemma}
\label{lem:regular-clusters}
Let $X\subseteq \vv(H_{f,k})$ be a strongly
regular cluster with parameters $n$, $p$, and $q$,
as specified in Definition~\ref{def:regular-clusters}. If $\Bridges{}{X} < 2k$,
then $\size{X} \leqslant \lceil (16 k^2 - 32 k + 13)/12 \rceil $.
\end{lemma}

\begin{proof}
By part 3 of Lemma~\ref{lem:restrictions-on-regular-clusters}, if
$\Bridges{}{X} < 2k$ and therefore $\Bridges{}{X} \neq 2k$, then $p\neq k-1$.
Because $X$ is strongly regular, it follows that $p< k-1$, by part 1 of
Lemma~\ref{lem:restrictions-on-regular-clusters}. Hence, by part 2 of
Lemma~\ref{lem:restrictions-on-regular-clusters}, we have:
\[
  \size{X}\ =\ p^2+pq+p+ (q/2)\quad\text{and}\quad
  \Bridges{}{X}\ =\ 2(p+1) + (q/2) .
\]
Let $c$ be a constant such that $\Bridges{}{X} = c$ which we keep fixed
throughout the proof. 
Relative to $c$, we determine how to set the values
of the parameters $p$ and $q$ in order to maximize $\size{X}$. We first express
$p$ in terms of $q$ and $c$, namely, if $2(p+1) + (q/2) = c$ then:
\[
     p\ =\ \frac{2c - q - 4}{4}
\]
which we substitute in $\size{X} = p^2+pq+p+ (q/2)$ to obtain a function
$f_c(q)$ depending on $q$ as a variable and $c$ as a constant:
\begin{alignat*}{5}
    & f_c(q)\ &&\triangleq\ \ && p \cdot (p + q + 1) + \frac{q}{2}
    \\
    & &&= &&
    \Big(\frac{2c - q - 4}{4}\Big)\cdot
    \Big(\frac{2c - q - 4}{4} + q + 1\Big) + \frac{q}{2}
    \\
    & &&= &&
    \frac{(2c - q - 4)\cdot (2c + 3q) + 8q}{16}
    \\
    & &&= &&
    -(3/16) q^2 + (1/4) (c-1) q + (1/4) (c^2 - 2c)
\end{alignat*}
where the last expression is obtained from the preceding one by straightforward
calculations. In the argument to follow, even though possible values of
$c = 2(p+1) + (q/2)$ are integers $\geqslant 5$, we deal with $q$ and $c$
as real numbers and we use the derivative of $f_c(q)$ relative to $q$
as such.

The function of $f_c(q)$ defines a parabola which is maximized at
its vertex, \ie, at the value of $q$ for which the derivative:
\[
       f'_c(q)\ \triangleq\ \frac{\text{d} f}{\text{d} q} = -(6/16) q + (1/4) (c-1)
\]
is zero. Hence, $f_c(q)$ as a function over the reals
is maximized at $\hat{q}$ given by:
\[
      \hat{q} \triangleq (2/3) (c-1).
\]
After substituting $\hat{q}$ in $f_c(q)$ and carrying out straightforward
calculations, we obtain:
\[
      f_c(\hat{q})\ =\ \frac{4c^2 - 8c + 1}{12}
\]
It is easily checked that $f_c(\hat{q})$ is monotonically increasing for
all $c\geqslant 1$. By the lemma hypothesis,
$c$ cannot exceed $2k-1\geqslant 1$.
Substituting $2k-1$ for $c$ in $f_c(\hat{q})$, we
obtain after simple calculations:
\[
     f_{2k-1}(\hat{q})\ =\ \frac{16 k^2 - 32 k + 13}{12}.
\]
We conclude that $\size{X}$ is an integer which
cannot exceed the real number $(16 k^2 - 32 k + 13)/12$.
\end{proof}

In the graph $H_{f,k}$ there are $k$ concentric cycles, such that two
consecutive cycles are connected by $f(k)$ ICE's. There is therefore
a total of $(k-1)\cdot f(k)$ ICE's in  $H_{f,k}$,
and since each ICE contributes two vertices, a total of
$2 (k-1)\cdot f(k)$ vertices. Hence, $\size{\vv(H_{f,k})} = 2 (k-1)\cdot f(k)$.

\begin{lemma}
\label{lem:no-binary-reassembling}
Consider the family ${\HH}_{f} = \Set{H_{f,k}}$ where $k\geqslant 2$ and
$f$ is any function such that the inequality:
\[
   \ 2 (k-1)\cdot f(k)\ >\ 4\cdot
   \Big(\frac{16 k^2 - 32 k + 13}{12}\Big) = \frac{16 k^2 - 32 k + 13}{3}
\]
is satisfied. Given any $H_{f,k}\in {\HH}_{f}$ there is no binary reassembling $\B$
of $H_{f,k}$ such that $\alpha(H_{f,k},\B) < 2k$.
\end{lemma}

\begin{proof}
If $\B$ is a binary reassembling of $H_{f,k}$ such that
$\alpha(H_{f,k},\B) < 2k$, then for every node/cluster $X\in\B$ it
holds that $\Bridges{}{X} < 2k$. Let $N = \size{\vv(H_{f,k})}$.
To prove that no such $\B$ can
exist, it suffices to show that for every cluster $X\in\B$, it must be
that $\size{X} < N/2$, \ie, $X$ contains fewer than half the vertices
of $H_{f,k}$, contradicting that $\B$ is a binary reassembling of
$H_{f,k}$.

Given any two disjoint clusters $X_1, X_2 \in\B$ such that
$\Bridges{}{X_i} < 2k$ and $\size{X_i} < N/2$, with $i = 1,2$, we show
that if $\Bridges{}{X_1\cup X_2} < 2k$, then again
$\size{X_1\cup X_2} < N/2$.
By Lemma~\ref{lem:regular-clusters}, every strongly regular cluster $X$
with $\Bridges{}{X} < 2k$ cannot include more than:
\[
   \frac{16 k^2 - 32 k + 13}{12} 
\]
vertices. Hence, by Lemmas~\ref{lem:full-clusters},
\ref{lem:boundaries-of-full-clusters},
\ref{lem:clean-boundaries}, \ref{lem:restrictions-on-regular-clusters},
and~\ref{lem:very-clean-boundaries}, any cluster $X$, 
strongly regular or not, with $\size{X} < N/2$ and $\Bridges{}{X} < 2k$
cannot include more than $(16 k^2 - 32 k + 13)/12$ vertices.
Hence,
\[
  \size{X_1\cup X_2} \ <\ \frac{16 k^2 - 32 k + 13}{6}\ <\ \frac{N}{2}
\]
The desired conclusion follows.
\end{proof}

\begin{remark}
\label{rem:simplify}
The inequality in the hypothesis of Lemma~\ref{lem:no-binary-reassembling} can
be simplified as follows. First, observe that:
\[
   (k-1)\cdot (16k - 13) = 16 k^2 - 29 k + 13 > 16 k^2 - 32 k + 13
\]
for all $k\geqslant 2$. Hence, if $f$ satisfies the inequality:
\[
   \ 2 (k-1)\cdot f(k)\ >\ 
   \frac{(k-1)\cdot (16k-13)}{3} > \frac{16 k^2 - 32 k + 13}{3}
\]
then the hypothesis of Lemma~\ref{lem:no-binary-reassembling} is satisfied.
Hence, it suffices to require that $f$ is any number-theoretic function such
that $f(k) > (16k-13)/6$.
\end{remark}

\begin{lemma}
\label{lem:optimality-of-KS}
For every graph $H_{f,k}\in {\HH}_{f}$ where $f$ is any
monotonically increasing or constant function $\geqslant 3$,
%
algorithm $\KS$ on input $H_{f,k}$ returns
a binary reassembling $\B=\KS(H_{f,k})$ such that $\alpha(H_{f,k},\B)= 2k$.
\end{lemma}

\begin{proof}
The function $f$ plays no role in this proof. Assuming that
$f(k)\geqslant 3$ for all $k$,
it is easy to see that the binary reassembling $\B=\KS(H_{f,k})$
returned by algorithm $\KS$ is such that $\alpha(H_{f,k},\B)= 2k$.
\end{proof}

\begin{theorem}[Optimality of Algorithm $\KS$]
\label{thm:optimality-of-KS}
For every graph in the family ${\HH}_{f}$ where $f(k) > (16k-13)/6$,
algorithm $\KS$ is an $\alpha$-optimal reassembling algorithm.
\end{theorem}

\begin{proof}
Immediate consequence of Lemma~\ref{lem:no-binary-reassembling},
Remark~\ref{rem:simplify}, and Lemma~\ref{lem:optimality-of-KS}.
\end{proof}

For the preceding theorem to hold, the number of ICE's must increase
as the edge-outerplanarity $k$ increases -- this is the
``density condition'' mentioned in the opening paragraph of
Section~\ref{sect:optimality-of-KS} -- otherwise our algorithm $\KS$
is not an $\alpha$-optimal reassembling algorithm as we explain next.

We define a family of $3$-regular plane graphs, called ${\G}_{\ell}$
where $\ell\geqslant 3$, which exhibits a low ``density of ICT's'' (or
``inter-cycle density'') in the sense that their number is bounded by
$\ell$ at every level.  For every graph $G\in {\G}_{\ell}$:
\begin{itemize}[itemsep=0pt,parsep=2pt,topsep=2pt,partopsep=0pt]
\item there is no restriction on $n = \size{\vv(G)}$
      or $k = \OutPlan{E}{G}$,
      and both can be unboundedly large, 
\item at every level of $G$, the number of ICT's is $\leqslant \ell$.
\end{itemize}
The second bullet point formalizes what we mean by low
``inter-cycle density''. For the definition of \emph{level}
in $3$-regular plane graphs, review Definition~\ref{def:levels}.

It is possible to show (not in this report) that there is linear-time 
reassembling algorithm which, given an arbitrary $G\in {\G}_{\ell}$,
returns an $\alpha$-optimal reassembling $(G,\B)$ such that
$\alpha(G,\B) \in \bigOO{\ell}$.
A consequence of this fact is that our algorithm
$\KS$ is not $\alpha$-optimal for the family ${\G}_{\ell}$. 
In the next proposition, we prove a particular case:
We consider the family ${\HH}_{f}$ as defined in the opening
paragraphs of Section~\ref{sect:optimality-of-KS}, where every
ICT is an inter-cycle edge (ICE), and choose a
particular function $f$, called $\bar{f}$, which makes ${\HH}_{\bar{f}}$
a proper subclass of $ {\G}_{\ell}$ for sufficiently large $\ell$, \ie,
${\HH}_{\bar{f}}\subsetneq {\G}_{\ell}$. The $\bar{f}$ in
question is a constant function, specifically here: $\bar{f}(k) = c$
for some constant $c\geqslant 3$ which makes ${\HH}_{\bar{f}}\subsetneq {\G}_{c}$.
Thus, for every $H_{\bar{f},k}\in{\HH}_{\bar{f}}$, the number of
inter-cycle one-edge trees at every level of $H_{\bar{f},k}$ is $c$
and independent of $k$

\begin{proposition} 
\label{thm:non-optimality-of-KS}
\label{prop:non-optimality-of-KS}
Let $c$ be a constant $\geqslant 3$.
There is linear-time $\alpha$-optimal reassembling algorithm $\A$ for
the family ${\HH}_{\bar{f}}$ where $\bar{f}(k) = c$.
Specifically, given an arbitrary $H_{\bar{f},k} \in {\HH}_{\bar{f}}$, algorithm $\A$
returns in time $\bigOO{\size{\vv(H_{\bar{f},k})}}$ an $\alpha$-optimal reassembling
$\B$ of $H_{\bar{f},k}$ such that $\alpha(H_{\bar{f},k},\B) = c+2$.
\end{proposition}

By contrast, given an arbitrary $H_{\bar{f},k} \in {\HH}_{\bar{f}}$ as
input, algorithm $\KS$ returns a reassembling $\B$ of $G$ such that
$\alpha(H_{\bar{f},k},\B) = 2\cdot \OutPlan{E}{H_{\bar{f},k}} = 2k$.
Hence, $\KS$ is not an $\alpha$-optimal reassembling algorithm for the
family ${\HH}_{\bar{f}}$.

\begin{figure}[H] 
\begin{custommargins}{-0.40cm}{-2.2cm}
\begin{center}
   \includegraphics[scale=.5]{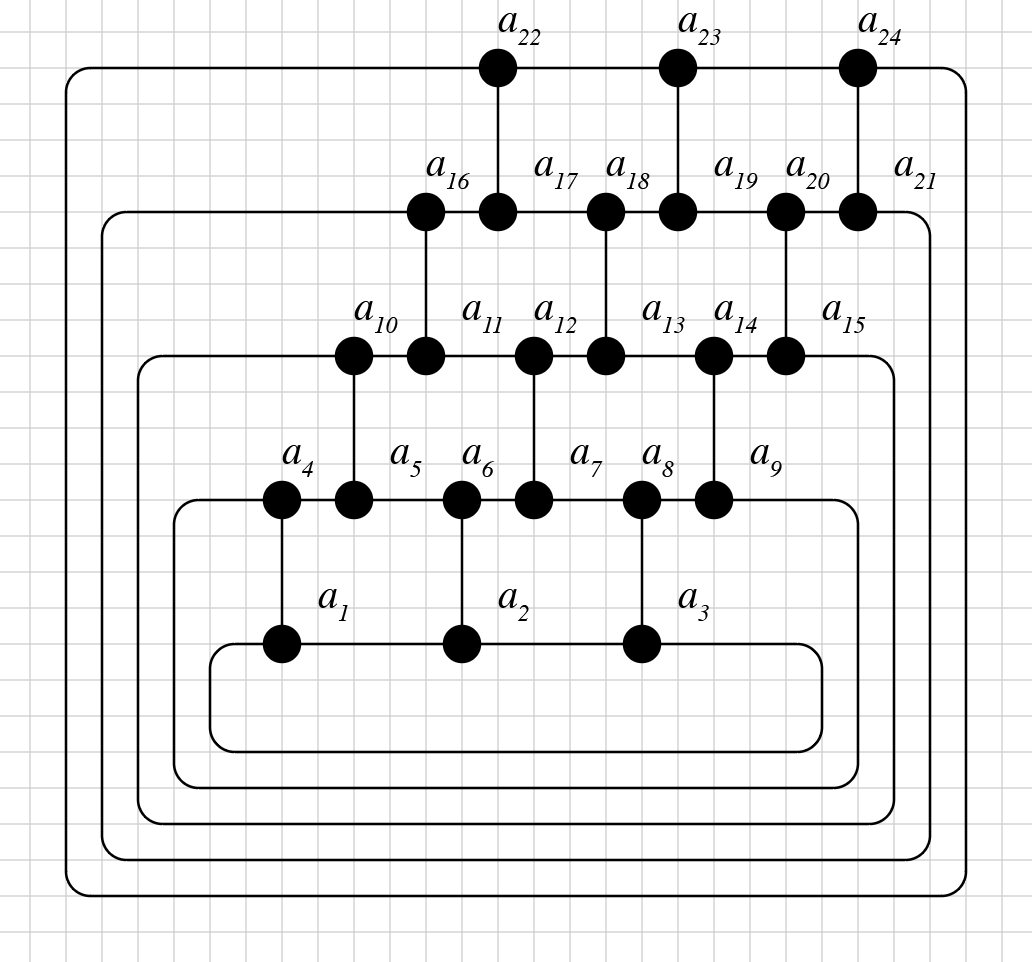}
   \vspace{.5in}
   \includegraphics[scale=.5]{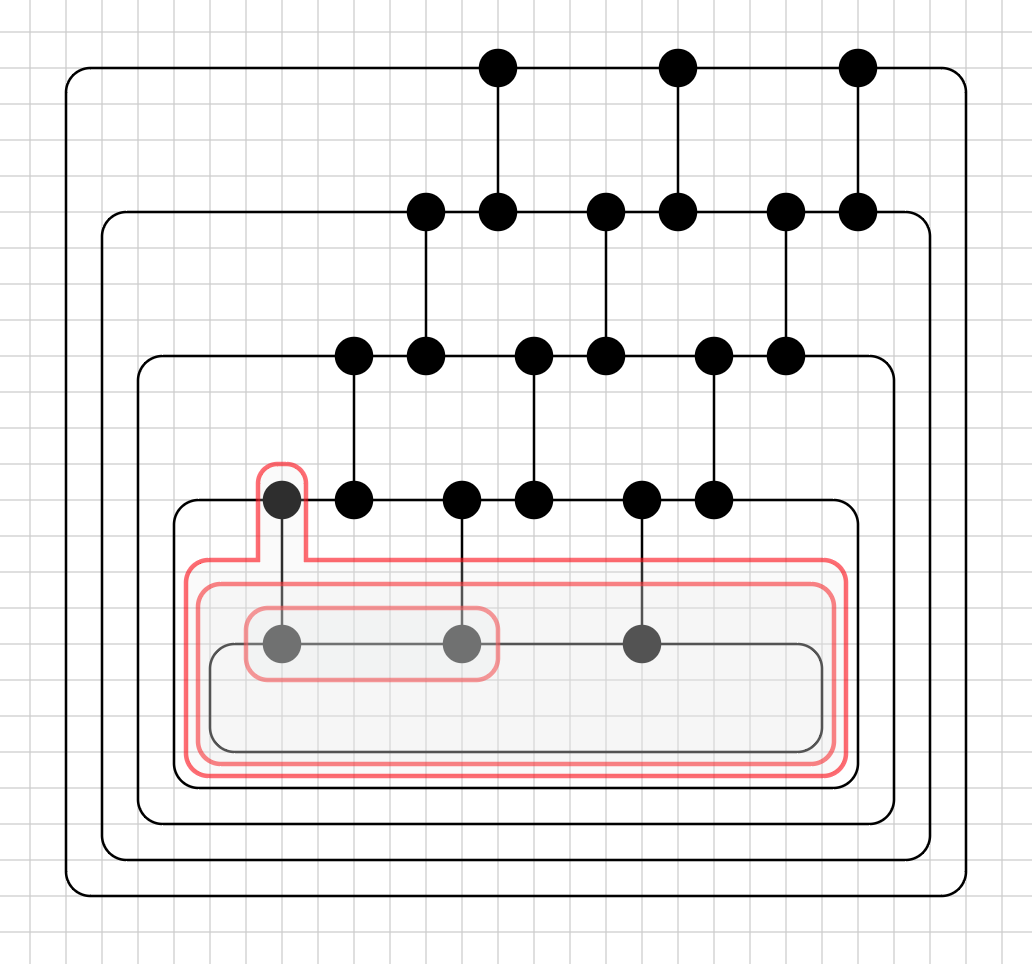}   
\end{center}   
\vspace{-.45in}
   \caption{For Proposition~\ref{prop:non-optimality-of-KS}:
   Graph $H_{\bar{f},k}$ when $k=4$ and $\bar{f}(k) = 3$.\\ The lower figure depicts
   the first three contractions of an $\alpha$-optimal reassembling:\\
   $\set{a_1\; a_2}$, \quad $\set{\set{a_1\; a_2}\; a_3}$,
   \quad and\quad $\set{\set{\set{a_1\; a_2}\; a_3}\; a_4}$.
   }
\label{fig:constant-intercycle-density}
\end{custommargins}
\end{figure}

\begin{sketch}
We restrict the proof to the case $c = 3$, so that $\bar{f}(k) = 3$ for
every $k = \OutPlan{E}{H_{\bar{f},k}}$; the proof
is easily generalized for any $c\geqslant 3$. With this restriction,
there is a total of $6k'$ vertices in $H_{\bar{f},k}\in {\HH}_{\bar{f}}$ where
we pose $k' = k-1$. Let $ \vv(H_{\bar{f},k}) = \Set{a_1,a_2,\ldots, a_{6k'}}$, 
with ``$a_1\; a_2\; a_3$'' be the clockwise sequence of vertices of the innermost 
cycle $L_{k-1}$. (We use the naming conventions for the nested cycles
introduced in Definition~\ref{def:modified}.) Continuing inside-out, 
``$a_4\ \cdots\ a_9$'' is the clockwise sequence of vertices of the next 
cycle $L_{k-2}$, etc., and ``$a_{6k'-2}\;a_{6k'-1}\; a_{6k'}$'' is the clockwise
sequence of vertices of the outermost cycle $L_0$. 
See the top of Figure~\ref{fig:constant-intercycle-density} for the case $k=5$.

It is easy to check that an $\alpha$-optimal reassembling of $H_{\bar{f},k}$
 (not the only one) proceeds as follows:
\[
     \SET{\SET{\SET{\SET{\ldots\Set{\Set{\Set{
       \Set{\Set{\Set{\Set{\Set{a_1, a_2}^4, a_3}^3, a_4}^4, a_5}^5, a_6}^4,
        a_7}^5, a_8}^4, a_9}^3 \ldots a_{6k'-3}}^3, a_{6k'-2}}^4, a_{6k'-1}}^3, a_{6k'}}^0
\]
where the superscript of every cluster is its edge-boundary degree. For example,
the contraction of the edge $\set{a_1\; a_2}$ produces a super vertex/cluster
of degree $4$, \ie, $\Bridges{}{\Set{a_1,a_2}} = 4$, and then the contraction
of the edge $\set{\set{a_1\; a_2}\; a_3}$ produces a super vertex/cluster
of degree $3$, \ie, $\Bridges{}{\Set{a_1,a_2,a_3}} = 3$, etc. The 
first three contractions of this reassembling are depicted
in the bottom of Figure~\ref{fig:constant-intercycle-density} 
for the case $k=5$. 

The resulting reassembling tree $\B$, as just described,
is such that $\alpha(H_{\bar{f},k},\B) = 5$,
whereas algorithm $\KS$ on input $H_{\bar{f},k}$ returns a reassembling tree 
$\KS(H_{\bar{f},k})= {\B}'$ such that $\alpha(H_{\bar{f},k},{\B}') = 2k$.
\end{sketch}

Whereas algorithm $\KS$ proceeds in an
`outside-in' fashion (starting
from the outermost cycles), the reassembling in the preceding proof proceeds
`inside-out' (starting from the innermost cycle). However, the order of the
latter can be reversed to work `outside-in' optimally too.


\clearpage


\section{Conclusion}
\label{sect:conclusion}

We developed a linear-time algorithm $\KS$ for the reassembling of 
$3$-regular plane graphs. Given such a graph $G$ as input, with
$\OutPlan{E}{G} = k \geqslant 2$,%
   \footnote{There is no $3$-regular plane graph $G$ such that
   $\OutPlan{E}{G} = 1$.
   }
algorithm $\KS$ returns a binary
reassembling $\B = \KS(G)$ of $G$ such that $\alpha(G,\B) = 2k$. 

In Section~\ref{sect:algorithm-KS} we proved the correctness of
algorithm $\KS$ and its time complexity, preceded by several examples
illustrating the progression of $\KS$ on several graphs exhibiting
different peculiarities (Section~\ref{sect:examples}).  In
Section~\ref{sect:optimality-of-KS} we showed that our algorithm is
optimal for the class of $3$-regular plane graphs that have a
sufficiently high ``inter-cycle density'' which increases as the
edge-outerplanarity increases (this informal description is made
precise in Section~\ref{sect:optimality-of-KS}).

\subsection{Related Work}

In Section~\ref{sect:graph-carving} we spelled out the connection between
\emph{graph carving} and \emph{graph reassembling}. The first of these two
notions has been studied for many years. A number of results about
\emph{graph carving} are readily translated into results about
\emph{graph reassembling}. We mention two of the most salient.

\begin{proposition}
\label{prep:cubic-time-algorithm}
There exists an algorithm which, given a plane graph $G$ (not necessarily
$3$-regular) as input
with $n$ vertices, returns in time $\bigO{n^3}$ a binary reassembling $\B$
of $\vv(G)$ such that $(G,\B)$ is $\alpha$-optimal.
\end{proposition}

\begin{proof}
This is an immediate consequence of Proposition~\ref{prop:reassembling+carving}
in this report and Theorem 2.1 in~\cite{qian-ping-gu2008}.
\end{proof}

The next proposition is about the existence of a \emph{fixed-parameter linear-time
algorithm} where the parameter-bound not to be exceeded is the $\alpha$-measure.
This is to be contrasted with our algorithm $\KS$, which can be recast (not here)
as a \emph{fixed-parameter linear-time algorithm} where the parameter-bound
not to be exceeded is the edge-outerplanarity. As usual, `linear time' means
`linear as a function of the number $n$ of vertices'.

\begin{proposition}
\label{prep:fixed-parameter-reassembling}
Let $p\geqslant 1$ be fixed. There exists a linear-time algorithm that
checks whether an input graph $G$ (not necessarily plane) has a binary
reassembling $\B$ of its vertices $\vv(G)$ such that
$\alpha(G,\B)\leqslant p$ and, if this is the case, the algorithm
returns a reassembling $\B$ such that $(G,\B)$ is $\alpha$-optimal.
\end{proposition}

\begin{proof}
This is a straightforward consequence of
Proposition~\ref{prop:reassembling+carving} in this report and the main result
in~\cite{thilikos+serna+bodlaender2000} (Theorem 1 and Theorem 2 in Section 6).
\end{proof}

We can also state, in reverse, implications from results on
\emph{graph reassembling}
to results on \emph{graph carving}, illustrated by
the following proposition. Review the definitions related to
\emph{graph carving} in Section~\ref{sect:graph-carving}.

\begin{proposition}
\label{prop:from-reassembling-to-carving}
There exists a linear-time algorithm which, given an arbitrary
$3$-regular plane graph $G$ as input with
$\OutPlan{E}{G} = k \geqslant 2$, returns a routing tree $T$ for $G$
such that $\width{G,T}\leqslant 2k$.

Note, in particular, the value of $\width{G,T}$ in the returned
routing tree $T$ is independent of $n = \size{\vv(G)}$.
\end{proposition}

Moreover, for the family of graphs ${\HH}_{f}$ defined at the beginning of
Section~\ref{sect:optimality-of-KS}, the bound $2k$ in
Proposition~\ref{prop:from-reassembling-to-carving}
is an optimal carving width, by Theorem~\ref{thm:optimality-of-KS}.

\begin{proof}
Straightforward consequence of
Proposition~\ref{prop:reassembling+carving},
Theorem~\ref{thm:algorithm-KS}, and Corollary~\ref{cor:algorithm-KS},
in this report.
\end{proof}

\subsection{Future Research}

One obvious task for the future is to extend our method for 
$3$-regular plane graphs to arbitrary plane graphs.
However, short of such an extension (which seems to require a
considerable effort to formulate with an equal loss in transparency),
we know there is at least one important application, discussed next,
for which there is no explicit need for extending our method beyond
its current form. Other applications are yet to be identified.

Consider an application involving the computation of maximum flows in
networks. We take a flow network to be a capacited directed graph $G$
with no self-loops and two distinguished vertices, a \emph{source
vertex} $s$ and a \emph{sink vertex} $t$, which we identify by the
quadruple $(G,c,s,t)$ where $c:\ee(G)\to\nreals$ is the (non-negative)
capacity function.  A flow $f:\ee(G)\to\nreals$ is feasible if
$f(e)\leqslant c(e)$ for every $e\in\ee(G)$ and $f$ satisfies the
usual flow-conservation condition at every vertex
$v\in\vv(G)- \Set{s,t}$. We write $\size{f}$ to denote the value of
the flow~$f$, which is the excess flow exiting $s$ or, equivalently,
the excess flow entering $t$.

\begin{figure}[H] 
\vspace{-.15in}
\begin{center}
   \includegraphics[scale=.45]{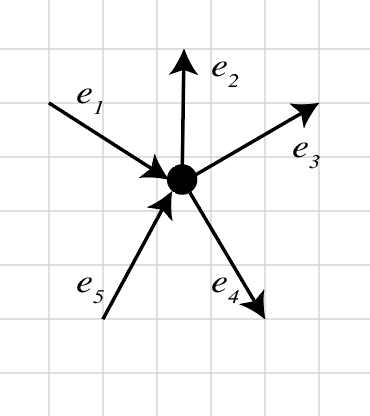}
   \hspace{2cm} 
   \includegraphics[scale=.45]{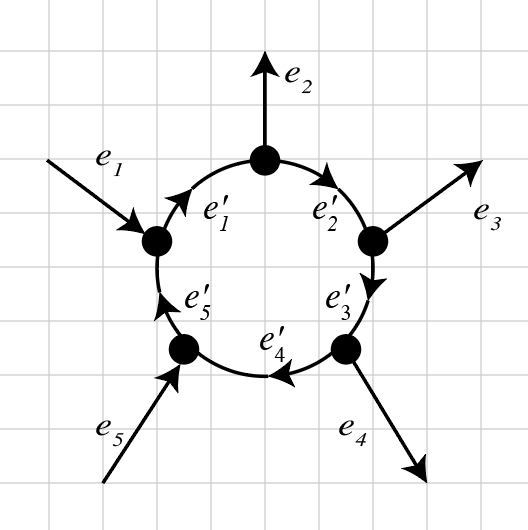}
\end{center}   
\vspace{-.25in}
\caption{
\label{fig:cycle-with-5-incident}
   A vertex of degree $=5$ (on the left) which is transformed into a cycle
   (on the right), \\
   with $5$ vertices each of degree $=3$ and the same incident edges
   $\Set{e_1,e_2,e_3,e_4,e_5}$. 
   }
\vspace{-.15in}
\end{figure}

Given a network $(G,c,s,t)$, we can transform it into an
equivalent $(G^*,c^*,s^*,t^*)$ in time $\bigOO{n}$ where
$n = \size{\vv(G)}$ and every vertex $v\in\vv(G^*)$ has degree $=3$.
Moreover, if $G$ is plane, then so is $G^*$ such that 
$\OutPlan{E}{G} = \OutPlan{E}{G^*}$. The crux of the transformation
$(G,c,s,t)\mapsto (G^*,c^*,s^*,t^*)$ is an expansion of
every $v\in\vv(G^*)$ of degree $\geqslant 4$;
Figure~\ref{fig:cycle-with-5-incident} shows the expansion
of a degree-$5$ vertex.

The details of the transformation $(G,c,s,t)\mapsto (G^*,c^*,s^*,t^*)$
will be in a separate report~\cite{kfoury2018b}, which will also
include a proof of the following result.

\begin{proposition} 
There is a fixed-parameter linear-time
algorithm to compute max flow in planar flow networks $(G,c,s,t)$,
where the parameter-bound not to be exceeded is
$k = \OutPlan{E}{G}$.
\end{proposition} 

\Hide{
\begin{proof}
Let $(G,c,s,t)$ be an arbitrary planar flow network. In linear time, we
can embed $G$ in the plane~\cite{patrignani2013}, and thus assume $G$ is plane.
Further, by Proposition~\ref{prop:max-flow}, in linear time we can transform
$(G,c,s,t)$ into an equivalent biconnected $3$-regular plane flow network,
and thus assume $G$ is a biconnected $3$-regular plane graph.

From the analysis in~\cite{kfoury2018}, given a reassembling $(G,\B)$ of $G$, we
can compute what is a called a \emph{principal typing} of $(G,c,s,t)$, and
therefore a max flow, in time $m\cdot 2^{\bigOO{\delta}}$ where $m = \size{\ee(G)}$ and
$\delta = \alpha(G,\B)$; this is from part 3 in Theorem 4, pages 7-8
in~\cite{kfoury2018}, where what is called the \emph{index} of a
reassembling is its $\alpha$-measure in this report.%
  \footnote{A more precise bound is
  $\bigO{n\cdot 2^{\delta}+ m\cdot\delta\cdot 2^{2\delta}}$ from the details
  at the end of Appendix C, page 24 in~\cite{kfoury2018}. Since $G$
  is a plane graph and $m = \bigOO{n}$, the bound can be further
  simplified as $\bigO{n\cdot\delta\cdot 2^{2\delta}}$.}
By Theorem~\ref{thm:algorithm-KS} above, there is a reassembling
$(G,\B)$ such that
$\alpha(G,\B) \leqslant 2 \cdot\OutPlan{E}{G}$. By Euler's formula
for planar graphs (Theorem 4.2.7 and its corollaries in~\cite{diestel2012}),
we have $m\leqslant 3n-6$. 
Hence, we can compute a max flow in $(G,c,s,t)$ in time
$(3n-1)\cdot 2^{\bigOO{2k}}$ where $k = \OutPlan{E}{G}$.
\end{proof}
}

We conclude with a conjecture which, we believe, can be examined
using the methodology developed in this paper. To make sense of the
conjecture statement, review how we classify and define the level
of \emph{inter-cycle trees} (ICT's) in Definition~\ref{def:modified},
Propositions~\ref{prop:partition} and~\ref{prop:partition-for-3-regular},
and the~\nameref{rem:about-typeA-typeB} on page~\pageref{rem:about-typeA-typeB}.

\begin{conjecture}
\label{conj:one}
There is a function $f: \nats\to\nats$ such that,
for every $3$-regular plane graph $G$, if for every level
$j \in \Set{0,\ldots,k-1}$ where $k =  \OutPlan{E}{G}$
the number of level-$j$ ICT's is $\geqslant f(k)$,
then $\KS$ is an $\alpha$-optimal reassembling algorithm for $G$,
\ie, $\KS$ returns a reassembling $\B = \KS(G)$ such
that $\alpha(G,\B) = 2k$ is optimal.
\end{conjecture}

If Conjecture~\ref{conj:one} is true, it defines a larger class of
$3$-regular plane graphs than Theorem~\ref{thm:optimality-of-KS} for which
$\KS$ is an $\alpha$-optimal reassembling algorithm.
(Theorem~\ref{thm:optimality-of-KS} is about $3$-regular plane
graphs where every ICT is an ICE, an \emph{inter-cycle edge}.)

A final open problem is: Characterize
the class of all $3$-regular plane graphs for which $\KS$ is an
$\alpha$-optimal reassembling algorithm.


\ifTR
\else
\fi


{\footnotesize 
\addcontentsline{toc}{section}{References} 
\bibliographystyle{plain} 
\bibliography{./reassembling}
}

\ifTR
\else
\fi

\newpage

\appendix

\section{Appendix: Pseudocode of Algorithm $\KS$}
  \label{appendix:pseudocode}
  
}

Below is the pseudocode of algorithm $\KS$, of which the full Python
code is an elaboration. The latter can be dowloaded from the website
\href{http://cs-people.bu.edu/bmsisson/}{Graph Reassembling}.
%
The pseudocode is based on the description in Sections~\ref{sect:informal}
\ref{sect:classification}, \ref{sect:contract-ICT-edges},
and~\ref{sect:contract-cycle-edges}. For easier cross-referencing, we 
use the following conventions in the pseudocode:
\begin{enumerate}[itemsep=0pt,parsep=0pt,topsep=0pt,partopsep=0pt]
\item
   Lines in black blodface appear verbatim in the Python code;
   these are headings of function definitions.
\item
   Lines in black Roman characters, which do not start with a hash \#,
   describe each an action that is implemented with several Python
   instructions; they are inserted as comments throughout the Python code
   in order to make the Python code more readable.
\item
   Lines in grey Roman characters, which start with a hash \#, are
   inserted as comments in the Python code; they describe the state of
   the execution at the points where they are inserted.
\item
   Lines in {\color{red}{red}} are related to the collapse operation, lines
   in {\color{green}{green}} 
   are related to the merge operation. We follow the same color convention
   in our examples in Section~\ref{sect:examples}.
\end{enumerate}
\vspace{-.2cm}
{\small
%
  \begin{alltt}
\textbf{def pre-process():}
    let the current E-outerplanarity level be 0
    
    \textcolor{gray}{# by marking edges at each E-outerplanarity layer one at a time}
    \textcolor{gray}{# the trees and cycles can be differentiated}
    initialize every vertex such that all vertices are unmarked
    while the graph has edges left unmarked
        starting at the upper leftmost unmarked vertex,
        traverse the graph by taking the next clockwise edge at each vertex encountered,
        marking each time a vertex is visited,
        until the traversal returns to the original vertex
       
        the edge of this particular E-outerplanarity are now
        defined to be all the edges traversed
        
        for all edges traversed
            if the edge was traversed once
                the edge is part of a cycle
            otherwise
                \textcolor{gray}{# the edge has been marked exactly twice}
                the edge is part of a cycle
             
        for all vertices that have two incident cycle edges
            \textcolor{gray}{# the vertex will have exactly one incident tree edge}
            if the tree edge is on the same E-outerplanarity as the vertex
                then it is an outward cycle vertex
            otherwise
                then it is an inward cycle vertex
            
        to determine the vertices of a particular cycle or tree
            begin a depth-first search at both endpoints of a tree or cycle edge
            and traverse every edge adjacent of the same type
            two cycles or two trees will never share the same vertex, 
            so the depth-first search will traverse the entire tree or cycle and then end
            \textcolor{gray}{# the entire graph will be traversed once as a whole}
            \textcolor{gray}{# all leaf vertices will be visited twice}
            \textcolor{gray}{# while other vertices will be visited once}
            
        increment the E-outerplanarity by 1

\textbf{def process():}
    at each E-outerplanarity level in the graph
        define a queue that holds trees to be collapsed and merged

    prepare_cycle(outermost cycle enclosing the whole graph)

    while any queue is not empty
        pop the oldest tree in the queue at the deepest E-outerplanarity level
        if the tree has not collapsed
            \textcolor{red}{collapse the tree}
        otherwise
            \textcolor{green}{merge the tree}

\textbf{def collapse_type_a(tree):}
    refer to the upper left most vertex of tree as v
    \textcolor{red}{collapse_tree(tree,v)}
    add tree to the queue for this layer so it merges as a type A tree

\textbf{def collapse_type_b(v):}
    refer to the tree v is part of as tree

    refer to the root of tree as u 
    \textcolor{gray}{# note that u and v will never be the same vertex}
       
    \textcolor{red}{collapse_tree(tree,v)}
    add tree to the queue for this layer so it merges as a type B tree


\textbf{def merge_type_a(tree):}
    refer to the cycle enclosing tree as the outer cycle
    refer to the super node that tree is part of as super

    if the outer cycle has no incident trees left uncollapsed
        merge_cycle(the outer cycle, super)
        return

    otherwise, if the outer cycle has a single incident tree left uncollapsed
        prep_incident_tree(the outer cycle, super)
        return

    find a vertex adjacent to tree on the outer cycle that 
        is part of a tree on the same layer as tree
        and is not in the same super vertex as tree

    if that vertex does not exist
        \textcolor{gray}{# there is nothing to do until the outer cycle is ready to merge}
        \textcolor{gray}{because the outer cycle is waiting on a tree that is not adjacent to this tree}
        return

    refer to that vertex as the successor vertex
    refer to the tree that the successor vertex is part of as the successor tree

    if the successor tree is Type-B
        if the successor tree has merged
            \textcolor{gray}{# there is nothing to do because there is nowhere for us to merge}
            \textcolor{gray}{that has not already merged}
            return
        \textcolor{gray}{# the successor vertex will never be the root vertex of the successor tree}
        \textcolor{green}{merge super and the super node of which the successor vertex is a part}

    otherwise, if the successor tree is Type-B or it has not collapsed yet
        \textcolor{green}{merge super and the successor vertex}
        if successor tree is ready to collapse
	\textcolor{gray}{# the successor tree may be either type A or type B}
            add the successor tree to the queue for this layer 
            so it merges with the proper type

    otherwise, 
        \textcolor{gray}{# the successor tree is a type A tree that has collapsed but not yet merged}
        \textcolor{green}{merge super and the super node of which the successor tree is a part}

\textbf{def merge_type_b(v):}
    refer to the super node that v is part of as the super
    refer to the cycle v is a on as cycle
    refer to the clockwise successor of v on cycle as the successor vertex
    refer to the tree v is part of as tree

    marked v as collapsed as it was not marked as collapsed earlier

    if the successor vertex is an outer vertex
        if there is nothing stored with the successor vertex
            refer to the tree that the successor vertex is part of 
            as the successor tree
            \textcolor{green}{merge the super and the successor vertex}
            if the successor tree is the only tree left on cycle uncollapsed
            and it is ready to collapse
                add the successor tree to the queue for this layer 
                so it merges as a type B tree

        otherwise, if there are any trees on cycle left uncollapsed
            \textcolor{green}{merge super and the super node of which the successor vertex is a part}
                

        otherwise,
            \textcolor{gray}{# cycle has been entirely collapsed and is ready to merge}
            merge_cycle(layer_states, layer, rs, c, super)

    otherwise
        \textcolor{green}{merge super and the successor vertex}
        if cycle has 0 or 1 incident trees left uncollapsed 
        and it has not been prepared
            prep_cycle(cycle)

\textbf{def merge_cycle(cycle, super):}
    refer to the cycle enclosing cycle as the outer cycle

    find the clockwise successor tree of this cycle on the layer of the outer cycle

    if there is such a tree
        \textcolor{green}{merge super and the vertex closest to this cycle}

    otherwise, if the outer cycle is the outermost cycle
        the graph has been fully collapsed and no more work is left

    otherwise, if the outer cycle has any incident trees left uncollapsed
        \textcolor{gray}{# it will have exactly one incident tree left uncollapsed}
        prepare_incident_tree(the outer cycle, super)

    otherwise, 
        \textcolor{gray}{# we must recurse upwards to continue}
        merge_cycle(the outer cycle, super)    

\textbf{def prep_incident_tree(cycle, super):}
    \textcolor{gray}{# there is exactly one tree uncollapsed incident to this cycle}
    \textcolor{green}{merge cycle and the root vertex of the incident tree on cycle}
    if the incident tree is ready to collapse
        add the incident tree to the queue for this layer so it as a type B tree

\textbf{def collapse_tree(tree,x):}
    refer to the vertex adjacent to x as v 
    note that x will always be leaf vertex of tree

    if v is a leaf vertex of tree
        assign an index to x
        assign an index to v
        tree is a tree consisting of two vertices
        return the super vertex of x and v

    while v != x:
        if v is a leaf vertex
            assign an index to v
            set v to the predecessor of v
            continue

        refer to the left child of v as l
        refer to the right child of v as l

        if l does not have an index
            v is the predecessor of l
            set v to l
            continue

        if r does not have an index
            v is the predecessor of l
            set v to l
            continue

        note that at this point, l and r may be super vertices, 
        but v is definitely not
        assign an index to v
        \textcolor{red}{collapse l and v into a super vertex}
        \textcolor{red}{collapse the result with r and store it under v}
        set v to the predecessor of v

    assign an index to x
    return the super vertex of x and v

  \end{alltt}
  }


\end{document}